\documentclass[a4paper,11pt]{article}
\usepackage{jheppub} 

\usepackage{jheppub} 
 \usepackage{amssymb,amsthm}
 \usepackage{amsmath,amssymb,amsfonts,bm,amscd}
 \usepackage{graphicx}
 \usepackage{graphics}
 \usepackage{float}
 \usepackage{tikz-cd} 
\usepackage{url}

 \usepackage{caption, subcaption}
 
 \usepackage{tikz}

\def\beq#1\eeq{\begin{align}#1\end{align}}

\title{Classification of  rank one 5d $\mathcal{N}=1$ and 6d $(1,0)$  SCFTs}

\author[a,b]{Dan Xie}

\affiliation[a]{Yau Mathematics science center, Tsinghua University, Beijing, 100084, China}
\affiliation[b]{Department of Mathematics, Tsinghua University, Beijing, 100084, China}

\abstract{This paper gives a classification of rank one 5d $\mathcal{N}=1$ and 6d $(1,0)$ SCFTs.  The idea is to compactify 5d theory on $S^1$ and 6d theory on $T^2$ to get effective 4d $\mathcal{N}=2$ theory.  These compactified theories all have 
 a 4d $\mathcal{N}=2$ Coulomb branch whose solution can be described by mixed Hodge module (MHM). In the rank one case every Coulomb branch solution is related to rational elliptic surface with a section, whose
classification is complete.  So the classification is then reduced to pick up a subset from the data base of rational elliptic surface by imposing various physical constraints. 
The crucial new input is that the singular fiber at infinity determines which dimension the UV theory lives.
Various physical properties such as  flavor symmetry, one form symmetry, BPS quiver and BPS spectrum, and RG flow are studied. 
D7 brane configurations for those theories are found which are very useful in studying them. The generalizations to higher rank theories will also be highlighted.}

\begin{document} 
\maketitle
\flushbottom

\section{Introduction}
There are lots of interests in studying supersymmetric theory with eight supercharges in various dimensions, such as 3d $\mathcal{N}=4$ theory \cite{Intriligator:1996ex}, 4d $\mathcal{N}=2$ theory \cite{Seiberg:1994rs,Seiberg:1994aj},
5d $\mathcal{N}=1$ theory \cite{Seiberg:1996bd}, and 6d $(1,0)$ theory \cite{Seiberg:1996qx}. Theories with eight supercharges are less rigid than theories with 16 supercharges \cite{Seiberg:1997ax}, and so they have more interesting 
dynamical properties while still remain  tractable.
These theories share some common features: such as the existence of moduli space of vacua like Higgs branch, Coulomb branch (tensor branch in 6d), and massive BPS states on the Coulomb branch, etc.
They do have some important differences, for example: the type of supersymmetric preserving deformations are different \cite{Cordova:2016emh}. 

One can relate theories in different dimensions by putting higher dimensional theory on compact manifold. To preserve all the
supersymmetry, one need to use the torus $T^d$ compactification, and  interesting properties of higher dimensional theory can be learned
by looking at the effective lower dimensional theory.
Let's put 5d $\mathcal{N}=1$ theory on $S^1$ \cite{Nekrasov:1996cz} and 6d $(1,0)$ theory on $T^2$ \cite{Ganor:1996pc} (these are called KK theories),  so the low energy theory has effective 4d $\mathcal{N}=2$ supercharges. 
The KK theory could have a $\mathcal{N}=2$ Coulomb branch where the low energy theory can be solved. While the Coulomb branch solution of any $\mathcal{N}=2$ theory have common structures, the details are affected by higher dimensional theory, for example,
 the effective photon couplings could receive contribution of higher dimensional BPS states, and the BPS particle also carry winding mode charge along the  $T^d$ direction, etc.
 This is the reason that one can use the Coulomb branch solution of the KK theory to classify higher dimensional theory.
 
 The main purpose of this paper is to use the Coulomb branch structure of the compactified theory to classify rank one 5d and 6d theories with eight supercharges. 
 The Coulomb branch solution for 4d $\mathcal{N}=2$ $SU(2)$ gauge theory has been solved elegantly by Seiberg-Witten (SW) \cite{Seiberg:1994rs,Seiberg:1994aj}. 
 The Coulomb branch for compactified higher dimensional theory were soon studied \cite{Nekrasov:1996cz,Ganor:1996pc}. The solutions are often described by a 
 family of algebraic varieties fibered over the Coulomb branch. In practice, the solutions are most easily found by using string theory motivated methods \cite{Witten:1997sc}, and the relation to integrable system \cite{Martinec:1995by,Donagi:1995cf}. 
 
\begin{figure}
\begin{center}

\tikzset{every picture/.style={line width=0.75pt}} 

\begin{tikzpicture}[x=0.55pt,y=0.55pt,yscale=-1,xscale=1]

\draw    (298,180) -- (309,191.72) ;
\draw    (298,192) -- (307,180.72) ;

\draw    (375,300) -- (386,311.72) ;
\draw    (375,312) -- (384,300.72) ;

\draw    (173,321) -- (184,332.72) ;
\draw    (173,333) -- (182,321.72) ;

\draw    (197.5,147.72) .. controls (177.5,170.72) and (243,281.72) .. (235,329.72) ;
\draw  [color={rgb, 255:red, 208; green, 2; blue, 27 }  ,draw opacity=1 ] (345,306.72) .. controls (345,295.67) and (361.57,286.72) .. (382,286.72) .. controls (402.43,286.72) and (419,295.67) .. (419,306.72) .. controls (419,317.76) and (402.43,326.72) .. (382,326.72) .. controls (361.57,326.72) and (345,317.76) .. (345,306.72) -- cycle ;
\draw    (357.5,281.72) .. controls (441.5,20.72) and (333.5,256.72) .. (382,306.72) ;
\draw  [color={rgb, 255:red, 208; green, 2; blue, 27 }  ,draw opacity=1 ] (266,184.75) .. controls (266,173.7) and (282.57,164.75) .. (303,164.75) .. controls (323.43,164.75) and (340,173.7) .. (340,184.75) .. controls (340,195.8) and (323.43,204.75) .. (303,204.75) .. controls (282.57,204.75) and (266,195.8) .. (266,184.75) -- cycle ;
\draw    (283,55.75) .. controls (223,47.75) and (298,138.75) .. (303,184.75) ;
\draw    (210,401) -- (221,412.72) ;
\draw    (210,413) -- (219,401.72) ;

\draw    (374,386) -- (385,397.72) ;
\draw    (374,398) -- (383,386.72) ;

\draw   (128.75,359) .. controls (128.75,262.76) and (206.76,184.75) .. (303,184.75) .. controls (399.24,184.75) and (477.25,262.76) .. (477.25,359) .. controls (477.25,455.24) and (399.24,533.25) .. (303,533.25) .. controls (206.76,533.25) and (128.75,455.24) .. (128.75,359) -- cycle ;

\draw (275,171.4) node [anchor=north west][inner sep=0.75pt]    {$\infty $};
\draw (114,115.4) node [anchor=north west][inner sep=0.75pt]    {$H\left( F^{\bullet } ,W^{\bullet },Q(\cdot,\cdot)\right)$};
\draw (361,100.4) node [anchor=north west][inner sep=0.75pt]    {$H_{s\ \ \ \ }\left( F^{\bullet } ,W^{\bullet }\right)$};
\draw (360,75.4) node [anchor=north west][inner sep=0.75pt]    {$H_{lim}\left( F^{\bullet } ,W^{\bullet }(N)\right)$};
\draw (360,125.4) node [anchor=north west][inner sep=0.75pt]    {$H_{van}\left( F^{\bullet } ,W^{\bullet }(N)\right)$};
\draw (427,305.4) node [anchor=north west][inner sep=0.75pt]    {$T$};
\draw (248,22.4) node [anchor=north west][inner sep=0.75pt]    {$H_{\infty }\left( F^{\bullet } ,W^{\bullet }(N_\infty)\right)$};
\draw (450,396.4) node [anchor=north west][inner sep=0.75pt]    {$u$};

\end{tikzpicture}
\end{center}
\caption{The structure of $\mathcal{N}=2$ Coulomb branch: 1): at generic point, one has a vector space with a mixed Hodge structure; 2): at special point, there are three vector spaces and all of them carry mixed Hodge structure; and there is a monodromy group $T$ acting on these vector spaces;  3): One can also have a vector space at $\infty$ of moduli space.}
\label{coulomb1}
\end{figure}
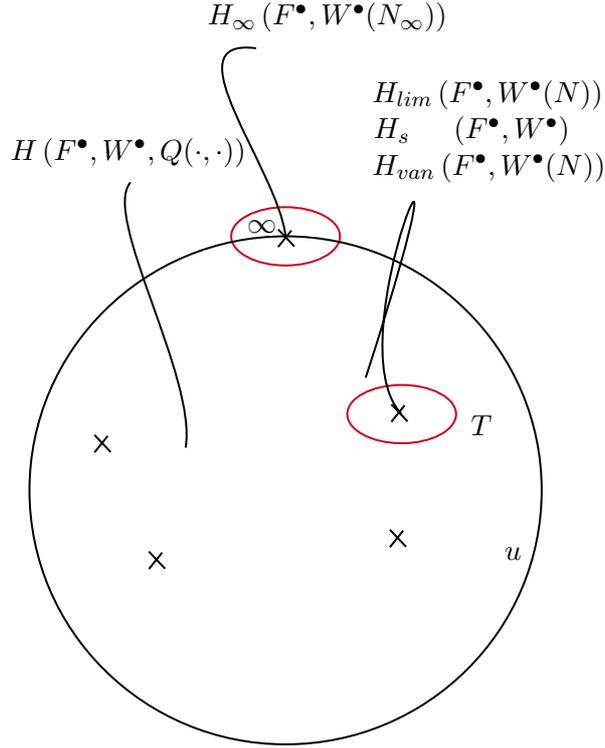

 More generally, It was noticed in \cite{Xie:2021hxd,Xie:2022} that the general Coulomb branch solution is given by the so-called mixed Hodge module (MHM) over the Coulomb branch. The geometric solutions 
 actually fit into this description by looking at the cohomology groups of the algebraic variety. Lets now  briefly review MHM \cite{saito1990mixed} and describe how to extract the physics at each vacua of Coulomb branch from MHM.
  Over the \textbf{generic} vacua, the MHM is described by variation of mixed Hodge structure \cite{steenbrink1985variation}; 
It is described by a flat holomorphic vector bundle with extra structures over very fiber $H$: a) a weight filtration $W^\bullet$, which ensure that  the flavor  and the electric-magnetic  charge can be distinguished;
 b) a Hodge filtration $F^\bullet$ and a polarization $Q(\cdot, \cdot)$, which ensure that  positive photon couplings $\tau_{ij}(u)$ will be derived. Over the \textbf{special} vacua, besides the vector space which is used to describe
the abelian part of the low energy theory, two extra vector spaces ( called nearby cycle and vanishing cycle) are defined using the limit behavior  of nearby vector spaces,
and both of them carry mixed Hodge structure; The vanishing cycle can be used to find the information of the interacting theory, i.e. the spectrum of Coulomb branch operator. 
Finally, at the \textbf{infinite} point of the Coulomb branch, a vector space together with 
a limit MHS can also be defined, from which one can find the properties of the UV theory. The structure of Coulomb branch solution is shown in figure. \ref{coulomb1}. 

Based on known Coulomb branch solutions, several assumptions on MHM for $\mathcal{N}=2$ Coulomb branch solution are proposed: First, the weight filtration has the structure $0=W_0\subset W_1\subset W_2=H$ for generic vacua, namely
there are only two  nontrivial weights; Second,  the monodromy group satisfies the relation $(T^k-I)^2=1$, namely, the maximal size of Jordan block is two; Finally, 
 the monodromy group acts trivially on the weight two part of generic fiber (the flavor 
charge): $T|{Gr_2^W}=I$. These assumptions simplified the analysis of the Coulomb branch solution of $\mathcal{N}=2$ theory.

 In this paper, the Coulomb branch is taken to be rank one, i.e. the weight one part of $H$ is two dimensional. Because the monodromy action on weight two part is trivial,
we could just look at weight one part on $u$ plane (with mass parameters fixed). So eventually there is a rank two  MHM over $\mathbb{P}^1$ (since we include the $\infty$
in our definition of the solution). The classification of the Coulomb branch solution then goes as follows:
\begin{enumerate}
\item \textbf{Local singularity}:  Let's first analyze the local behavior of MHM. First, the monodromy group (topological data) is classified by a conjugacy class of $SL(2,Z)$ group satisfying the condition $(T^2-I)^2=0$. All such conjugacy classes are classified and actually coincide with the list of Kodaira's singularity in the study of
degeneration of elliptic surface. These singularities
are labeled as $I_n, I_n^*, II,III, IV, II^*, III^*, IV^*$ \cite{ barth2015compact}. The holomorphic data is given by the period mapping, and in the rank one case, there is a holomorphic function $\tau(u)$ and the related $J$ invariant $j(u)$ \cite{carlson2017period}. 
\item \textbf{Global constraints}: There are global constraints on the monodromy and holomorphic data. In the rank one case, there is a nice correspondence with rational elliptic surface \cite{schutt2019elliptic}. This correspondence is extremely useful
as the rational elliptic surface has been classified \cite{persson1990configurations,miranda1990persson}, so there is a data base to work with.
\item \textbf{Singular fiber at infinity}: The next question is the singular type appearing at \textbf{$\infty$} which reflects the UV properties. This data reflects what dimension the UV theory lives: 
\begin{itemize}
\item For 4d theory, only type $ I_n^*,II, III, IV, II^*, III^*, IV^*$ singularities can be put at infinity.
\item For 5d  KK theory, only type $ I_n,~n\geq1$ singularities can be put at infinity.
\item For 6d KK  theory,  only type $ I_0$ singularities can be put at infinity.
\end{itemize}
\item  \textbf{Generic deformations}:  For the purpose of classification, the \textbf{bulk} singularity is restricted to have just one dimensional Coulomb branch deformations (These include the expectation values of Coulomb branch operator, mass parameters, and relevant deformations). Essentially we consider  co-dimensional one singularity at the full generalized Coulomb branch. This constraint
removes the appearance of type  $II, III, IV$ singularities at the bulk, as the scaling dimension of Coulomb branch operator associated with them is less than two, and so it has at least two dimensional deformation spaces. 
Those singularities appear in the generic deformations are called undeformable singularities \footnote{One must be careful that just specifying the singularity type does not specify the low energy theory. For example, just specifying a $I_n$ singularity does not tell us what is the low energy theory, one need more data, i.e.e the weight two part of the MHM to determine the IR theory.}.

\item \textbf{Dirac quantization}: Let's now assume that only type $I_n$  (undeformable) singularities appear in the bulk, and assume that there is a massless BPS hypermultiplet with charge $\sqrt{n}(1,0)$ (in proper duality frame) at the singularity. This BPS particle is stable at any given point of the Coulomb branch.
The Dirac quantization condition requires that the Dirac pairing between these massive BPS particles to be integral.  This condition means for any two $I_{n_i}$ and $I_{n_j}$ singular fibers at bulk, the ratio ${n_i\over n_j}$ should be square. 

\item \textbf{Base change}:  The base change method \cite{Caorsi:2018ahl} of rational elliptic surface (for 4d theory this is interpreted as discrete gauging \cite{Argyres:2016yzz}) is used to get other type of undeformable singularities at the bulk. Here we start with configuration $B$ with only type $I_n$ bulk singularities, and use a $Z_n$ action of $B$
to get another configurations $B^{'}$ satisfying: a) the singular fiber of $B^{'}$ is related to $B$  in a specific way, see formula. \ref{4dbase} for 4d case, formula. \ref{5dbase} for 5d case; b): no appearance of type $II,III,IV$ singular fiber in the bulk. 
\end{enumerate}
Using above guidelines and the data base \cite{persson1990configurations} for rational elliptic surface and the base change maps \cite{karayayla2012classification},  the complete list of theories are reported in table. \ref{4d} for 4d theory, and table. \ref{5d} for 5d theory, and table. \ref{6d} for 6d theory.
The physical properties of these theories are summarized in table. \ref{4dfull} for 4d theory, table. \ref{5dfull} for 5d theory, and table. \ref{6dfull} for 6d theory.

While the properties of rational elliptic surface is quite useful in classifying the theories, we find that the D7 brane configuration \cite{DeWolfe:1998eu,Fukae:1999zs} can play an amazing role in further analyzing those theories. 
A D7 brane configuration for each of the theory is constructed, and the string junctions of them are used to study the flavor symmetry, BPS spectrum, BPS quiver, etc for those theories. 

This paper is organized as follows: in section two we review the mixed Hodge structure for $\mathcal{N}=2$ Coulomb branch solution; In section 3 we classify rank one 4d $\mathcal{N}=2$ theory by using the correspondence with rational elliptic surface (see the known results in \cite{Argyres:2015ffa,Argyres:2015gha,Argyres:2016xmc,Argyres:2016xua,Caorsi:2018ahl}); section 4
discusses the D7 brane configurations for 4d theory, and  string junctions are used to compute the flavor symmetry, BPS quiver, BPS spectrum, RG flow etc; section 5 gives a classification for rank one 5d $\mathcal{N}=1$ SCFT; section 6 gives 
a classification for rank one 6d $(1,0)$ theory. Finally, a conclusion is given.

\newpage
\section{Mixed Hodge module and $\mathcal{N}=2$ Coulomb branch solution}
There are several important goals of solving the Coulomb branch of a $\mathcal{N}=2$ theory:
\begin{enumerate}
\item At a generic point, the low energy theory is described by $U(1)^r$ abelian gauge theory, free hypermultiplets, and possibly interacting SCFT whose 
Coulomb branch deformation is trivial \footnote{This picture of Coulomb branch might be found by looking at Type IIB string theory on a CY manifold $X$: $X$ usually has non-trivial three cycles and so one get free vector multiplets; If $X$ has non-trivial four cycles, then one get free  hypermultiplets; if $X$ has a rigid singularity (singularity admits no complex deformation), then one get interacting SCFT with no Coulomb branch. The generic structure on Higgs branch is found in \cite{Xie:2019vzr}.}.  We'd like to determine those three components and their physical properties; An important goal is to  determine the effective coupling for the $U(1)^r$ gauge theory, here $r$ is called the rank of the theory.
\item At a special point, new massless degrees of freedom appear and we'd like to determine the effective low energy theory, which could be IR free gauge theory, or SCFT, or the direct sum of
them.
\item The new massless degrees of freedom at singularity come from massive BPS particles at the generic point, and it is important to find the spectrum of stable BPS particles and their central charges. 
\end{enumerate}

The Coulomb branch solution for $SU(2)$ gauge theory was solved in an elegant way by Seiberg and Witten \cite{Seiberg:1994rs,Seiberg:1994aj}.  They 
solved the theory by finding a family of algebraic curves $F(x,y, u,m, \Lambda)=0$ (Here $u$ parameterizes the Coulomb branch, $m$ the mass parameters,  and $\Lambda$ the dynamical generated scale.), and a SW differential $\lambda$ is also needed.
The physical information is extracted as follows: a):  At the generic point of the $u$ plane, the SW curve $F_u$ is smooth and the low energy theory is just $U(1)$ gauge theory; The photon coupling is given by the complex structure of the curve $F_u$; 
b) At the special point, the $SW$ curve $F_u$ becomes singular, and  the low energy theory is $U(1)$ gauge theory coupled with one massless hypermultiplet, which comes from the massive BPS particle at generic point; c) Finally, the central charge for the 
BPS particle is given as $Z=na+an_Da_D+\sum S_i m_i$, here $a, a_D, m_i$ are defined by doing period integral of the SW differential: $a=\int_A \lambda, a_D=\int_B \lambda,m_i=\int_{\Omega_i} \lambda$ ($A,B,\Omega_i$ are one cycles on Riemann surface $F_u$).
One of the crucial insights of  \cite{Seiberg:1994rs,Seiberg:1994aj} is the electric-magnetic duality of the $U(1)$ gauge theory, which is encoded in the geometry automatically: the complex structure of the elliptic curve $F_u$ has the  $SL(2,Z)$ invariance.

\subsection{Mixed Hodge module}
Generally one solved the Coulomb branch of a $\mathcal{N}=2$ theory by finding a family of algebraic varieties $F(z_i, m_i,u_i,\lambda_i)=0$ and a SW differential $\lambda$, here $u_i$ denotes the expectation value of Coulomb branch operators,
$m_i$ the mass parameters, and $\lambda_i$ the coupling constants including the exact marginal deformations and the relevant deformations.  It was observed in \cite{Xie:2021hxd,Xie:2022} that one need the so-called mixed Hodge structure 
to understand the low energy theory at the non-generic vacua.  In fact, the Coulomb branch solution could be represented as a mixed Hodge module over the  generalized Coulomb branch (parameter space including Coulomb branch operators, masses,
relevant and marginal couplings). 

Let's now describe aspects of the mixed Hodge module which are relevant for $\mathcal{N}=2$ solution, here the rank of the theory is $r$ and the flavor symmetry has rank $f$.
First, at the generic points of the Coulomb branch, there is a \textbf{flat} \footnote{The flat structure gives an integrable connection which is required for the definition of the mixed Hodge module.} holomorphic vector bundle whose rank is $2r+f$. To get the information of the low energy theory, two extra structures
are needed on the fiber $H$: 
\begin{itemize}
\item  A mixed Hodge structure, namely a Hodge filtration and a weight filtration; The weight filtration is an increasing filtration which takes the following form \footnote{If the SW geometry is given by a three dimensional variety \cite{Xie:2015rpa}, then the maximal weight is 4. These MHS could be brought to the form presented here by doing a Tate twist.}:
\begin{equation}
\{0\}=W^0\subset W^1\subset W^2= H;
\end{equation}
so we have two quotient spaces $Gr_1^W=W^1/W^0,~~Gr_2^W=W^2/W^1$, with dimension $dim(Gr_1^W)=2r,~dim(Gr_2^W)=f$. The weight filtration is needed so that we can separate the electric-magnetic part and the flavor part of the central charge: $Gr_1^W$ gives the
electric-magnetic charge, and $Gr_2^W$ gives the flavor charge. 
The Hodge filtration is a decreasing filtration and takes the following form
\begin{equation}
H=F^0\supset F^1;
\end{equation}
So in our case,  two holomorphic sub-bundles  $W^1$ and $F^1$ are needed.  The weight filtration and Hodge filtration together defines a so-called Mixed Hodge structure, and Hodge decomposition takes 
the form $Gr_1^W=H^{1,0}\oplus H^{0,1} $ and $Gr_2^W= H^{1,1}$, with dimension $h^{1,0}=h^{0,1}=r$ and $h^{1,1}=f$;

\item A polarization $Q(\cdot, \cdot)$ (which satisfies Riemann-Hodge bilinear relations on $Gr_1^W$ and acts trivially on $Gr_2^W$) on  $H$ so that  positive definite coupling constants can be defined. In fact, 
 a period matrix $Z_{ij}$ which is symmetric and satisfies the condition $Im(Z)>0$ can be defined using the polarization.
\end{itemize}

At the singular point \footnote{The Coulomb branch is not singular, but the physics is different from that of the generic point of the Coulomb branch.}, there is also a vector space $H_s$ whose dimension is 
smaller than $H$, so the mathematical structure is not the vector bundle which is more familiar to physicists. The physics of the abelian gauge theory at singular point is described by $H_s$.
The crucial point of the mixed Hodge module is that one can define two more vector spaces at the singular point. The first 
is the so-called nearby cycle $H_{lim}$ which can be thought of as the limiting objects for the nearby vector spaces. There is a mixed Hodge structure on $H_{lim}$ which is quite different from that of the generic fiber described earlier.
The weight filtration is now determined by the nilpotent part $N$ of the monodromy group $T$ around the singularity. 

For the known solution, the monodromy group $T$ satisfies the following condition 
\begin{equation}
(T^k-1)^2=1.
\label{monodromytheorem}
\end{equation}
Namely the maximal size of the Jordan block is two, and the eigenvalue satisfies $\lambda^k=1$. We conjecture that this is true for the Coulomb branch solution of any $\mathcal{N}=2$ field theory. Furthermore, if we restrict the monodromy on the weight two part of the generic fiber, 
its action is trivial
\begin{equation}
T|_{Gr_2^W}=I.
\end{equation}
What this implies is that the monodromy matrix takes the form
\begin{equation*}
T=\left[\begin{array}{cc}
I&0\\
*&M\\
\end{array}\right].
\end{equation*}
and $M$ acts on weight one part. Finally, $H_s$ and $H_{lim}$ can be used to define a third vector space called vanishing cycle $H_{van}$. All of these three spaces carry mixed Hodge structure, and they form an exact sequence of mixed Hodge structure.  

Using the limit mixed Hodge structure (let's assume $H_s=0$), one can define 
a set of rational numbers $(\alpha_1,\alpha_2,\ldots, \alpha_s)$ called spectrum \cite{kulikov1998mixed}, and its relation to the eigenvalue is given as 
\begin{equation*}
\lambda_i=\exp(2\pi i \alpha_i).
\end{equation*}
The monodromy group acts on the vector space $H_{lim}$, and so it has the decomposition $H_{lim}=\oplus_\lambda H^{lim}_\lambda$. The limit Hodge filtration defines a filtration on $H_\lambda$: $F^0(H_\lambda)\supset F^{1}(H_\lambda)$. Now for
a basis element $e_i$ in $H_\lambda$, a spectral number $\alpha$ is defined as
\begin{equation*}
\begin{cases}
& e_i\in F^1(H_\lambda),~~~~~~~~~~~-1<\alpha\leq 0 \\
& e_i \in F^0(H_\lambda)/F^1(H_\lambda),~~~~0<\alpha\leq 1
\end{cases}
\end{equation*}
here $\exp(2\pi \alpha)=\lambda$.
An important consistent condition is that the spectral numbers are in pair
\begin{equation*}
\alpha_i+\alpha_j=0.
\end{equation*}
One can find the Coulomb branch spectrum from the spectral numbers as follows. Assume there is a minimal spectrum number $\alpha_{min}$, then for a spectral number $\alpha_i$, one associate a Coulomb branch scaling dimension as \cite{Xie:2015xva}
\begin{equation}
[u_i]={1+\alpha_{min}-\alpha_i\over 1+\alpha_{min}}.
\label{scale}
\end{equation}
So if $\alpha_i=0$, then $[u_i]=1$ which gives a mass parameter. The maximal scaling dimension is given as ${1\over 1+\alpha_{min}}$. 

While the above analysis is carried for the finite points of the Coulomb branch, it is possible to do the similar computation for the $\infty$ point on the moduli space, and the MHS at $\infty$ is useful to extract 
information for UV theory. 
The general structure of the Coulomb branch solution is summarized in figure. \ref{coulomb1}.

\textbf{SW differential and BPS spectrum}: 
To study the massive BPS spectrum,  a $Z$ module $H_Z$ at each generic point is needed, so that 
$H_Z\otimes \mathbb{C}=H$. The monodromy group $T$ on weight one part is now an element of $Sp(2r, \mathbb{Z})$. Given a SW differential $\lambda$ and
 period integral can be defined, so that the central charge for the BPS particles can be found. However, the determination of  BPS spectrum is a much more difficult problem.

The SW differential $\lambda$ is a section of the holomorphic vector bundle.  Because of the Hodge filtration, the weight one part $Gr_1^W$ has a Hodge decomposition $Gr_1^W=H^{1,0}\oplus H^{0,1}$. 
Let's choose the coordinate of the Coulomb branch as $u_i$ (these are the expectation values of Coulomb branch operators), $\lambda$ has to satisfy the condition \cite{Xie:2021hxd}:
\begin{equation}
\partial_{u_i} \lambda \subset H^{1,0}.
\label{differential}
\end{equation}
so that a special Khaler structure on the Coulomb branch can be defined.

\section{Classification of rank one 4d $\mathcal{N}=2$ theory}
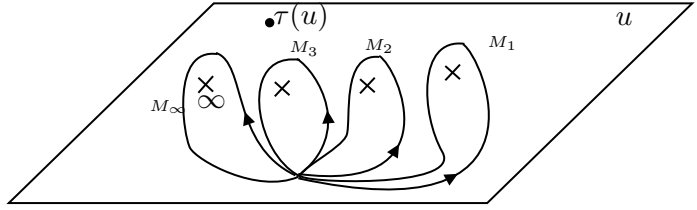
\begin{figure}
 \begin{center}

\tikzset{every picture/.style={line width=0.75pt}} 

\begin{tikzpicture}[x=0.55pt,y=0.55pt,yscale=-1,xscale=1]

\draw   (208.5,160.22) -- (534,160.22) -- (394.5,297.22) -- (69,297.22) -- cycle ;
\draw    (198,211.21) -- (208,220.76) ;
\draw    (198,221.59) -- (208,210.14) ;

\draw    (264,278.22) .. controls (220,257.22) and (231,201.22) .. (210,194.22) ;
\draw [shift={(229.55,234.67)}, rotate = 70.54] [fill={rgb, 255:red, 0; green, 0; blue, 0 }  ][line width=0.08]  [draw opacity=0] (8.93,-4.29) -- (0,0) -- (8.93,4.29) -- cycle    ;
\draw    (266,279.22) .. controls (238,291.22) and (196.84,267.94) .. (192.32,259.39) .. controls (187.79,250.84) and (178.96,190.37) .. (211,195.05) ;
\draw    (250,214.22) -- (260,223.22) ;
\draw    (250,224) -- (260,213.22) ;

\draw    (265,279.22) .. controls (305,249.22) and (279,207.22) .. (265,198.22) ;
\draw [shift={(285.76,233.64)}, rotate = 84.99] [fill={rgb, 255:red, 0; green, 0; blue, 0 }  ][line width=0.08]  [draw opacity=0] (8.93,-4.29) -- (0,0) -- (8.93,4.29) -- cycle    ;
\draw    (266,278.22) .. controls (258.89,274.43) and (252.84,267.65) .. (248.32,259.6) .. controls (240.06,244.91) and (236.87,225.96) .. (241.51,213.13) .. controls (245.05,203.33) and (253.14,197.09) .. (267,199) ;
\draw    (308,212.22) -- (318,221.22) ;
\draw    (308,222) -- (318,211.22) ;

\draw    (266,277.22) .. controls (370,291.22) and (335,205.22) .. (321,196.22) ;
\draw [shift={(334.59,256.78)}, rotate = 120.56] [fill={rgb, 255:red, 0; green, 0; blue, 0 }  ][line width=0.08]  [draw opacity=0] (8.93,-4.29) -- (0,0) -- (8.93,4.29) -- cycle    ;
\draw    (268,280.22) .. controls (260.89,276.43) and (296,260.22) .. (299,250.22) .. controls (302,240.22) and (295,194.22) .. (321,197) ;
\draw    (366,203.22) -- (376,212.22) ;
\draw    (366,213) -- (376,202.22) ;

\draw    (266,280.22) .. controls (443,319.22) and (392,197.22) .. (378,188.22) ;
\draw [shift={(374.56,277.18)}, rotate = 151.14] [fill={rgb, 255:red, 0; green, 0; blue, 0 }  ][line width=0.08]  [draw opacity=0] (8.93,-4.29) -- (0,0) -- (8.93,4.29) -- cycle    ;
\draw    (268,278.22) .. controls (263,283.22) and (378.55,288.12) .. (364,262.22) .. controls (349.45,236.32) and (346.96,183.59) .. (379,188) ;
\draw  [color={rgb, 255:red, 0; green, 0; blue, 0 }  ,draw opacity=1 ][fill={rgb, 255:red, 0; green, 0; blue, 0 }  ,fill opacity=1 ] (244,173.5) .. controls (244,172.12) and (245.12,171) .. (246.5,171) .. controls (247.88,171) and (249,172.12) .. (249,173.5) .. controls (249,174.88) and (247.88,176) .. (246.5,176) .. controls (245.12,176) and (244,174.88) .. (244,173.5) -- cycle ;

\draw (195,221.4) node [anchor=north west][inner sep=0.75pt]    {$\infty $};
\draw (247,157.4) node [anchor=north west][inner sep=0.75pt]    {$\tau ( u)$};
\draw (480,164.4) node [anchor=north west][inner sep=0.75pt]    {$u$};
\draw (163,224.4) node [anchor=north west][inner sep=0.75pt]  [font=\tiny]  {$M_{\infty}$};
\draw (258,184.4) node [anchor=north west][inner sep=0.75pt]  [font=\tiny]  {$M_{3}$};
\draw (309,182.4) node [anchor=north west][inner sep=0.75pt]  [font=\tiny]  {$M_{2}$};
\draw (393,180.4) node [anchor=north west][inner sep=0.75pt]  [font=\tiny]  {$M_{1}$};

\end{tikzpicture}

 \end{center}
 \caption{The moduli space of rank one $\mathcal{N}=2$ theory, here the mass parameters are fixed.}
 \label{rankone}
 \end{figure}

Let's now study 4d rank one $\mathcal{N}=2$ theory, i.e. $r=1$.  For the SW solution, since the monodromy action acts trivially on weight two part, we might
 focus on weight one part, so there are: a): a dimension two complex vector space at generic point with Hodge filtration and polarization; b): a monodromy group around each singularity, which is an element of $SL(2,Z)$ group;
 c): A period function $\tau(u)$ at each generic point.  For  fixed mass parameters $m_i$, the Coulomb branch solution is shown in figure. \ref{rankone}. Since  the point $\infty$ is included in the Coulomb branch, the $u$ plane 
 is now a compact space $\mathbb{P}^1$. 
 
 \textbf{Remark}:  Since the weight two part of the Coulomb branch solution is ignored,  a lot of information is lost, for example,  the full low energy theory can not be decided
 by just looking at the weight one part. However, we will try to recover the weight two part by looking at extra structures from the weight one part of mixed Hodge module.
 
Let's now study the local monodromy group, which is actually a conjugacy class of $SL(2,\mathbb{Z})$ group. Because of the monodromy theorem \ref{monodromytheorem},
the monodromy group is constrained by the relation $(T^k-1)^2=0$, and such conjugacy classes have been classified, see table. \ref{kodaira}.  

\begin{table}[htp]
\begin{center}
\resizebox{4in}{!}{\begin{tabular}{| l | l|l | l | l | l | l | l | l| }
\hline 
Name & Algebra & Monodromy &  Eigenvalues & Euler number & Scaling dimension \\ \hline
$I_n$ & $A_{n-1}$ & $(\begin{array}{cc}1 &n \\0&1 \end{array}) $& $(\exp(2\pi i), \exp(2\pi i))$& $n$ &  1  \\ \hline
$I_n^*$& $D_{n+4}$ &  $(\begin{array}{cc}-1&-n \\0&-1 \end{array}) $& $(\exp(2\pi i{1\over2}), \exp(2\pi i{1\over2}))$ & $n+6$ & 2\\ \hline
$II$ & $\emptyset$ &  $(\begin{array}{cc}1&1 \\-1&0 \end{array}) $ & $(\exp(2\pi i{1\over6}), \exp(2\pi i{5\over6}))$&$2$  & ${6\over 5}$\\ \hline
$III$ & $A_1$ &  $(\begin{array}{cc}0&1 \\-1&0 \end{array}) $ & $(\exp(2\pi i{1\over4}), \exp(2\pi i{3\over4}))$&$3$ & ${4\over 3}$\\ \hline
$IV$ & $A_2$ &  $(\begin{array}{cc}0&1 \\-1&-1 \end{array}) $ &$(\exp(2\pi i{1\over3}), \exp(2\pi i{2\over3}))$ &$4$& ${3\over 2}$ \\ \hline
$II^*$ & $E_8$ &  $(\begin{array}{cc}0&-1 \\1&1 \end{array}) $ &$(\exp(2\pi i{1\over6}), \exp(2\pi i{5\over6}))$& $10$ &$6$ \\ \hline
$III^*$ & $E_7$ &  $(\begin{array}{cc}0&-1 \\1&0 \end{array}) $ & $(\exp(2\pi i{1\over4}), \exp(2\pi i{3\over4}))$&$9$ &$4$\\ \hline
$IV^*$ & $E_6$ &  $(\begin{array}{cc}-1&-1 \\1&0 \end{array}) $ & $(\exp(2\pi i{1\over3}), \exp(2\pi i{2\over3}))$&$8$ & $3$\\ \hline
\end{tabular}}
\end{center}
\caption{The basic data for local singularity associated with rank one theory.}
\label{kodaira}
\end{table}%

After determining the local monodromy group, we then further impose following constraints in doing classification:
\begin{enumerate}
\item \textbf{Global constraints}: Since the moduli space is now a compact space, the total monodromy around the singularities should be trivial, i.e.
\begin{equation*}
M_1M_2\ldots M_\infty=1.
\end{equation*}
There are other global constraints due to the consistency of the period mapping. The constraints are best understood by using the correspondence between the Coulomb branch data shown in figure. \ref{rankone} and 
the rational elliptic surface. Coulomb branch solution provides the  monodromy representation and the period function $\tau(u)$ (which is in turn gives a $j$ invariant $j(u)$). 
The $j$ invariant together with the monodromy data determines a rational elliptic surface \cite{miranda} with a section. Fortunately, the singular fiber types of all the rational elliptic surface have been classified, see \cite{persson1990configurations,miranda1990persson}.  So
there is a  complete data base from which the rank one $\mathcal{N}=2$ theory can be studied.
\item \textbf{UV completeness}:  Let's now constrain the possible singular fiber at $\infty$. The main constraint is the UV completeness, which means that the scaling dimension associated for $\infty$ is bigger than one. 
Using the relation between the scaling dimension and the spectrum of limit mixed Hodge structure (see formula. \ref{scale}),   the dimension is given in table. \ref{kodaira}. The conclusion is that $I_n$ singularity can not appear at  $\infty$. Notice that here
 the loop around the singularity at infinity is also chosen to be counterclockwise. The actual local monodromy group at $\infty$ should be computed using loop in clockwise direction.

\item \textbf{Generic deformations}:  We look at the generic mass deformations, namely, the local singularity at the bulk (the space $\mathbb{P}^1/\infty$) can not be deformed. This condition means the deformation space of 
the physical theory at the singularity is only one dimensional. 
This condition immediately excludes
type $II, III, IV$ singularities, since the Coulomb branch spectrum associated with them is less than two, so there is at least one relevant deformation and one Coulomb branch deformation, which means that the physical theory has 
at least two dimensional physical deformations. 
\item \textbf{Dirac quantization}: For the $I_n$ singularity with just one dimensional deformation, the basic 
assumption is that there is a massless hypermultiplet with electric-magnetic charge $\sqrt{n}(1,0)$ (in appropriate electric-duality frame). If there are only $I_n$ type singularities for the generic deformation, 
the Dirac pairing between  BPS particles which become massless at the singularities are $z_1\cdot z_2= p_1 q_1-p_2 q_2$ (here the charges are $z_1=(p_1,q_1), z_2=(p_2, q_2)$). The Dirac quantization condition implies that the Dirac pairing should be integral. 
This condition then means that for the bulk configuration $\{I_{n_1},I_{n_2},\ldots\}$, the ratio ${n_i \over n_j}$ should be square for any pair of $i,j$. 
\end{enumerate}

\textbf{Remark 1}: Let's now explain how we can find the scaling dimension from the eigenvalue of the monodromy group. To really determine the scaling dimension by using formula. \ref{scale}, we need to know 
the structure of the limit mixed structure. However, the spectrum is constrained in finite set. Let's take $II, II^*$ singularity as an example. The set of eigenvalues are 
$(\lambda_1, \lambda_2)=(\exp(2\pi i{1\over6}), \exp(2\pi i{5\over6}))$, 
and the singularity spectrum$(\alpha_1, \alpha_2)$ (here we shift the singularity spectrum by one so the smallest spectrum number is bigger than one) are constrained by two conditions: a):  $\exp(2\pi i \alpha_i)=\lambda_i$; b): $\alpha_1+\alpha_2=2$, then the two choices are 
\begin{equation*}
(\alpha_1, \alpha_2)=(\frac{1}{6}, \frac{11}{6}),~~(\frac{5}{6}, \frac{7}{6}).
\end{equation*}
Now use the formula \ref{scale}, we get the possible scaling dimensions $\Delta=6,6/5$. To determine the scaling dimension for particular monodromy group, one need other inputs from the holomorphic data. However, for $I_n$ and $I_n^*$ singularity, 
the scaling dimension must be $1$ and $2$ respectively, this then means that $I_n$ can not be the singular fiber at $\infty$ for the Coulomb branch of pure 4d theory.

Another question is  whether the UV theory is asymptotical free or conformal from the monodromy data. Here we conjecture that if the monodromy group is semi-simple, then the theory is superconformal, otherwise it is asymptotic free.
This implies that if $I_n^*$ is put at $\infty$, the theory would be asymptotical free, otherwise it would be superconformal.

\textbf{Remark 2}: The undeformable singularities of type $I_0^*, II^*, III^*, IV^*$ are studied in the context of F theory, and are called frozen singularity in \cite{Tachikawa:2015wka}. In fact,  \cite{Tachikawa:2015wka} also considers 
the partial frozen singularity from which one can actually get new $\mathcal{N}=2$ SCFTs found in \cite{Argyres:2015ffa,Argyres:2015gha,Argyres:2016xmc,Argyres:2016xua} by using D3 brane probing those singularities \cite{Banks:1996nj}.
In fact, all of those theories which has a pure Coulomb branch (i.e. there is no free hypermultiplets at the generic point of Coulomb branch) can be found in this way, please compare table 3 of \cite{Tachikawa:2015wka} and table. \ref{4dfull}.
 
\textbf{Remark 3}: Here we gave a comment on Dirac quantization which might be related to the Dirac quantization condition for the branes in $M$ theory. It was noticed in \cite{Martinec:1995by} that the SW curve for non-simply laced gauge group $\mathfrak{g}$ is not 
identified with the spectral curve of Toda chain of type $\mathfrak{g}$, and they found the correct one should be the spectral curve of the twisted type. One might wonder what is the problem with the spectral curve for the Toda chain of the non-simply 
laced group. We'd like to point out that the problem with the spectral curve is that it does not obey the Dirac quantization condition. In fact, one can get the SW curve for those gauge theory by using Type IIA brane systems \cite{Landsteiner:1997vd}, and the special 
role is played by configuration of orientifold. These orientifolds are strongly constrained by Dirac quantization condition \cite{Hori:1998iv}, which actually agrees with the Dirac quantization condition derived from SW curve.

Now using above four rules,  we can easily classify all the configuration with just $I_n$ type singularities at the bulk by scanning the data set in \cite{persson1990configurations}, see  table. \ref{4d}.  

\begin{figure}
\begin{center}

\tikzset{every picture/.style={line width=0.75pt}} 

\begin{tikzpicture}[x=0.35pt,y=0.35pt,yscale=-1,xscale=1]

\draw   (231.7,10.72) -- (371,10.72) -- (311.3,82) -- (172,82) -- cycle ;
\draw   (278,52.36) .. controls (278,49.4) and (280.46,47) .. (283.5,47) .. controls (286.54,47) and (289,49.4) .. (289,52.36) .. controls (289,55.32) and (286.54,57.72) .. (283.5,57.72) .. controls (280.46,57.72) and (278,55.32) .. (278,52.36) -- cycle ; \draw   (279.61,48.57) -- (287.39,56.15) ; \draw   (287.39,48.57) -- (279.61,56.15) ;
\draw   (225,51.36) .. controls (225,48.4) and (227.46,46) .. (230.5,46) .. controls (233.54,46) and (236,48.4) .. (236,51.36) .. controls (236,54.32) and (233.54,56.72) .. (230.5,56.72) .. controls (227.46,56.72) and (225,54.32) .. (225,51.36) -- cycle ; \draw   (226.61,47.57) -- (234.39,55.15) ; \draw   (234.39,47.57) -- (226.61,55.15) ;
\draw    (229,109) -- (192.57,230.8) ;
\draw [shift={(192,232.72)}, rotate = 286.65] [color={rgb, 255:red, 0; green, 0; blue, 0 }  ][line width=0.75]    (10.93,-3.29) .. controls (6.95,-1.4) and (3.31,-0.3) .. (0,0) .. controls (3.31,0.3) and (6.95,1.4) .. (10.93,3.29)   ;
\draw    (287,110.72) -- (362.93,230.03) ;
\draw [shift={(364,231.72)}, rotate = 237.53] [color={rgb, 255:red, 0; green, 0; blue, 0 }  ][line width=0.75]    (10.93,-3.29) .. controls (6.95,-1.4) and (3.31,-0.3) .. (0,0) .. controls (3.31,0.3) and (6.95,1.4) .. (10.93,3.29)   ;
\draw    (372,108.72) -- (524.43,228.48) ;
\draw [shift={(526,229.72)}, rotate = 218.16] [color={rgb, 255:red, 0; green, 0; blue, 0 }  ][line width=0.75]    (10.93,-3.29) .. controls (6.95,-1.4) and (3.31,-0.3) .. (0,0) .. controls (3.31,0.3) and (6.95,1.4) .. (10.93,3.29)   ;
\draw   (133.8,242.72) -- (250,242.72) -- (200.2,304.72) -- (84,304.72) -- cycle ;
\draw   (187,283.36) .. controls (187,280.4) and (189.46,278) .. (192.5,278) .. controls (195.54,278) and (198,280.4) .. (198,283.36) .. controls (198,286.32) and (195.54,288.72) .. (192.5,288.72) .. controls (189.46,288.72) and (187,286.32) .. (187,283.36) -- cycle ; \draw   (188.61,279.57) -- (196.39,287.15) ; \draw   (196.39,279.57) -- (188.61,287.15) ;
\draw   (131,282.36) .. controls (131,279.4) and (133.46,277) .. (136.5,277) .. controls (139.54,277) and (142,279.4) .. (142,282.36) .. controls (142,285.32) and (139.54,287.72) .. (136.5,287.72) .. controls (133.46,287.72) and (131,285.32) .. (131,282.36) -- cycle ; \draw   (132.61,278.57) -- (140.39,286.15) ; \draw   (140.39,278.57) -- (132.61,286.15) ;
\draw   (318.5,243.72) -- (434,243.72) -- (384.5,307.72) -- (269,307.72) -- cycle ;
\draw   (374,286.36) .. controls (374,283.4) and (376.46,281) .. (379.5,281) .. controls (382.54,281) and (385,283.4) .. (385,286.36) .. controls (385,289.32) and (382.54,291.72) .. (379.5,291.72) .. controls (376.46,291.72) and (374,289.32) .. (374,286.36) -- cycle ; \draw   (375.61,282.57) -- (383.39,290.15) ; \draw   (383.39,282.57) -- (375.61,290.15) ;
\draw   (318,285.36) .. controls (318,282.4) and (320.46,280) .. (323.5,280) .. controls (326.54,280) and (329,282.4) .. (329,285.36) .. controls (329,288.32) and (326.54,290.72) .. (323.5,290.72) .. controls (320.46,290.72) and (318,288.32) .. (318,285.36) -- cycle ; \draw   (319.61,281.57) -- (327.39,289.15) ; \draw   (327.39,281.57) -- (319.61,289.15) ;
\draw   (484.1,244.72) -- (608,244.72) -- (554.9,304.72) -- (431,304.72) -- cycle ;
\draw   (548,285.36) .. controls (548,282.4) and (550.46,280) .. (553.5,280) .. controls (556.54,280) and (559,282.4) .. (559,285.36) .. controls (559,288.32) and (556.54,290.72) .. (553.5,290.72) .. controls (550.46,290.72) and (548,288.32) .. (548,285.36) -- cycle ; \draw   (549.61,281.57) -- (557.39,289.15) ; \draw   (557.39,281.57) -- (549.61,289.15) ;
\draw   (492,284.36) .. controls (492,281.4) and (494.46,279) .. (497.5,279) .. controls (500.54,279) and (503,281.4) .. (503,284.36) .. controls (503,287.32) and (500.54,289.72) .. (497.5,289.72) .. controls (494.46,289.72) and (492,287.32) .. (492,284.36) -- cycle ; \draw   (493.61,280.57) -- (501.39,288.15) ; \draw   (501.39,280.57) -- (493.61,288.15) ;
\draw [color={rgb, 255:red, 208; green, 2; blue, 27 }  ,draw opacity=1 ]   (43,361) -- (616,359.72) ;
\draw   (110.7,396.72) -- (250,396.72) -- (190.3,468) -- (51,468) -- cycle ;
\draw   (157,438.36) .. controls (157,435.4) and (159.46,433) .. (162.5,433) .. controls (165.54,433) and (168,435.4) .. (168,438.36) .. controls (168,441.32) and (165.54,443.72) .. (162.5,443.72) .. controls (159.46,443.72) and (157,441.32) .. (157,438.36) -- cycle ; \draw   (158.61,434.57) -- (166.39,442.15) ; \draw   (166.39,434.57) -- (158.61,442.15) ;
\draw   (104,437.36) .. controls (104,434.4) and (106.46,432) .. (109.5,432) .. controls (112.54,432) and (115,434.4) .. (115,437.36) .. controls (115,440.32) and (112.54,442.72) .. (109.5,442.72) .. controls (106.46,442.72) and (104,440.32) .. (104,437.36) -- cycle ; \draw   (105.61,433.57) -- (113.39,441.15) ; \draw   (113.39,433.57) -- (105.61,441.15) ;
\draw    (123,480.72) -- (120.07,559.72) ;
\draw [shift={(120,561.72)}, rotate = 272.12] [color={rgb, 255:red, 0; green, 0; blue, 0 }  ][line width=0.75]    (10.93,-3.29) .. controls (6.95,-1.4) and (3.31,-0.3) .. (0,0) .. controls (3.31,0.3) and (6.95,1.4) .. (10.93,3.29)   ;
\draw   (81.8,577.72) -- (198,577.72) -- (148.2,639.72) -- (32,639.72) -- cycle ;
\draw   (135,618.36) .. controls (135,615.4) and (137.46,613) .. (140.5,613) .. controls (143.54,613) and (146,615.4) .. (146,618.36) .. controls (146,621.32) and (143.54,623.72) .. (140.5,623.72) .. controls (137.46,623.72) and (135,621.32) .. (135,618.36) -- cycle ; \draw   (136.61,614.57) -- (144.39,622.15) ; \draw   (144.39,614.57) -- (136.61,622.15) ;
\draw   (79,617.36) .. controls (79,614.4) and (81.46,612) .. (84.5,612) .. controls (87.54,612) and (90,614.4) .. (90,617.36) .. controls (90,620.32) and (87.54,622.72) .. (84.5,622.72) .. controls (81.46,622.72) and (79,620.32) .. (79,617.36) -- cycle ; \draw   (80.61,613.57) -- (88.39,621.15) ; \draw   (88.39,613.57) -- (80.61,621.15) ;
\draw   (278.7,400.72) -- (418,400.72) -- (358.3,472) -- (219,472) -- cycle ;
\draw   (325,442.36) .. controls (325,439.4) and (327.46,437) .. (330.5,437) .. controls (333.54,437) and (336,439.4) .. (336,442.36) .. controls (336,445.32) and (333.54,447.72) .. (330.5,447.72) .. controls (327.46,447.72) and (325,445.32) .. (325,442.36) -- cycle ; \draw   (326.61,438.57) -- (334.39,446.15) ; \draw   (334.39,438.57) -- (326.61,446.15) ;
\draw   (272,441.36) .. controls (272,438.4) and (274.46,436) .. (277.5,436) .. controls (280.54,436) and (283,438.4) .. (283,441.36) .. controls (283,444.32) and (280.54,446.72) .. (277.5,446.72) .. controls (274.46,446.72) and (272,444.32) .. (272,441.36) -- cycle ; \draw   (273.61,437.57) -- (281.39,445.15) ; \draw   (281.39,437.57) -- (273.61,445.15) ;
\draw    (291,484.72) -- (288.07,563.72) ;
\draw [shift={(288,565.72)}, rotate = 272.12] [color={rgb, 255:red, 0; green, 0; blue, 0 }  ][line width=0.75]    (10.93,-3.29) .. controls (6.95,-1.4) and (3.31,-0.3) .. (0,0) .. controls (3.31,0.3) and (6.95,1.4) .. (10.93,3.29)   ;
\draw   (249.8,581.72) -- (366,581.72) -- (316.2,643.72) -- (200,643.72) -- cycle ;
\draw   (303,622.36) .. controls (303,619.4) and (305.46,617) .. (308.5,617) .. controls (311.54,617) and (314,619.4) .. (314,622.36) .. controls (314,625.32) and (311.54,627.72) .. (308.5,627.72) .. controls (305.46,627.72) and (303,625.32) .. (303,622.36) -- cycle ; \draw   (304.61,618.57) -- (312.39,626.15) ; \draw   (312.39,618.57) -- (304.61,626.15) ;
\draw   (247,621.36) .. controls (247,618.4) and (249.46,616) .. (252.5,616) .. controls (255.54,616) and (258,618.4) .. (258,621.36) .. controls (258,624.32) and (255.54,626.72) .. (252.5,626.72) .. controls (249.46,626.72) and (247,624.32) .. (247,621.36) -- cycle ; \draw   (248.61,617.57) -- (256.39,625.15) ; \draw   (256.39,617.57) -- (248.61,625.15) ;
\draw   (443.7,400.72) -- (583,400.72) -- (523.3,472) -- (384,472) -- cycle ;
\draw   (490,442.36) .. controls (490,439.4) and (492.46,437) .. (495.5,437) .. controls (498.54,437) and (501,439.4) .. (501,442.36) .. controls (501,445.32) and (498.54,447.72) .. (495.5,447.72) .. controls (492.46,447.72) and (490,445.32) .. (490,442.36) -- cycle ; \draw   (491.61,438.57) -- (499.39,446.15) ; \draw   (499.39,438.57) -- (491.61,446.15) ;
\draw   (437,441.36) .. controls (437,438.4) and (439.46,436) .. (442.5,436) .. controls (445.54,436) and (448,438.4) .. (448,441.36) .. controls (448,444.32) and (445.54,446.72) .. (442.5,446.72) .. controls (439.46,446.72) and (437,444.32) .. (437,441.36) -- cycle ; \draw   (438.61,437.57) -- (446.39,445.15) ; \draw   (446.39,437.57) -- (438.61,445.15) ;
\draw    (456,484.72) -- (453.07,563.72) ;
\draw [shift={(453,565.72)}, rotate = 272.12] [color={rgb, 255:red, 0; green, 0; blue, 0 }  ][line width=0.75]    (10.93,-3.29) .. controls (6.95,-1.4) and (3.31,-0.3) .. (0,0) .. controls (3.31,0.3) and (6.95,1.4) .. (10.93,3.29)   ;
\draw   (414.8,581.72) -- (531,581.72) -- (481.2,643.72) -- (365,643.72) -- cycle ;
\draw   (468,622.36) .. controls (468,619.4) and (470.46,617) .. (473.5,617) .. controls (476.54,617) and (479,619.4) .. (479,622.36) .. controls (479,625.32) and (476.54,627.72) .. (473.5,627.72) .. controls (470.46,627.72) and (468,625.32) .. (468,622.36) -- cycle ; \draw   (469.61,618.57) -- (477.39,626.15) ; \draw   (477.39,618.57) -- (469.61,626.15) ;
\draw   (412,621.36) .. controls (412,618.4) and (414.46,616) .. (417.5,616) .. controls (420.54,616) and (423,618.4) .. (423,621.36) .. controls (423,624.32) and (420.54,626.72) .. (417.5,626.72) .. controls (414.46,626.72) and (412,624.32) .. (412,621.36) -- cycle ; \draw   (413.61,617.57) -- (421.39,625.15) ; \draw   (421.39,617.57) -- (413.61,625.15) ;

\draw (280,22.4) node [anchor=north west][inner sep=0.75pt]    [font=\tiny] {$I_{0}$};
\draw (226,21.4) node [anchor=north west][inner sep=0.75pt]    [font=\tiny]  {$F_{\infty }$};
\draw (225,179.4) node [anchor=north west][inner sep=0.75pt]    [font=\tiny]  {$n=2$};
\draw (348,175.4) node [anchor=north west][inner sep=0.75pt]    [font=\tiny]  {$n=3$};
\draw (481,168.4) node [anchor=north west][inner sep=0.75pt]    [font=\tiny]  {$n=4$};
\draw (188,251.4) node [anchor=north west][inner sep=0.75pt]   [font=\tiny]   {$I_{0}^{*}$};
\draw (135,252.4) node [anchor=north west][inner sep=0.75pt]    [font=\tiny]  {$F'_{\infty }$};
\draw (375,253.4) node [anchor=north west][inner sep=0.75pt]    [font=\tiny]  {$IV^{*}$};
\draw (320,252.4) node [anchor=north west][inner sep=0.75pt]   [font=\tiny]   {$F_{\infty }^{'}$};
\draw (548,254.4) node [anchor=north west][inner sep=0.75pt]     [font=\tiny] {$III^{*}$};
\draw (496,254.4) node [anchor=north west][inner sep=0.75pt]    [font=\tiny]  {$F_{\infty }^{'}$};
\draw (159,408.4) node [anchor=north west][inner sep=0.75pt]   [font=\tiny]   {$I_{2}$};
\draw (105,407.4) node [anchor=north west][inner sep=0.75pt]    [font=\tiny]  {$F_{\infty }$};
\draw (129,511.4) node [anchor=north west][inner sep=0.75pt]     [font=\tiny] {$n=2$};
\draw (136,586.4) node [anchor=north west][inner sep=0.75pt]     [font=\tiny] {$I_{1}^{*}$};
\draw (83,587.4) node [anchor=north west][inner sep=0.75pt]    [font=\tiny]  {$F'_{\infty }$};
\draw (327,412.4) node [anchor=north west][inner sep=0.75pt]   [font=\tiny]   {$I_{4}$};
\draw (273,411.4) node [anchor=north west][inner sep=0.75pt]    [font=\tiny]  {$F_{\infty }$};
\draw (297,515.4) node [anchor=north west][inner sep=0.75pt]    [font=\tiny]  {$n=2$};
\draw (304,590.4) node [anchor=north west][inner sep=0.75pt]    [font=\tiny]  {$I_{2}^{*}$};
\draw (251,591.4) node [anchor=north west][inner sep=0.75pt]    [font=\tiny]  {$F'_{\infty }$};
\draw (492,412.4) node [anchor=north west][inner sep=0.75pt]  [font=\tiny]    {$I_{4}$};
\draw (438,411.4) node [anchor=north west][inner sep=0.75pt]    [font=\tiny]  {$F_{\infty }$};
\draw (462,515.4) node [anchor=north west][inner sep=0.75pt]    [font=\tiny]  {$n=4$};
\draw (469,590.4) node [anchor=north west][inner sep=0.75pt]   [font=\tiny]   {$I_{1}^{*}$};
\draw (416,591.4) node [anchor=north west][inner sep=0.75pt]    [font=\tiny]  {$F'_{\infty }$};

\end{tikzpicture}
\end{center}
\caption{The change of singular fiber at $z=0$ in using a base change map induced by the  $z^n$ group. Upper: The fiber at $z=0$ before the base change is $I_0$ fiber; Botton: The fiber at $z=0$ is 
the undeformable $I_n$ singularity.}
\label{basechange}
\end{figure}
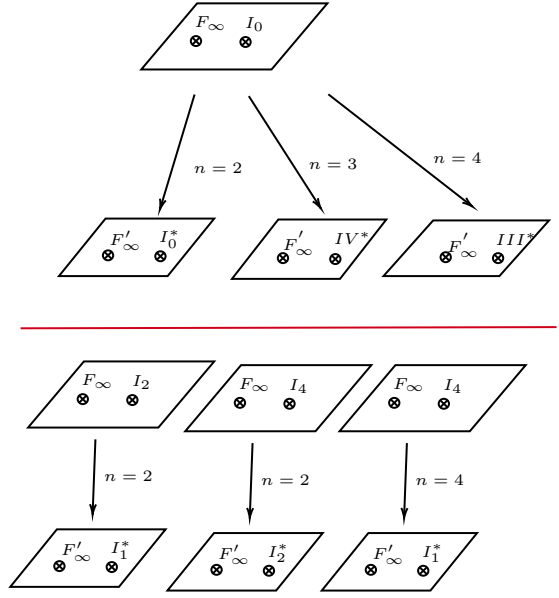

One can also have $I_n^*,II^*,III^*,IV^*$ type singularity appearing in the bulk. 
The low energy theory associated with those singularities has just one dimensional Coulomb branch deformation. Those theories are not familiar, so we'd like to find them by using the geometric method. 
The idea is that we start with the configuration with just $I_n$ type singularities (with generic deformations) at the bulk, and then using the so-called base change method.   This method works as follows.
 Let $g_n$ be the order $n$ automorphism on $P^1$ which is the base of the elliptic surface $B$:
\begin{equation}
g_n:P^1\to P^1,~~z\to z^n.
\end{equation}
We can get a rational elliptic surface $B^{'}$ from $B$ from following diagram:
\begin{equation}
 \begin{tikzcd}
B \arrow{r}{\varphi} \arrow[swap]{d}{\beta} & B^{'} \arrow{d}{\beta^{'}} \\%
P^1\arrow{r}{g_n}& P^1
\end{tikzcd}
\end{equation}
In the process, the elliptic fibers are changed as follows: a) The $j$ invariant for the smooth fiber is not changed; b) The singular fiber at points $z\neq 0, \infty$ are identified by the $\mathbb{Z}_n$ action;  c): 
The  fiber at point $z=0,\infty$ is changed, and the rule for the change is given inn \cite{karayayla2012classification}, and we summarize the relevant facts in figure. \ref{basechange}.  For us, the important thing is that if 
the singular fiber at $z=0$ of $B$ is undeformable,  then the singular fiber at $z=0$ for $B^{'}$ is also undeformable.  

To find new configurations with other type of undeformable singularities, we impose following two conditions for the pair $B,B^{'}$: a): the singular fiber at $z=0$ for $B$ is undeformable; b):
the scaling dimensions of fiber at infinity of $B$ and $B^{'}$ are related as 
\begin{equation}
\Delta(B_\infty^{'})= n \Delta(B_\infty).
\label{4dbase}
\end{equation}
A quite useful result for us is that all the pairs $B$ and $B^{'}$ have been classified in \cite{karayayla2012classification}. By scanning the tables of \cite{karayayla2012classification} and  imposing above two conditions, 
all other theories with undeformable $I_0^*, II^*, III^*, IV^*$ fibers are listed in table. \ref{4d}.

\begin{table}
\caption{Singular fiber configurations for  rank one 4d $\mathcal{N}=2$ SCFTs.}
\begin{center}
\resizebox{6.5in}{!}{\begin{tabular}{| l | l | l | l | l | }
\hline 
$I_1$ series. & $(II,I_1^{10})$ & $(III ,I_1^9)$& $(IV,I_1^8)$ & $(I_0^*, I_1^6)$ \\ \hline
~& $(IV^*,I_1^4)$ & $(III^*,I_1^3)$&$(II^*,I_1^2)$&\\ \hline
$I_2$ series & $(I_0^*, I_2^3)$ \\ \hline
$I_4$ series & $(II, I_4I_1^5)$ & $(III, I_4I_1^4)$ & $(IV, I_4I_1^3)$ & $(I_0^*, I_4I_1^2)$   \\ \hline
~&$(II,I_4^2I_1)$ \\ \hline
$Z_4$ covering &$(IV^*, I_1^4) \to (II, III^* I_1)$ \\ \hline
$Z_3$ covering &$(I_0^*, I_1^6) \to (II, IV^* I_1^2)$  &$(III^*, I_1^3) \to (III, IV^* I_1)$ &  $(I_0^*, I_2^3) \to (II,  IV^* I_2 )$ \\ \hline
$Z_2$ covering &$(IV, I_4I_1^4) \to (II, I_2^* I_1^2)$ &$(IV, I_1^8) \to (II, I_0^* I_1^4)$ &$(I_0^*, I_4I_1^2) \to (III, I_2^* I_1)$ &$(I_0^*, I_2^3) \to (III, I_1^*{~}_{Q=\sqrt{2}} I_2)$\\ \hline
~&$(I_0^*, I_1^6) \to (III, I_0^*I_1^3)$&$(IV^*, I_1^4) \to (IV, I_0^* I_1^2)$  \\ \hline
\end{tabular}}
\caption{Singular fiber configurations for  rank one 4d $\mathcal{N}=2$  asymptotical free theories.}
\resizebox{3in}{!}{\begin{tabular}{| l | l | l | l | l |l|}
\hline 
$I_1$ series & $(I_4^*,I_1^2)$ & $(I_3^* ,I_1^3)$ & $(I_2^*,I_1^4)$ & $(I_1^*, I_1^5)$ \\ \hline
$I_2$ series & $(I_2^*, I_2^2)$ &~&~ &~ \\ \hline
$I_4$ series & $(I_1^*, I_4 I_1)$&~&~&~ \\ \hline
\end{tabular}}
\end{center}
\label{4d}
\end{table}

\textbf{Example}: Let's give an example for the base change map of $B=(IV^*, I_1^4) \to B^{'}=(II, III^* I_1)$. Here $B=(IV^*, I_1^4)$ and has a $Z_4$ automorphism, and at $z=0$ there is a trivial $I_0$ fiber. Under
 $Z_4$ action, the $IV^*$ type fiber at infinity is changed to $II$, while the $I_0^*$ fiber is changed to $III^*$ fiber; and four $I_1$ fiber of $B$ is under a single orbit of $Z_4$ action, and becomes 
 a single $I_1$ fiber of $B^{'}$.

\textbf{Remark 1}:  Some of these new theories can be found  from class $S$ construction \cite{Xie:2012hs}. For example, it was found in \cite{Xie:2017obm} that one 
can find rank one theory with $G_2$ and $F_4$ flavor symmetry (and other theories). In that paper, the scaling dimension is found as the one before the discrete gauging.
However, after further scrutiny, the theories found in \cite{Xie:2017obm} could be the same as  those found in \cite{Argyres:2015ffa,Argyres:2015gha,Argyres:2016xmc,Argyres:2016xua}. 
Some of the other theories can also be found by scanning theories in \cite{Xie:2012hs}.

\textbf{Remark 2}: In \cite{Argyres:2015ffa}, they suggest that one could also interpret  $I_n^*$ type theories as $SU(2)$ gauge theory coupled with 
hypermultiplets in certain representation so that mass deformation is not possible, and several class of theories are given in \cite{Argyres:2015ffa} (such as $I_1^*$ series). Interestingly, 
one could also find those singular fiber configurations by using the base change method \cite{Caorsi:2019vex}; however, in this case one need to start with $I_n$ configurations which 
violate Dirac quantization condition. This might suggest that those theories$I_1^*$ series studied in \cite{Argyres:2015ffa} might also violate Dirac quantization. It would be 
interesting to further clarify this issue.

\subsection{Flavor symmetry and Mordell-Weil lattice}
We have classified the weight one part of the Coulomb branch solution of rank one theory by studying the generic deformation of a theory. 
Since the weight two part is discarded, we seem to lose lots of information, i.e. the charge lattice, flavor symmetry, etc. Fortunately, 
one can find a lattice from the rational elliptic surface: Mordell-Weil lattice. It was shown in \cite{Caorsi:2018ahl} one can find the flavor symmetry by finding out
a root lattice of the Mordell-Weil lattice (see also \cite{Argyres:2018taw} the discussion of finding flavor symmetry by using the SW geometry.).

The physical reason for finding the charge lattice from purely the weight one data might be the following:  BPS particle would become 
massless at the singularity, and those BPS particles also carry flavor charge. If there is a way of finding the charge lattice $\Gamma$ formed by these BPS particles, 
we can find out the flavor symmetry by looking at the sub lattice $\Gamma_0$ (namely the part of the lattice with trivial electric-magnetic charge).

Let's first look at  the theories with only $I_n$ type singularities at the bulk. There is one hypermultiplet associated with each singularity, and 
the charges of them \textbf{span} the full charge lattice, which has the rank
\begin{equation}
rank(\Gamma)=\# sing  =2+f\to f=\sum sing-2.
\label{fla}
\end{equation}
Here the summation is over the singular fiber at the bulk. Let's now relate this number to the topological data associated with local singularities (see data listed in table. \ref{kodaira}).

The Euler numbers of the singular fibers of a rational elliptic surface satisfy the condition 
\begin{equation}
e(f_\infty)+\sum e_i =12.
\end{equation}
The Euler number of the singular fiber at $\infty$ satisfies the condition $e(f_\infty)=2+rank(g_\infty)$, where $g_\infty$ is  the algebra
associated with the fiber at $\infty$. For a $I_n$ type singularity at the bulk, we have $e_i=rank(g_i)+1$, so the above equation becomes
\begin{equation*}
2+rank(g_\infty)+\sum_{bulk} rank(g_i)+\# sing=12\to \# sing=10-\sum_v rank(g_v).
\end{equation*}
Here the summation is over all the singular fibers; so the rank of the flavor symmetry (see \ref{fla}) is 
\begin{equation}
f=\# sing-2=8-\sum_v rank(g_v).
\label{latticerank}
\end{equation}
For the general case, if there are \textbf{two} charge vectors associated with type $I_n^*, II^*, III^*, IV^*$ singularities 
at bulk, the  formula \ref{latticerank} for the rank of flavor symmetry is still true.
 
Interestingly, the rank of  flavor symmetry group (see formula. \ref{latticerank}) is  the same as that of the so-called Mordell-Weil lattice associated with a rational elliptic surface.
In fact, the flavor symmetry can be extracted from this lattice as shown in \cite{Caorsi:2018ahl}. 
 For the rational elliptic surface $X$ with a section ( there is a fiberation $\pi: X\to \mathbb{P}^1$, where the generic fiber is an elliptic surface),
the rational sections form a group called Mordell-Weil group $MW(X)$, with the zero section the identity of 
the group. Furthermore one can define a pairing on $MW(X)$, which makes it into a lattice called Mordell-Weil lattice \cite{schutt2019elliptic}.
For section $P, Q$, one can define a symmetric pairing (here we use the fact that the surface is a rational surface)
\begin{equation*}
(P,Q)=1+(P\cdot O)+(Q\cdot O)-(P\cdot Q)-\sum_v contr_v(P,Q).
\end{equation*}
Here $O$ is the zero section and $contr_v(P,Q)$ is defined as 
\begin{equation*}
contr_v(P,Q)=\begin{cases}
-(A_v^{-1})_{ij},~~i\geq 1,~j\geq 1\\
0
\end{cases}
\end{equation*}
Here $i$ means that $P$ intersects $i$th component of the singular fiber $v$, and $Q$ intersect the $j$th component;  $A_v$ is the Cartan matrix of Lie algebra attached to the singular fiber.
In particular, the self-intersection form is given as
\begin{equation*}
(P,P)=2+2P\cdot O-\sum local.
\end{equation*}
which is also called the height of $P$. 

Let's now summarize  basic properties of Mordell-Weil lattice. Firstly, the rank of $MW(X)$ is given by the data of singular fibers as follows 
\begin{equation*}
rank(MW(X))=8-\sum_v rank(g_v).
\end{equation*} 
and $g_v$ is the  algebra associated with the singular fiber; secondly there is  a subalgebra $T=\oplus \mathfrak{g}_v$ for $X$, and the  root lattice of $T$ has an embedding 
into root lattice of $\mathfrak{e}_8$ algebra. One can define a lattice $L$ which is the orthogonal part of $T$ inside $\mathfrak{e}_8$ lattice.
The torsion free part of $MW(X)$ is then given by the dual $L^\vee$ of $L$.

The flavor symmetry can be found by doing following computations on MW(X): 
\begin{enumerate}
\item The rank of the flavor symmetry is given by the rank of MW(X).
\item The semi-simple part of the flavor symmetry is identified by finding a restricted Mordell-Weil root system. First of all, the \textbf{short} roots 
are easily found, i.e they are given by the height two  elements of the lattice $L$, which can be easily  read from the intersection form of $L$, see the table in \cite{schutt2019elliptic}. 
The \textbf{long} roots are more complicated: 
one need to find element $S$ of $L^\vee$ satisfying following two conditions:
\begin{enumerate}
\item $S$ does not intersect with the zero section, i.e. $S\cdot O=0$ (such section is also called integral section).
\item $S$ intersects the zero component of the fiber  $F_\infty$.
\item $h(S)k(S)=2$, here $h(S)$ is the height of the section $S$, and $k(S)$ is the smallest integer such that $k(S) S$ is an element of  $L$.   The length of the long root is 
given as $2k(S)$.
\end{enumerate}
The semi-simple part of the flavor symmetry can be found from the above root system. The abelian part of flavor symmetry can be found using the rank of Mordell-Weil group. The results are shown  in table. \ref{4dfull}.
\end{enumerate}

\begin{table}[htp]
\begin{center}
\resizebox{6in}{!}{
\begin{tabular}{|l|l|l|l| l|l| l|l|l|}
\hline 
Theory & $G_F$&$h$&Torsion&Theory&$G_F$&$h$ & Torsion\\ \hline
\textbf{$I_1$ series}    \\ \hline
$(II^*,I_1^2)$& $\emptyset$ &0&0&$(III^*,I_1^3)$ & $\mathfrak{su}(2)$&0&0 \\ \hline
$(IV^*,I_1^4)$&$\mathfrak{su}(3)$ &0&0& $(I_0^*, I_1^6)$ & $\mathfrak{so}(8)$ & 0&0 \\ \hline
 $(IV, I_1^8)$& $\mathfrak{e}(6)$ & 0&0& $(III, I_1^9)$ & $\mathfrak{e}(7)$ &0&0 \\ \hline
  $(II,I_1^{10})$ &$\mathfrak{e}(8)$ & 0 &0& ~ & ~&~&~\\ \hline
\textbf{$I_2$ series}  \\ \hline
$(I_0^*, I_2^3)$ & $\mathfrak{su}(2)$ & 1& $Z_2\times Z_2$& ~&~&~&~\\ \hline
\textbf{$I_4$ series}&  \\ \hline
$(II, I_4I_1^6)$& $\mathfrak{sp}(10)$&5&0&$(III, ~I_4I_1^5)$ &$\mathfrak{sp}(6)\times \mathfrak{sp}(2)$&3&0\\ \hline
 $(IV, I_4I_1^4)$&$\mathfrak{sp}(4)\times \mathfrak{u}(1)$ & 2 &0&$(I_0^*, I_4I_1^2)$& $\mathfrak{su}(2)$ & 1&$Z_2$ \\ \hline
 $(II, I_4^2I_1^2)$ &   $\mathfrak{sp}(4)$  &2 &0 &&&&\\ \hline
$Z_4$ covering &~&~&~&\\ \hline
$ (II, III^* I_1)$ & $\mathfrak{su}(2)$&0&0&& \\ \hline
$Z_3$ covering &~&~&~\\ \hline
$ (II, IV^* I_1^2)$& $\mathfrak{g}_2$  &0&0 & $(III, IV^* I_1)$& $\mathfrak{su}(2)$ & 0&0  \\ \hline
$(II, IV^* I_2)$ & $\mathfrak{su}(2)$ & 1&0 \\ \hline
$Z_2$ covering & \\ \hline
$ (II, I_2^* I_1^2)$ & $\mathfrak{sp}(4)$&2&0&$ (II, I_0^* I_1^4)$ & $\mathfrak{f}(4)$& 0 &0\\ \hline
$ (III, I_2^* I_1)$ &  $\mathfrak{su}(2)$ &1&$Z_2$& $(III, I_1^*{~}_{Q=\sqrt{2}} I_2)$& $\mathfrak{su}(2)$&1 &$Z_2$ \\ \hline
$(III, I_0^*I_1^3)$&$\mathfrak{spin}(7)$&0&0&$(IV, I_0^* I_1^2)$ & $\mathfrak{su}(3)$ & 0 &0\\ \hline
\end{tabular}}
\end{center}
\caption{Flavor symmetry, free hypermultiplets at generic point of Coulomb branch, and one form symmetry (which is identified with the torsion subgroup of the Mordeil-Weil lattice) for 4d rank one $\mathcal{N}=2$ SCFTs.}
\label{4dfull}
\end{table}

Let's now give some details for the computations of flavor symmetry.

\textbf{Example 1}: Let's look at the example 
$(IV, I_0^* I_1^2)$ (here $T=A_2\times D_4$, and so $L^\vee$ has rank two), the intersection form on $L^\vee$ is given as $\Lambda_{(23)}={1\over6}\left(\begin{array}{cc} 2&1\\1&2\end{array} \right)$. There is no short root as there 
are no height two elements.
An element 
in the lattice $L^\vee$ has the pairing (here $P=n_1e_1+n_2e_2$, with $e_i$ the basis of the $L^\vee$ )
\begin{equation*}
(P,P)=n^T\Lambda_{23} n={1\over 3} n_1^2+ {n_1 n_2 \over 3}+ {1\over 3} n_2^2.
\end{equation*}
Now for an integral section which intersects the zeroth component of $IV$ fiber, the height for an integral section has the form
\begin{equation*}
(P,P)=2-\delta_i(I_0^*).
\end{equation*}
Here $\delta_i$ denotes the $i$th diagonal value for the inverse of the Cartan matrix of $D_4$ algebra (namely $P$ intersects the $i$th component of $I_0^*$ singularity), 
and it can take value$ (\frac{1}{2}, 1, \frac{3}{4},\frac{3}{4})$ with level $(4,2,4,4)$. So
\begin{equation*}
h(P)=(P,P)=\begin{cases}
{3\over 2},~~k=4 \\
1,~~k=2  \\
{5\over 4},~~k=4 \\
\end{cases}
\end{equation*}
To have $h(P)k(P)=2$, the only solution is $h(P)=1, k(P)=2$, 
and the long root have length $2k(P)=4$.  The number of the long roots are given by the solutions of following equation:
 \begin{equation*}
 {1\over 3} n_1^2+ {n_1 n_2 \over 3}+ {1\over 3} n_2^2=1.
 \end{equation*}
 and the solutions are $(n_1, n_2)=\pm (1,1), \pm(1,-2), \pm(2,-1)$, so there are a total of 6 roots with length $4$. Since the flavor symmetry has rank 2, and 
the root system has the equal length and should be the simply laced Lie algebra $\mathfrak{su}(3)$. 

\textbf{Example 2}: Let's look at example $(II, I_4^2 I_1^2)$ (here $T=A_3\oplus A_3$, so $L^\vee$ has rank two). The intersection form on $L^\vee$ is given as $({1\over 4}) \oplus ({1\over 4})$, and so the pairing for a section $P=n_1 e_1+n_2 e_2$ is given as 
\begin{equation*}
(P,P)={1\over 4}n_1^2+{1\over 4}n_2^2.
\end{equation*}
Now for an integral section which intersects the zeroth component of $II$ fiber, the height takes the form
\begin{equation*}
(P,P)=2-\delta_i(I_4)-\delta_j(I_4).
\end{equation*}
Here $\delta_i(I_4)$ denotes the $i$th diagonal value for the inverse of the Cartan matrix of $A_3$ algebra, and it can take value$ (\frac{3}{4},  1,\frac{3}{4})$, and the level is $(4,2,4)$. So for 
an integral section $P$, it can take value: 
\begin{equation*}
(P,P)=\begin{cases}
{5\over 4},~~k=4 \\
1,~~k=2  \\
{1\over 2},~~k=4 \\
{1\over 4},~~k=4 \\
\end{cases}
\end{equation*}
and so we can get  long roots for integral section with height $1$ (root length $4$) and $\frac{1}{2}$ (root length $8$). Therefore, we need to count solutions for the following equations:
\begin{align*}
\begin{cases}
{1\over 4}n_1^2+{1\over 4}n_2^2=1 ~~~\text{short root, length four}  \\
 {1\over 4}n_1^2+{1\over 4}n_2^2={1\over 2}~~~\text{long root, length eight}
\end{cases}
\end{align*}
The solutions are $(n_1, n_2)=\pm (2,0), \pm(0,2)$ (short root), $(n_1, n_2)=\pm (1,1), \pm(1,-1)$ (long root). So the root system has four short roots and eight long roots, which
gives the root system $\mathfrak{sp}(4)$. 

\textbf{Example 3}: Finally, let's do the computation for the configuration $(III, I_0^*I_1^3)$ (here $T=A_1\oplus D_4$, so $L^\vee$ has rank three). The intersection form on $L^\vee$ is given as $A_1^*\oplus A_1^* \oplus A_1^* $, here $A_1^*$ is the 
dual of the $A_1$ root lattice. There are 6 short roots from the roots of three $A_1$. 
The height for a section $P=n_1 e_1+n_2 e_2+n_3 e_3$ is given as 
\begin{equation*}
(P,P)={1\over 2}n_1^2+{1\over 2}n_2^2+{1\over 2}n_3^2.
\end{equation*}
Now for an integral section which intersects the zeroth component of $III$ fiber, the height takes the form
\begin{equation*}
(P,P)=2-\delta_i(I_0^*).
\end{equation*}
Using the data summarized in example 1, an integral section $P$ has height
\begin{equation*}
(P,P)=\begin{cases}
{3\over 2},~~k=4 \\
1,~~k=2  \\
{5\over 4},~~k=4 \\
\end{cases}
\end{equation*}
so the long roots are given by $h(P)=1, k(P)=2$, and the number is given by the number of solutions of equation:
\begin{equation*}
{1\over 2}n_1^2+{1\over 2}n_2^2+{1\over 2}n_3^2=1.
\end{equation*}
and the solutions are $(n_1, n_2, n_3)=\pm (1,1,0), \pm (1,-1,0), \pm (1,0,1), \pm (1,0,-1),~\pm (0,1,1), ~\pm (0,1,-1)$. 
So there are a total of 12 long roots. In summary, we find a root system with 6 short roots and 12 long roots, 
which can be identified as the root system of $\mathfrak{sp}(7)$ Lie algebra.

\subsection{One form symmetry}
The one form symmetry acts on the line operators of a theory \cite{Gaiotto:2014kfa}. The line operators on the other hand can be studied 
in the IR limit \cite{Gaiotto:2009hg}, and in particular its charge can be found from the BPS quiver. Since the charge lattice of the theory is 
determined by the Mordell-Weil lattice. It might be natural to identify the one form symmetry as the torsion part of the Mordell-Weil group \cite{Caorsi:2019vex}.
This seems so by looking at the familiar $\mathcal{N}=2^*$ theory with $SU(2)$ gauge group. This theory has two realizations: one  is $(I_0^*, I_2^3)$ whose $MW(X)_{torsion}$ is $Z_2$, 
and the other is $(I_0^*, I_4I_1^2)$ whose $MW(X)_{torsion}$ is $Z_2\times Z_2$. Interestingly, for $\mathcal{N}=2^*$ $SU(2)$ theory, there are two choices of line operators which form separate $SL(2,Z)$ orbits \cite{Aharony:2013hda,Xie:2013vfa}: a): for the first one, in one duality frame
the minimal Wilson line is in fundamental representation (the charge is $(n,0)$), and the Dirac quantization condition implies that the minimal 't hooft line operator is in adjoint representation (the charge is $(0,2n)$); The $Z_2$ one form symmetry 
could be identified as the center of the gauge group and acts on 't hooft line operators; b): In the other description, the minimal line operator has electric-magnetic charge $(1,1)$, and so the Wilson line has charge $(2n,0)$ and 't hooft line has charge $(0,2n)$; 
the one form symmetry in this case is $Z_2\times Z_2$. At least for above example, 
the one form symmetry can be identified with $MW(X)_{torsion}$. It would be interesting to study one form symmetry for other theories.

\subsection{SW curve and Weierstrass model}
In previous sections, the rational elliptic surface is used  to classify  4d $\mathcal{N}=2$ theory and study the flavor symmetry. 
Now let's use the correspondence with the Weierstrass model to write down the SW curve and SW differential.  

There are two useful models for the rational elliptic surface $X$. In the first description, the surface is taken to be smooth and minimal. Let's take the elliptic fibration $\pi: X\to \mathbb{P}^1$ with a section $O$, and the generic fiber is 
an elliptic surface (a torus with one marked point determined by zero section). The minimal condition means that there is no $(-1)$ curve in the vertical direction. The singular fiber is described by a a set of rational curves whose 
form is determined by the  algebra associated with fiber, see \cite{miranda}.  The other model is called Weierstrass model, which is given by the form
\begin{equation*}
y^2=x^3+fx+g.
\end{equation*}
Here $f \in H^0(L^4)$ and $g\in H^0(L^6)$ with $L$ (${\cal O}(-1)$) a line bundle on $\mathbb{P}^1$. The total space is in general singular. We consider the so-called minimal Weierstrass model, namely, the singularity of the total space is just $ADE$ singularity.
Given a minimal rational elliptic surface (which is in turn determined by the monodromy data and holomorphic data), one can associate a minimal Weierstrass model \cite{miranda}. It is natural to identify this Weierstrass model 
as the SW curve of the theory.  The  SW differential can be found  by imposing the condition \ref{differential}. We leave the detailed study to a separate publication.

\subsection{Low energy theory at other special vacua}
In  previous discussion, we  consider only the Coulomb branch with generic  mass deformations. The low energy theory at undeformable singularity is a $U(1)$ gauge theory coupled 
with a massless hypermultiplets for $I_n$ singularity, and an interacting theory for type $I_n^*, II^*, III^*, IV^*$ singularities.  If we allow non-generic deformation, the singularities of generic deformations 
would merge and new type of singularities appear. One should be clear that the label for the singularity is still in the same class listed in table. \ref{kodaira}. The physical interpretation of other singularities are the following:
a): If the singularities are formed by merging type $I_1$  singularities, the low energy theory is simple and well studied, see table. \ref{simple}. b): If the singularities are formed by merging other type undeformable singularities, 
the low energy theory can be found from table. \ref{4d} by looking at the bulk singularities.  However, we do not know whether several singularities can merge or not by simply looking at the singularity type. In the next section, we will 
associate a brane configuration for each theory and it is then easy to determine whether several singularities can merge or not.

\begin{table}[htp]
\begin{center}
\resizebox{1.5in}{!}{
\begin{tabular}{|c|c|}
\hline 
Merging &Theory  \\ \hline
$II^*(I_1^{10})$ & $E_8$ SCFT\\ \hline
$III^*(I_1^9)$ & $E_7$ SCFT \\ \hline
$IV^*(I_1^8)$ & $E_6$ SCFT \\ \hline
$II(I_1^2)$ & $H_0$ SCFT \\ \hline
$III(I_1^3)$ &  $H_1$ SCFT \\ \hline
$IV(I_1^4)$ &  $H_2$ SCFT \\ \hline
$I_n(I_1^n)$ & $U(1)-n$ \\ \hline
$I_n^*(I_1^{6+n})$ & $SU(2)-(n+4)$\\ \hline
\end{tabular}}
\end{center}
\caption{The theory on generic special vacua formed by merging $I_1$ singularities. The $E_n$ type theories were found in \cite{Minahan:1996cj}, and the $H_i$ type theories were found in \cite{Argyres:1995xn}.}
\label{simple}
\end{table}

\newpage
\section{Brane construction}
In the last section, the classification of rational elliptic surface is used to classify 4d $\mathcal{N}=2$ rank one theories. While 
the geometric tools are quite useful, it is desirable to have a more physical approach to study these theories. Indeed, it was shown in \cite{DeWolfe:1998bi} that one can associate 
a D7 brane configuration for $\mathcal{N}=2$ theories found in \cite{Minahan:1996cj,Argyres:1995xn}. Here we give a general treatment and one of our insight is to associate a collapsed brane system 
for the singular fiber at $\infty$. These brane configurations are quite useful in studying many aspects of the theory, such as the Mordell-Weil lattice, BPS quiver, BPS spectrum, etc. 

\subsection{D7 branes}
The fundamental building blocks are $(p,q)$ seven branes in Type IIB string theory. 
There is elementary  $(p,q)$ (here $p,q$ is coprime) D7 brane around which the monodromy matrix is given as 
\begin{equation*}
K_{[p,q]}=\left[\begin{array}{cc}
1+pq & -p^2 \\
q^2& 1-pq
\end{array}\right].
\end{equation*}
Here the loop for computing the monodromy is taken in the counter-clockwise direction around the brane. Notice that a $(p,q)$ brane 
is the same as a $-(p,q)$ brane.

\begin{figure}
\begin{center}
\tikzset{every picture/.style={line width=0.75pt}} 

\begin{tikzpicture}[x=0.35pt,y=0.35pt,yscale=-1,xscale=1]

\draw  [color={rgb, 255:red, 0; green, 0; blue, 0 }  ,draw opacity=1 ][fill={rgb, 255:red, 0; green, 0; blue, 0 }  ,fill opacity=1 ] (157,140.5) .. controls (157,139.12) and (158.12,138) .. (159.5,138) .. controls (160.88,138) and (162,139.12) .. (162,140.5) .. controls (162,141.88) and (160.88,143) .. (159.5,143) .. controls (158.12,143) and (157,141.88) .. (157,140.5) -- cycle ;
\draw  [dash pattern={on 4.5pt off 4.5pt}]  (159.5,140.5) -- (158.5,290.72) ;

\draw    (98,139) -- (323,140.72) ;
\draw  [color={rgb, 255:red, 0; green, 0; blue, 0 }  ,draw opacity=1 ][fill={rgb, 255:red, 0; green, 0; blue, 0 }  ,fill opacity=1 ] (199,141.5) .. controls (199,140.12) and (200.12,139) .. (201.5,139) .. controls (202.88,139) and (204,140.12) .. (204,141.5) .. controls (204,142.88) and (202.88,144) .. (201.5,144) .. controls (200.12,144) and (199,142.88) .. (199,141.5) -- cycle ;
\draw  [dash pattern={on 4.5pt off 4.5pt}]  (201.5,141.5) -- (200.5,291.72) ;

\draw  [color={rgb, 255:red, 0; green, 0; blue, 0 }  ,draw opacity=1 ][fill={rgb, 255:red, 0; green, 0; blue, 0 }  ,fill opacity=1 ] (240,141.5) .. controls (240,140.12) and (241.12,139) .. (242.5,139) .. controls (243.88,139) and (245,140.12) .. (245,141.5) .. controls (245,142.88) and (243.88,144) .. (242.5,144) .. controls (241.12,144) and (240,142.88) .. (240,141.5) -- cycle ;
\draw  [dash pattern={on 4.5pt off 4.5pt}]  (242.5,141.5) -- (241.5,291.72) ;

\draw  [color={rgb, 255:red, 0; green, 0; blue, 0 }  ,draw opacity=1 ][fill={rgb, 255:red, 0; green, 0; blue, 0 }  ,fill opacity=1 ] (278,142.5) .. controls (278,141.12) and (279.12,140) .. (280.5,140) .. controls (281.88,140) and (283,141.12) .. (283,142.5) .. controls (283,143.88) and (281.88,145) .. (280.5,145) .. controls (279.12,145) and (278,143.88) .. (278,142.5) -- cycle ;
\draw  [dash pattern={on 4.5pt off 4.5pt}]  (280.5,142.5) -- (279.5,292.72) ;

\draw (149,112.4) node [anchor=north west][inner sep=0.35pt]    {$X_{1}$};
\draw (187,112.4) node [anchor=north west][inner sep=0.35pt]    {$X_{2}$};
\draw (226,113.4) node [anchor=north west][inner sep=0.35pt]    {$X_{3}$};

\end{tikzpicture}
\end{center}
\caption{D7 brane configuration. Here there is a branch cut represented by slash lines attached to each D7 brane.}
\label{d7}
\end{figure}
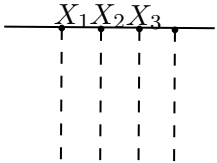

Let's put a sequence of D7 branes with label $[p_i, q_i]$ on a real line (with ordering from left to right) $X_1X_2\ldots X_n$ (see figure. \ref{d7}),  and the total monodromy around the brane configuration is 
\begin{equation*}
K_{[p_n,q_n]}K_{[p_{n-1},q_{n-1}]}\ldots K_{[p_1,q_1]}.
\end{equation*}

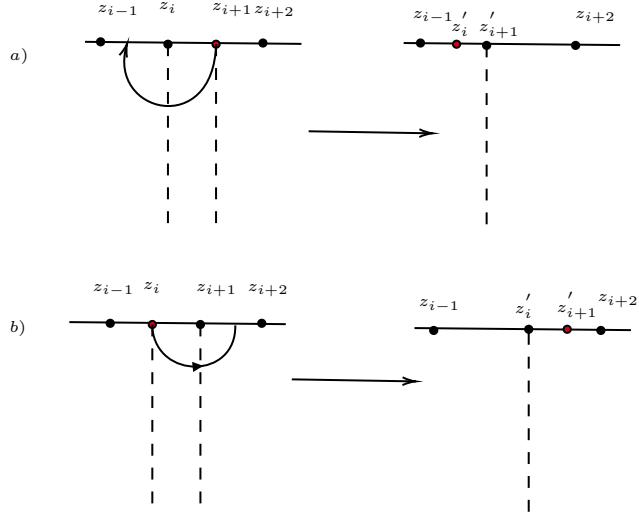
\begin{figure}
\begin{center}

\tikzset{every picture/.style={line width=0.75pt}} 

\begin{tikzpicture}[x=0.45pt,y=0.45pt,yscale=-1,xscale=1]

\draw    (111,98) -- (291,99.72) ;
\draw  [fill={rgb, 255:red, 0; green, 0; blue, 0 }  ,fill opacity=1 ] (177,100) .. controls (177,98.34) and (178.34,97) .. (180,97) .. controls (181.66,97) and (183,98.34) .. (183,100) .. controls (183,101.66) and (181.66,103) .. (180,103) .. controls (178.34,103) and (177,101.66) .. (177,100) -- cycle ;
\draw  [fill={rgb, 255:red, 208; green, 2; blue, 27 }  ,fill opacity=1 ] (217,100) .. controls (217,98.34) and (218.34,97) .. (220,97) .. controls (221.66,97) and (223,98.34) .. (223,100) .. controls (223,101.66) and (221.66,103) .. (220,103) .. controls (218.34,103) and (217,101.66) .. (217,100) -- cycle ;
\draw  [dash pattern={on 4.5pt off 4.5pt}]  (180,100) -- (180,258.72) ;
\draw  [dash pattern={on 4.5pt off 4.5pt}]  (220,100) -- (220,258.72) ;
\draw    (220,100) .. controls (218.02,178.92) and (129.79,158.15) .. (145.49,101.45) ;
\draw [shift={(146,99.72)}, rotate = 107.24] [color={rgb, 255:red, 0; green, 0; blue, 0 }  ][line width=0.75]    (10.93,-3.29) .. controls (6.95,-1.4) and (3.31,-0.3) .. (0,0) .. controls (3.31,0.3) and (6.95,1.4) .. (10.93,3.29)   ;
\draw    (98,333) -- (278,334.72) ;
\draw  [fill={rgb, 255:red, 208; green, 2; blue, 27 }  ,fill opacity=1 ] (164,335) .. controls (164,333.34) and (165.34,332) .. (167,332) .. controls (168.66,332) and (170,333.34) .. (170,335) .. controls (170,336.66) and (168.66,338) .. (167,338) .. controls (165.34,338) and (164,336.66) .. (164,335) -- cycle ;
\draw  [fill={rgb, 255:red, 0; green, 0; blue, 0 }  ,fill opacity=1 ] (204,335) .. controls (204,333.34) and (205.34,332) .. (207,332) .. controls (208.66,332) and (210,333.34) .. (210,335) .. controls (210,336.66) and (208.66,338) .. (207,338) .. controls (205.34,338) and (204,336.66) .. (204,335) -- cycle ;
\draw  [dash pattern={on 4.5pt off 4.5pt}]  (167,335) -- (167,493.72) ;
\draw  [dash pattern={on 4.5pt off 4.5pt}]  (207,335) -- (207,493.72) ;
\draw    (236,335.72) .. controls (237,381.16) and (171,381.72) .. (167,338) ;
\draw [shift={(209.4,369.68)}, rotate = 174.22] [fill={rgb, 255:red, 0; green, 0; blue, 0 }  ][line width=0.08]  [draw opacity=0] (8.93,-4.29) -- (0,0) -- (8.93,4.29) -- cycle    ;
\draw    (376,99) -- (556,100.72) ;
\draw  [fill={rgb, 255:red, 0; green, 0; blue, 0 }  ,fill opacity=1 ] (442,101) .. controls (442,99.34) and (443.34,98) .. (445,98) .. controls (446.66,98) and (448,99.34) .. (448,101) .. controls (448,102.66) and (446.66,104) .. (445,104) .. controls (443.34,104) and (442,102.66) .. (442,101) -- cycle ;
\draw  [fill={rgb, 255:red, 208; green, 2; blue, 27 }  ,fill opacity=1 ] (417,100) .. controls (417,98.34) and (418.34,97) .. (420,97) .. controls (421.66,97) and (423,98.34) .. (423,100) .. controls (423,101.66) and (421.66,103) .. (420,103) .. controls (418.34,103) and (417,101.66) .. (417,100) -- cycle ;
\draw  [dash pattern={on 4.5pt off 4.5pt}]  (445,101) -- (445,259.72) ;
\draw    (385,338) -- (565,339.72) ;
\draw  [fill={rgb, 255:red, 0; green, 0; blue, 0 }  ,fill opacity=1 ] (537,340) .. controls (537,338.34) and (538.34,337) .. (540,337) .. controls (541.66,337) and (543,338.34) .. (543,340) .. controls (543,341.66) and (541.66,343) .. (540,343) .. controls (538.34,343) and (537,341.66) .. (537,340) -- cycle ;
\draw  [fill={rgb, 255:red, 0; green, 0; blue, 0 }  ,fill opacity=1 ] (477,339) .. controls (477,337.34) and (478.34,336) .. (480,336) .. controls (481.66,336) and (483,337.34) .. (483,339) .. controls (483,340.66) and (481.66,342) .. (480,342) .. controls (478.34,342) and (477,340.66) .. (477,339) -- cycle ;
\draw  [dash pattern={on 4.5pt off 4.5pt}]  (480,342) -- (480,500.72) ;
\draw  [fill={rgb, 255:red, 0; green, 0; blue, 0 }  ,fill opacity=1 ] (256,99) .. controls (256,97.34) and (257.34,96) .. (259,96) .. controls (260.66,96) and (262,97.34) .. (262,99) .. controls (262,100.66) and (260.66,102) .. (259,102) .. controls (257.34,102) and (256,100.66) .. (256,99) -- cycle ;
\draw  [fill={rgb, 255:red, 0; green, 0; blue, 0 }  ,fill opacity=1 ] (121,98) .. controls (121,96.34) and (122.34,95) .. (124,95) .. controls (125.66,95) and (127,96.34) .. (127,98) .. controls (127,99.66) and (125.66,101) .. (124,101) .. controls (122.34,101) and (121,99.66) .. (121,98) -- cycle ;
\draw  [fill={rgb, 255:red, 0; green, 0; blue, 0 }  ,fill opacity=1 ] (387,99) .. controls (387,97.34) and (388.34,96) .. (390,96) .. controls (391.66,96) and (393,97.34) .. (393,99) .. controls (393,100.66) and (391.66,102) .. (390,102) .. controls (388.34,102) and (387,100.66) .. (387,99) -- cycle ;
\draw  [fill={rgb, 255:red, 0; green, 0; blue, 0 }  ,fill opacity=1 ] (516,101) .. controls (516,99.34) and (517.34,98) .. (519,98) .. controls (520.66,98) and (522,99.34) .. (522,101) .. controls (522,102.66) and (520.66,104) .. (519,104) .. controls (517.34,104) and (516,102.66) .. (516,101) -- cycle ;
\draw    (297,173) -- (398,174.69) ;
\draw [shift={(400,174.72)}, rotate = 180.96] [color={rgb, 255:red, 0; green, 0; blue, 0 }  ][line width=0.75]    (10.93,-3.29) .. controls (6.95,-1.4) and (3.31,-0.3) .. (0,0) .. controls (3.31,0.3) and (6.95,1.4) .. (10.93,3.29)   ;
\draw  [fill={rgb, 255:red, 0; green, 0; blue, 0 }  ,fill opacity=1 ] (255,334) .. controls (255,332.34) and (256.34,331) .. (258,331) .. controls (259.66,331) and (261,332.34) .. (261,334) .. controls (261,335.66) and (259.66,337) .. (258,337) .. controls (256.34,337) and (255,335.66) .. (255,334) -- cycle ;
\draw  [fill={rgb, 255:red, 0; green, 0; blue, 0 }  ,fill opacity=1 ] (129,334) .. controls (129,332.34) and (130.34,331) .. (132,331) .. controls (133.66,331) and (135,332.34) .. (135,334) .. controls (135,335.66) and (133.66,337) .. (132,337) .. controls (130.34,337) and (129,335.66) .. (129,334) -- cycle ;
\draw  [fill={rgb, 255:red, 0; green, 0; blue, 0 }  ,fill opacity=1 ] (398,340) .. controls (398,338.34) and (399.34,337) .. (401,337) .. controls (402.66,337) and (404,338.34) .. (404,340) .. controls (404,341.66) and (402.66,343) .. (401,343) .. controls (399.34,343) and (398,341.66) .. (398,340) -- cycle ;
\draw  [fill={rgb, 255:red, 208; green, 2; blue, 27 }  ,fill opacity=1 ] (509,339) .. controls (509,337.34) and (510.34,336) .. (512,336) .. controls (513.66,336) and (515,337.34) .. (515,339) .. controls (515,340.66) and (513.66,342) .. (512,342) .. controls (510.34,342) and (509,340.66) .. (509,339) -- cycle ;
\draw    (283,381) -- (384,382.69) ;
\draw [shift={(386,382.72)}, rotate = 180.96] [color={rgb, 255:red, 0; green, 0; blue, 0 }  ][line width=0.75]    (10.93,-3.29) .. controls (6.95,-1.4) and (3.31,-0.3) .. (0,0) .. controls (3.31,0.3) and (6.95,1.4) .. (10.93,3.29)   ;

\draw (170,62.4) node [anchor=north west][inner sep=0.75pt]   [font=\tiny] {$z_{i}$};
\draw (214,63.4) node [anchor=north west][inner sep=0.75pt]   [font=\tiny]   {$z_{i+1}$};
\draw (157,297.4) node [anchor=north west][inner sep=0.75pt]     [font=\tiny] {$z_{i}$};
\draw (201,298.4) node [anchor=north west][inner sep=0.75pt]    [font=\tiny]  {$z_{i+1}$};
\draw (436,70.4) node [anchor=north west][inner sep=0.75pt]    [font=\tiny]  {$z_{i+1}^{'}$};
\draw (466,306.4) node [anchor=north west][inner sep=0.75pt]    [font=\tiny]  {$z_{i}^{'}$};
\draw (501,305.4) node [anchor=north west][inner sep=0.75pt]   [font=\tiny]   {$z_{i+1}^{'}$};
\draw (46,100.4) node [anchor=north west][inner sep=0.75pt]    [font=\tiny]  {$a)$};
\draw (46,328.4) node [anchor=north west][inner sep=0.75pt]   [font=\tiny]   {$b)$};
\draw (516,69.4) node [anchor=north west][inner sep=0.75pt]    [font=\tiny]  {$z_{i+2}$};
\draw (249,65.4) node [anchor=north west][inner sep=0.75pt]    [font=\tiny]  {$z_{i+2}$};
\draw (119,64.4) node [anchor=north west][inner sep=0.75pt]    [font=\tiny]  {$z_{i-1}$};
\draw (381,69.4) node [anchor=north west][inner sep=0.75pt]    [font=\tiny]  {$z_{i-1}$};
\draw (413,70.4) node [anchor=north west][inner sep=0.75pt]    [font=\tiny]  {$z_{i}^{'}$};
\draw (115,298.4) node [anchor=north west][inner sep=0.75pt]    [font=\tiny]  {$z_{i-1}$};
\draw (244,298.4) node [anchor=north west][inner sep=0.75pt]   [font=\tiny]   {$z_{i+2}$};
\draw (386,311.4) node [anchor=north west][inner sep=0.75pt]   [font=\tiny]   {$z_{i-1}$};
\draw (535,307.4) node [anchor=north west][inner sep=0.75pt]    [font=\tiny]  {$z_{i+2}$};

\end{tikzpicture}

\end{center}
\caption{Two ways of exchanging the position of two adjacent D7 branes. Upper: move $(i+1)$th brane; Bottom: move $i$th brane.}
\label{D7move}
\end{figure}

There are some equivalence relations for the D7 brane configuration.  There is one simple equivalence, namely, one can do a global $SL(2,Z)$ transformation $g$, so that the brane charge vector $z_i=[p_i,q_i]$ and the monodromy matrix are changed as 
\begin{equation}
z_i^{'}=gz_i, ~~~K_{z_i^{'}}=gK_{z_i}g^{'}.
\end {equation}
There is another more complicated equivalence relation, i.e. one can move the branes around, and the brane charge would be changed accordingly.
The basic move is to exchange the order of two adjacent D7 branes with charges $z_i=[p_i, q_i]$ and $z_{i+1}=[p_{i+1}, q_{i+1}]$:
\begin{equation*}
X_{z_i}X_{z_{i+1}}=X_{z_i^{'}}X_{z_{i+1}^{'}}.
\end{equation*}
The brane changes are related as:
\begin{align}
&a): \text{if we move the $i$th brane}: z_i^{'}=z_{i+1},~~z_{i+1}^{'}=z_{i}+z_i\cdot z_{i+1}z_{i+1} \nonumber\\
&b):  \text{if we move the $(i+1)$th brane}:   z_i^{'}=z_{i+1}+(z_i\cdot z_{i+1}) z_i ,                     ~~z_{i+1}^{'}=z_i.
\label{moverule}
\end{align}
Here $z_i\cdot z_j=p_iq_j-p_j q_i$ is the symplectic pairing of $z_i,z_j$. The two brane moves are shown in figure. \ref{D7move}. Notice that 
the brane moves does not depend on using the charge $z_i$ or $-z_i$.

There are several useful invariants for the brane configuration (independent of above two equivalence relations). First, there is the following
trace formula for the total monodromy $K$:
\begin{equation*}
Tr(K)=2+\sum_{k=2}^n \sum_{i_1<i_2<\ldots<i_k}(z_{i_1}\times z_{i_2})(z_{i_2}\times z_{i_3})\ldots (z_{i_{k}}\times z_{i_1}).
\end{equation*}
Another important invariant is defined as 
\begin{equation*}
l=gcd(z_i\cdot z_j),~~\text{for all $i$ and $j$}
\end{equation*}
Finally one can define a symmetric matrix as follows
\begin{equation*}
A_{ij}=\begin{cases}
(a_{ij}),~~~a_{ij}=-{1\over2}(p_i q_j-q_i p_j),~~i< j \\
2,~~i=j
\end{cases}
\end{equation*}
and there is a relation:
\begin{equation*}
\det(A)={1\over4}TrK+{1\over2}.
\end{equation*}

There are three basic D7 branes labeled as $A=[1,0],~B=[1,-1], C=[1,1]$. 
One of the most interesting results of \cite{DeWolfe:1998eu} are that the
Kodaira singularities can be represented by configurations of D7 branes. The  brane configurations are listed in table. \ref{kodairabrane}.  Here we want to point out that: in the convention of \cite{DeWolfe:1998eu}, the monodromy for 
the brane configuration is the negative of that in the Kodaira's list! To match the monodromy group, one simply need to define the monodromy of 
the fundamental $[p,q]$ brane as $K^{'}_{[p,q]}=-K_{[p,q]}$.

 \begin{table}[htp]
\begin{center}
\resizebox{5in}{!}{\begin{tabular}{|c|c|c|c|c|c|}
\hline  
Name & Brane configuration & $n$ &$K$ &$f_K(p,q)$  \\ \hline
$I_n $ & $A^n$ & $n\geq 0$ &$\left( \begin{array}{cc} 1& -n \\ 0&1\end{array}\right)$& $-{1\over n}p^2$\\ \hline
$I_n^*$ & $A^{n+4}X_{[1,-1]}C=A^{n+4} BC$ & $n\geq 0$&$\left( \begin{array}{cc} -1& n \\ 0&-1\end{array}\right)$& ${n\over 4}q^2$  \\ \hline
$II,III,IV$  & $A^{n}X_{[0,-1]}C=A^{n+1}C$ &$n=0,1,2$ &$\left( \begin{array}{cc} 1& -n-1 \\ 1&-n\end{array}\right)$& ${1\over n+1}(-p^2+(n+1)pq-(n+1)q^2)$\\ \hline
$II^*,III^*,IV^*$ & $A^{n}X_{[2,-1]}C=A^{n-1}BC^2$&  $n=8,7,6$&$\left( \begin{array}{cc} -3& 3n-11 \\ -1&n-4\end{array}\right)$& ${1\over 9-n}(p^2+(1-n)pq+(3n-11)q^2)$ \\ \hline
\end{tabular}}
\end{center}
\caption{Brane configurations for the Kodaira singularity. Here the monodromy is computed using the brane configuration $A^nX_{[a,-1]}C$ representation. $f_K(p,q)$ is the quadratic form associated with the singularity.}
\label{kodairabrane}
\end{table}

There is a  remarkable brane configurations \cite{DeWolfe:1998pr}:
\begin{equation*}
A^8 B CBC.
\end{equation*}
and the total monodromy around  it is trivial. For later purpose, it can  be shown that the above brane system is equivalent to \cite{Fukae:1999zs}:
\begin{equation*}
A^9X_{[2,-1]}X_{[1,-2]}C.
\end{equation*}
For every such brane configuration, there is also a cyclic equivalence due to the trivial total monodromy, i.e. $X_1\ldots X_{11} X_{12} \sim X_{12} X_1 \ldots X_{11}$.

\textbf{Example 1}: In this example, we show that for the brane system $X_1X_2$, if the charge pairing $|z_1\cdot z_2|=1$,  they are
all equivalent by using global $SL(2,Z)$ transformation.  

\textbf{Proof}: It is always possible to use the $SL(2,Z)$ transformation to set the charge of $X_1$ to be $[1,0]$, so 
let's consider configuration $AX_{[a,1]}$ (we also use the fact $\pm [p,q]$ brane describes the same type of brane, so the $q$ charge of $X_2$ is set to be non-negative). 
We then use the $SL(2,Z)$ transformation
\begin{equation*}
K_A=\left(\begin{array}{cc}
1&-1\\
0&1 
\\
\end{array}\right).
\end{equation*}
which leaves the charge of $A$ brane invariant. Then let's use the transformation  $(K_A)^b$ to get the equivalence
\begin{equation*}
AX_{[a,1]}\sim AX_{[a^{'},1]}.
\end{equation*} 
Here $b=a-a^{'}$.

\textbf{Example 2}: Here we use the brane moves to prove the equivalence of $A^8BCBC=A^9X_{[2,-1]}X_{[1,-2]}C$. The moves are
\begin{align*}
& A^8BC(\underline{B}C)=A^7 (\underline{A} B)C^2X_{[3,1]}=A^7B(X_{[0,1]} \underline{C}^2) X_{[3,1]} \nonumber\\
& =A^7 (\underline{B} A^2) X_{[0,1]} X_{[3,1]}=A^9 X_{[3,-1]} (X_{[0,1]} \underline{X_{[3,1]}})=A^9 X_{[3,-1]} X_{[3,-2}] X_{[0,1]} \nonumber \\
&\rightarrow^{K_A^{-1}} A^9 X_{[2,-1]} X_{[1,-2]} X_{[1,1]} = A^9X_{[2,-1]}X_{[1,-2]}C.
\end{align*}
In every step, we move the brane with the line beneath it, and in the last step, we use a global $SL(2,Z)$ transformation $K_A$. 

 \begin{table}[htp]
\begin{center}
\resizebox{5in}{!}{\begin{tabular}{|l|l|l|l|}
\hline  
Theory & Brane configuration &Theory&Brane configuration \\ \hline
$(II,I_1^{10})$ &$A^8X_{[2,-1]}C ( X_{[-3,-1]}X_{[4,1]})$ &$(II,I_4 I_1^4)$ &$(A^4) A^4X_{[2,-1]}C ( X_{[-3,-1]}X_{[4,1]})$  \\ \hline
$(III,I_1^9)$ &$A^7X_{[2,-1]}C ( X_{[-3,-1]}^2X_{[4,1]})$& $(III,I_4 I_1^4)$ &$(A^4) A^3X_{[2,-1]}C ( X_{[-3,-1]}X_{[4,1]})$  \\ \hline
$(IV,I_1^8)$ &$A^6X_{[2,-1]}C ( X_{[-3,-1]}^3X_{[4,1]})$& $(IV,I_4 I_1^4)$ &$(A^4) A^2X_{[2,-1]}C ( X_{[-3,-1]}X_{[4,1]})$ \\ \hline
$(I_0^*,I_1^6)$ &$(A^4BC) A^4 BC$ &$(I_0^*,I_4I_1^2)$ &$(A^4BC) (A^4) BC$ \\ \hline
$(II^*,I_1^2)$ &$(A^8X_{[2,-1]}C) X_{[-3,-1]}X_{[4,1]}$&$(II,I_4^2I_1^2)$ &$(A^4) (A^4)X_{[2,-1]}C ( X_{[-3,-1]}X_{[4,1]})$\\ \hline
$(III^*,I_1^3)$ &$(A^7X_{[2,-1]}C) X_{[-3,-1]}^2X_{[4,1]}$&$(I_0^*,I_2^3)$ &$(A^4BC)(A^2)(X_{[0,1]}^2)(C^2)$  \\ \hline
$(IV^*,I_1^4)$ &$(A^6X_{[2,-1]}C)X_{[-3,-1]}^3X_{[4,1]}$&& \\ \hline
\end{tabular}}
\quad
\bigskip
\resizebox{4in}{!}{\begin{tabular}{|l|l|l|l|}
\hline  
Theory& Brane configuration &Theory& Brane configuration \\ \hline
$(II,III^*I_1)$ &$(A^6BC^2)A(AC)$ &$(II,I_0^* I_1^4)$ &$(A^4BC)A^3B(X_{[0,1]}A)$  \\ \hline
$(II,IV^*I_1^2)$ &$(A^5BC^2)A^2(CA)$& $(III,I_2^* I_1)$ &$(A^6BC)B(X^2_{[0,1]}C)$  \\ \hline
$(III,IV^*I_1)$ &$(A^5BC^2)(A^2 C)A$& $(III,I_1^* I_2)$ &$(A^5BC)(B^2) (X_{[0,1]}^2A)$ \\ \hline
$(II,IV^*I_2)$ &$(A^5BC^2)(A^2)(CA)$ &$(III,I_0^*I_1^3)$ &$(A^4BC) A^2B  (X_{[0,1]}^2A) $ \\ \hline
$(II,I_2^*I_1^2)$ &$(A^6BC)AB(X_{[0,1]}C)$&$(IV,I_0^*I_1^2)$ &$(A^4BC) AB  (X_{[0,1]}^3A)$\\ \hline
\end{tabular}}
\quad
\resizebox{4in}{!}{\begin{tabular}{|l|l|l|l|}
\hline  
Theory& Brane configuration &Theory& Brane configuration \\ \hline
$(I_4^*,I_1^2)$ &$(A^8BC)BC$ &$(I_2^*,I_2^2)$ &$(A^6BC)(B^2)(X_{[0,1]}^2)$  \\ \hline
$(I_3^*,I_1^3)$ &$(A^7BC)ABC$& $(I_1^*,I_4 I_1)$ &$(A^5BC)(B^4)X_{[1,-2]}$  \\ \hline
$(I_2^*,I_1^4)$ &$(A^6BC)A^2BC$& & \\ \hline
$(I_1^*,I_1^5)$ &$(A^5BC)A^3BC$& & \\ \hline
\end{tabular}}
\end{center}
\caption{Brane configurations for rank one $\mathcal{N}=2$ theory. The convention is: the brane within the parentheses can not be separated, i.e. they either represent the fiber at $\infty$ or the undeformable fiber at the bulk.}
\label{4dbranes}
\end{table}

Using above brane configuration $A^8BCBC$ and brane moves, we can 
realize all the possible singular fiber configuration of rational elliptic surface \cite{Fukae:1999zs}. The brane configurations for four dimensional $\mathcal{N}=2$ theories 
are listed in table. \ref{4dbranes}.

\newpage
\subsection{Mordell-Weil lattice and string junctions}
For each D7 brane configuration, one can define a charge lattice $\Gamma$ by using string junctions. These string junctions  are orthogonal to the collapsed branes.
The total dimension of the string junction lattice $\Gamma$ is equal to $2+f$, where $f$ is the rank of the flavor symmetry. A string junction has the charge $(p,q; w_i)$, where $w_i$ are 
the flavor charges. There is a sub-lattice  $\Gamma_0$ which is defined as the space of string junctions with zero $(p,q)$ charge, and 
so $\Gamma_0$ has dimension $rank(f)$. This lattice $\Gamma_0$ is closed related to the   Mordell-Weil lattice from which one can find out the flavor symmetry.

 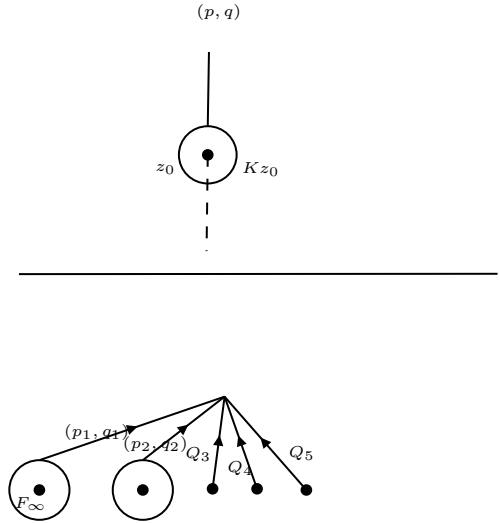
\begin{figure}[h]
\begin{center}
\tikzset{every picture/.style={line width=0.75pt}} 

\tikzset{every picture/.style={line width=0.75pt}} 

\begin{tikzpicture}[x=0.45pt,y=0.45pt,yscale=-1,xscale=1]

\draw  [fill={rgb, 255:red, 0; green, 0; blue, 0 }  ,fill opacity=1 ] (255,161) .. controls (255,158.79) and (256.79,157) .. (259,157) .. controls (261.21,157) and (263,158.79) .. (263,161) .. controls (263,163.21) and (261.21,165) .. (259,165) .. controls (256.79,165) and (255,163.21) .. (255,161) -- cycle ;
\draw   (235,161) .. controls (235,147.75) and (245.75,137) .. (259,137) .. controls (272.25,137) and (283,147.75) .. (283,161) .. controls (283,174.25) and (272.25,185) .. (259,185) .. controls (245.75,185) and (235,174.25) .. (235,161) -- cycle ;
\draw    (259,137) -- (260,74.72) ;
\draw  [dash pattern={on 4.5pt off 4.5pt}]  (259,161) -- (258,241.72) ;
\draw    (102,261) -- (500,260.72) ;
\draw   (180,442) .. controls (180,428.19) and (191.19,417) .. (205,417) .. controls (218.81,417) and (230,428.19) .. (230,442) .. controls (230,455.81) and (218.81,467) .. (205,467) .. controls (191.19,467) and (180,455.81) .. (180,442) -- cycle ;
\draw  [fill={rgb, 255:red, 0; green, 0; blue, 0 }  ,fill opacity=1 ] (115,442) .. controls (115,439.79) and (116.79,438) .. (119,438) .. controls (121.21,438) and (123,439.79) .. (123,442) .. controls (123,444.21) and (121.21,446) .. (119,446) .. controls (116.79,446) and (115,444.21) .. (115,442) -- cycle ;
\draw  [fill={rgb, 255:red, 0; green, 0; blue, 0 }  ,fill opacity=1 ] (201,442) .. controls (201,439.79) and (202.79,438) .. (205,438) .. controls (207.21,438) and (209,439.79) .. (209,442) .. controls (209,444.21) and (207.21,446) .. (205,446) .. controls (202.79,446) and (201,444.21) .. (201,442) -- cycle ;
\draw    (205,417) -- (273,363.72) ;
\draw [shift={(242.94,387.28)}, rotate = 141.92] [fill={rgb, 255:red, 0; green, 0; blue, 0 }  ][line width=0.08]  [draw opacity=0] (8.93,-4.29) -- (0,0) -- (8.93,4.29) -- cycle    ;
\draw  [fill={rgb, 255:red, 0; green, 0; blue, 0 }  ,fill opacity=1 ] (259,441) .. controls (259,438.79) and (260.79,437) .. (263,437) .. controls (265.21,437) and (267,438.79) .. (267,441) .. controls (267,443.21) and (265.21,445) .. (263,445) .. controls (260.79,445) and (259,443.21) .. (259,441) -- cycle ;
\draw  [fill={rgb, 255:red, 0; green, 0; blue, 0 }  ,fill opacity=1 ] (296,441) .. controls (296,438.79) and (297.79,437) .. (300,437) .. controls (302.21,437) and (304,438.79) .. (304,441) .. controls (304,443.21) and (302.21,445) .. (300,445) .. controls (297.79,445) and (296,443.21) .. (296,441) -- cycle ;
\draw  [fill={rgb, 255:red, 0; green, 0; blue, 0 }  ,fill opacity=1 ] (337,442) .. controls (337,439.79) and (338.79,438) .. (341,438) .. controls (343.21,438) and (345,439.79) .. (345,442) .. controls (345,444.21) and (343.21,446) .. (341,446) .. controls (338.79,446) and (337,444.21) .. (337,442) -- cycle ;
\draw    (273,363.72) -- (263,441) ;
\draw [shift={(268.83,395.91)}, rotate = 97.37] [fill={rgb, 255:red, 0; green, 0; blue, 0 }  ][line width=0.08]  [draw opacity=0] (8.93,-4.29) -- (0,0) -- (8.93,4.29) -- cycle    ;
\draw    (273,363.72) -- (300,441) ;
\draw [shift={(284.36,396.22)}, rotate = 70.74] [fill={rgb, 255:red, 0; green, 0; blue, 0 }  ][line width=0.08]  [draw opacity=0] (8.93,-4.29) -- (0,0) -- (8.93,4.29) -- cycle    ;
\draw    (273,363.72) -- (341,442) ;
\draw [shift={(302.74,397.95)}, rotate = 49.02] [fill={rgb, 255:red, 0; green, 0; blue, 0 }  ][line width=0.08]  [draw opacity=0] (8.93,-4.29) -- (0,0) -- (8.93,4.29) -- cycle    ;
\draw   (94,442) .. controls (94,428.19) and (105.19,417) .. (119,417) .. controls (132.81,417) and (144,428.19) .. (144,442) .. controls (144,455.81) and (132.81,467) .. (119,467) .. controls (105.19,467) and (94,455.81) .. (94,442) -- cycle ;
\draw    (119,417) -- (273,363.72) ;
\draw [shift={(200.73,388.72)}, rotate = 160.92] [fill={rgb, 255:red, 0; green, 0; blue, 0 }  ][line width=0.08]  [draw opacity=0] (8.93,-4.29) -- (0,0) -- (8.93,4.29) -- cycle    ;

\draw (247,32.4) node [anchor=north west][inner sep=0.75pt]  [font=\tiny]  {$( p,q)$};
\draw (213,166.4) node [anchor=north west][inner sep=0.75pt]   [font=\tiny]   {$z_{0}$};
\draw (285,164.4) node [anchor=north west][inner sep=0.75pt]  [font=\tiny]    {$Kz_{0}$};
\draw (96,445.4) node [anchor=north west][inner sep=0.75pt]     [font=\tiny] {$F_{\infty }$};
\draw (186,394.4) node [anchor=north west][inner sep=0.75pt]    [font=\tiny]  {$(p_2,q_2)$};
\draw (137,384.4) node [anchor=north west][inner sep=0.75pt]    [font=\tiny]  {$(p_1, q_1)$};
\draw (238,403.4) node [anchor=north west][inner sep=0.75pt]   [font=\tiny]   {$Q_{3}$};
\draw (273,415.4) node [anchor=north west][inner sep=0.75pt]  [font=\tiny]    {$Q_{4}$};
\draw (324,402.4) node [anchor=north west][inner sep=0.75pt]    [font=\tiny]  {$Q_{5}$};

\end{tikzpicture}

\end{center}
\caption{Up: A string junction around the branes, and this junction carries zero charge of the gauge algebra on the branes. Bottom: A string junction gives rise to
an element of Mordell-Weil lattice, and the asymptotical $(p,q)$ charge is zero.}
\label{string}
\end{figure}

In fact, Mordell-Weil lattice can be computed by looking at the string junction ending on D7 branes \cite{Fukae:1999zs}. Here a brief review will be given. 
The elements of Mordell-Weil group are represented by string junction with trivial $(p,q)$ charge, see figure. \ref{string}. 
For the collapsed branes (see the brane configuration listed in table. \ref{kodairabrane}: we use  $A^n$ for $I_n$ singularity, $A^{n+4}BC$ for $I_n^*$ singularity, and $A^{n-1}BC^2$ for $E_n$ singularity.),  one need to use the string loops around the branes, see upper diagram in figure. \ref{string}. The asymptotic charge for the loops around ADE singularity is given as follows:
\begin{align}
& A_n:~~\delta_{r,s}=-s(s_1+\ldots+s_{n+1}),~~\nonumber  \\
&~~~(p,q)=(-(n+1)s,0) \nonumber\\
&D_n:~\delta_{r,s}=-(s_1+\ldots+s_n)-(r-(n-1)s)s_{n+1}-(r-(n-3)s)s_{n+2} \nonumber\\
&~~~~~~~~~(p,q)=(-2r,-2s) \nonumber\\
&E_n:~\delta_{r,s}=-s(s_1+\ldots+s_{n-1})-(r-(n-2)s)s_{n} \nonumber\\
&~~~~~~~~-(r-(n-4)s)(s_{n+1}+s_{n+2}), \nonumber \\
&~~~~~~~~(p,q)=(-3r+(2n-9)s,-r+(n-6)s).
\label{stringloop}
\end{align}
Here $s_i$ represents the basic string going out of the $i$th seven brane, and has $(p,q)$ charge of the $i$th seven brane.
Notice that for $I_n$ type singularity, there is only one independent asymptotical charge,  while there are two for other type of singularities.
These junctions are singlet of the gauge algebra on the brane. The $s_i$ represents the basic string of the $i$th basic D7 brane, and there is a symmetric pairing for the basic string:
\begin{equation*}
(s_i,s_i)=-1,~~~(s_i,s_j)={1\over2}(p_iq_j-p_jq_i)~\text{for}~i<j.
\label{pair}
\end{equation*}
With this pairing, the string junctions form a lattice which is identified with the Mordell-Weil lattice.

\begin{figure}
\begin{center}

\tikzset{every picture/.style={line width=0.75pt}} 

\begin{tikzpicture}[x=0.65pt,y=0.65pt,yscale=-1,xscale=1]

\draw   (234,218) .. controls (234,199.77) and (248.77,185) .. (267,185) .. controls (285.23,185) and (300,199.77) .. (300,218) .. controls (300,236.23) and (285.23,251) .. (267,251) .. controls (248.77,251) and (234,236.23) .. (234,218) -- cycle ;
\draw    (283,189) -- (351,135.72) ;
\draw [shift={(320.94,159.28)}, rotate = 141.92] [fill={rgb, 255:red, 0; green, 0; blue, 0 }  ][line width=0.08]  [draw opacity=0] (8.93,-4.29) -- (0,0) -- (8.93,4.29) -- cycle    ;
\draw    (351,135.72) -- (419,208) ;
\draw [shift={(380.55,167.13)}, rotate = 46.75] [fill={rgb, 255:red, 0; green, 0; blue, 0 }  ][line width=0.08]  [draw opacity=0] (8.93,-4.29) -- (0,0) -- (8.93,4.29) -- cycle    ;
\draw    (351,135.72) -- (358,210.72) ;
\draw [shift={(353.9,166.75)}, rotate = 84.67] [fill={rgb, 255:red, 0; green, 0; blue, 0 }  ][line width=0.08]  [draw opacity=0] (8.93,-4.29) -- (0,0) -- (8.93,4.29) -- cycle    ;
\draw   (147,219) .. controls (147,200.22) and (162.22,185) .. (181,185) .. controls (199.78,185) and (215,200.22) .. (215,219) .. controls (215,237.78) and (199.78,253) .. (181,253) .. controls (162.22,253) and (147,237.78) .. (147,219) -- cycle ;
\draw    (181,185) -- (351,135.72) ;
\draw [shift={(270.8,158.97)}, rotate = 163.83] [fill={rgb, 255:red, 0; green, 0; blue, 0 }  ][line width=0.08]  [draw opacity=0] (8.93,-4.29) -- (0,0) -- (8.93,4.29) -- cycle    ;

\draw (250,170.4) node [anchor=north west][inner sep=0.75pt]   [font=\tiny]  {$(-4Q_{1},0)$};
\draw (190,152.4) node [anchor=north west][inner sep=0.75pt]   [font=\tiny]  {$(-2r,-2s)$};
\draw (257,210.4) node [anchor=north west][inner sep=0.75pt]  [font=\tiny]   {$A^{4}$};
\draw (377,184.4) node [anchor=north west][inner sep=0.75pt]  [font=\tiny]   {$Q_{4}$};
\draw (168,205.4) node [anchor=north west][inner sep=0.75pt]  [font=\tiny]  {$A^{4} BC$};
\draw (351,215.4) node [anchor=north west][inner sep=0.75pt]   [font=\tiny]  {$B$};
\draw (411,211.4) node [anchor=north west][inner sep=0.75pt]   [font=\tiny]  {$C$};
\draw (326,168.4) node [anchor=north west][inner sep=0.75pt]  [font=\tiny]   {$Q_{3}$};

\end{tikzpicture}

\end{center}
\caption{A string junction with zero $(p,q)$ charge for the brane configuration $(A^4BC) (A^4) BC$. If all the charge numbers are integral, it gives rise to the narrow Mordell-Weil lattice. If we allow the fractional $r,s, Q_1$ (so that the asymptotical charge is still integral), then this gives the torsion free
part of the MW lattice. To find out the torsion part of the lattice, we require the self-intersection number of the junction to be zero; and the common divisor of these fractional charge gives the torsion part.}
\label{sutwostar}
\end{figure}
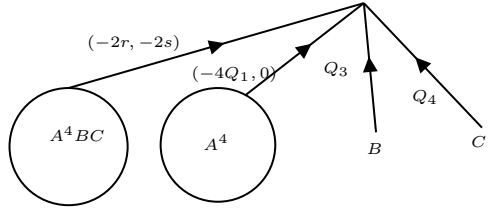

\textbf{Example 1}:  Let's consider the configuration $(A^4BC)(A^4)BC$, which gives the $SU(2)$ $\mathcal{N}=2^*$ gauge theory. A generic junction which is orthogonal to the gauge algebra on the collapsed branes are represented in figure. \ref{sutwostar}, and $J$ has the representation
\begin{equation*}
J=-s(s_1+s_2+s_3+s_4)-(r-3s)s_5-(r-s)s_{6}-Q_1(s_7+s_8+s_9+s_{10})+Q_3 s_{11}+Q_4 s_{12}.
\end{equation*}
Here $(r,s)$ is the charge $z_0$ for $D_4$ singularity,  and the asymptotical charge is $(-2r,-2s)$, see formula. \ref{stringloop};
$(r,Q_1)$ is the $z_0$ charge for the $I_4$ singularity, so the asymptotical charge is $(-4Q_1,0)$. 
The total $(p,q)$ charge of $J$ has to vanish:
\begin{equation*}
(-2r,-2s)+(-4Q_1,0)+Q_3(1,-1)+Q_4(1,1)=0.
\end{equation*}
We find the equations
\begin{align}
&-2s-Q_3+Q_4=0, \nonumber\\
&-2r-4Q_1+Q_3+Q_4=0.
\label{trivialpq}
\end{align}
There is a total of five variables and two equations, so we find a three dimensional space. 

There are also string junctions whose self-intersection number are zero, and these gave the torsion part of the Mordell-Weil lattice. The dimension of such null junction is two. In the current example,
(see  formula. \ref{pair} for the definition of  \textbf{symmetric} pairing ):
\begin{align*}
&J^2=-4s^2-(r-3s)^2-(r-s)^2-(r-3s)4s+(r-s)4s+2(r-s)(r-3s)\nonumber\\
&-4Q_1^2-Q_3^2-Q_4^2+4Q_1((r-3s)-(r-s))+Q_3(4s+2(r-s)+4Q_1)+ \nonumber\\
&Q_4(-4s-2(r-3s)-4Q_1+2Q_3).
\end{align*}
Substitute the result of $r,s$ of equations \ref{trivialpq} into above equation, we get
\begin{equation}
J^2=(2 Q_1 - Q_3 + Q_4)^2.
\end{equation}
So null junction span a two dimensional space.  Junctions with non-zero self-intersection number form a lattice, which is identified with the lattice $L$ (which is called narrow Mordell-Weil lattice). 
The generator for the lattice $L$ is
\begin{equation*}
Q_1=-1,~~Q_3=Q_4=-2,~\to~J=(s_7+s_8+s_9+s_{10})-2s_{11}-2s_{12},~~J^2=-4.
\end{equation*}

To find out the full Mordell-Weil lattice (the torsion free part), we need to relax the condition for the loop junction: the charge $z_0$ could be fractional as long as the external charge is integral.  For example, for a $I_{n}$ singularity, the charge $s$ can be taken as ${a\over n},~~a=0,\ldots, n$, 
and the external charge $(p,q)=(-a,0),~~~a=0,\ldots,n$. These fractional junctions form the full Mordell-Weil lattice. For the current example, the above condition implies that $Q_1$ could take values in ${1\over 4}$, while $r, s$ could take value in ${1\over 2}$. In fact, the lattice is generated by the charges
\begin{equation*}
Q_1=-{1\over4},~~~Q_3=-1,Q_4=0,~r=0,s={1\over2}.
\end{equation*}

The torsion part is represented by fractional null junctions.  The charge for null junction satisfies the condition
\begin{equation*}
2 Q_1 - Q_3 + Q_4=0.
\end{equation*}
so a null junction is represented as (here $Q_3, Q_4$ is used as the coordinate):
\begin{align*}
&J_{null}=-{Q_4-Q_3\over2}(s_1+s_2+s_3+s_4)-Q_3s_5 \nonumber\\
&-Q_4s_{6}-{Q_3-Q_4\over2}(s_7+s_8+s_9+s_{10})+Q_3 s_{11}+Q_4 s_{12}.
\end{align*}
The integral condition for the null junction is $2(Q_3-Q_4)=2n$ (to get a junction with integral charge), and so  a $Z_2$ torsion group is derived.

\subsection{BPS quiver and BPS states}
\begin{figure}[h]
\begin{center}

\tikzset{every picture/.style={line width=0.75pt}} 

\tikzset{every picture/.style={line width=0.75pt}} 

\begin{tikzpicture}[x=0.45pt,y=0.45pt,yscale=-1,xscale=1]

\draw   (234,218) .. controls (234,199.77) and (248.77,185) .. (267,185) .. controls (285.23,185) and (300,199.77) .. (300,218) .. controls (300,236.23) and (285.23,251) .. (267,251) .. controls (248.77,251) and (234,236.23) .. (234,218) -- cycle ;
\draw    (283,189) -- (351,135.72) ;
\draw [shift={(320.94,159.28)}, rotate = 141.92] [fill={rgb, 255:red, 0; green, 0; blue, 0 }  ][line width=0.08]  [draw opacity=0] (8.93,-4.29) -- (0,0) -- (8.93,4.29) -- cycle    ;
\draw    (351,135.72) -- (422,219) ;
\draw [shift={(382.28,172.41)}, rotate = 49.55] [fill={rgb, 255:red, 0; green, 0; blue, 0 }  ][line width=0.08]  [draw opacity=0] (8.93,-4.29) -- (0,0) -- (8.93,4.29) -- cycle    ;
\draw    (351,135.72) -- (359,223) ;
\draw [shift={(354.41,172.89)}, rotate = 84.76] [fill={rgb, 255:red, 0; green, 0; blue, 0 }  ][line width=0.08]  [draw opacity=0] (8.93,-4.29) -- (0,0) -- (8.93,4.29) -- cycle    ;
\draw  [fill={rgb, 255:red, 0; green, 0; blue, 0 }  ,fill opacity=1 ] (177,223) .. controls (177,220.79) and (178.79,219) .. (181,219) .. controls (183.21,219) and (185,220.79) .. (185,223) .. controls (185,225.21) and (183.21,227) .. (181,227) .. controls (178.79,227) and (177,225.21) .. (177,223) -- cycle ;
\draw  [fill={rgb, 255:red, 0; green, 0; blue, 0 }  ,fill opacity=1 ] (263,222) .. controls (263,219.79) and (264.79,218) .. (267,218) .. controls (269.21,218) and (271,219.79) .. (271,222) .. controls (271,224.21) and (269.21,226) .. (267,226) .. controls (264.79,226) and (263,224.21) .. (263,222) -- cycle ;
\draw  [fill={rgb, 255:red, 0; green, 0; blue, 0 }  ,fill opacity=1 ] (355,223) .. controls (355,220.79) and (356.79,219) .. (359,219) .. controls (361.21,219) and (363,220.79) .. (363,223) .. controls (363,225.21) and (361.21,227) .. (359,227) .. controls (356.79,227) and (355,225.21) .. (355,223) -- cycle ;
\draw  [fill={rgb, 255:red, 0; green, 0; blue, 0 }  ,fill opacity=1 ] (418,219) .. controls (418,216.79) and (419.79,215) .. (422,215) .. controls (424.21,215) and (426,216.79) .. (426,219) .. controls (426,221.21) and (424.21,223) .. (422,223) .. controls (419.79,223) and (418,221.21) .. (418,219) -- cycle ;

\draw (276,157.4) node [anchor=north west][inner sep=0.75pt]   [font=\tiny] {$Q_{2}$};
\draw (400,166.4) node [anchor=north west][inner sep=0.75pt]   [font=\tiny]   {$Q_{4}$};
\draw (326,168.4) node [anchor=north west][inner sep=0.75pt]    [font=\tiny]  {$Q_{3}$};
\draw (166,185.4) node [anchor=north west][inner sep=0.75pt]   [font=\tiny]   {$F_{\infty }$};

\end{tikzpicture}

\end{center}
\caption{Open string junction ending between D3 and D7 brane gives the BPS particle. The black dots represent $D7$ brane.}

\label{bpsone}
\end{figure}
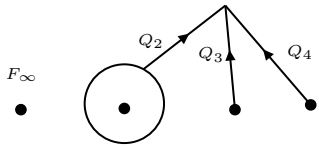

One can use  D3 brane probe to study the properties of $\mathcal{N}=2$ system represented by 
D7 brane systems \cite{Banks:1996nj}. In particular, the BPS particles are represented by open string ending between
D3 brane and D7 brane \cite{Sen:1996sk,Mikhailov:1998bx}, more generally, one could use the open string junction to get BPS states, see figure. \ref{bpsone}.

On the other hand, to find out the spectrum of BPS particles, a very useful tool is the so-called BPS quiver from which one could derive the spectrum by using green mutation \cite{Alim:2011kw,Xie:2012gd}. 
It is possible to find out the BPS quiver from the brane configuration for the corresponding 4d $\mathcal{N}=2$ theories.  Let's fix a generic 
point at $\mathbb{P}^1$ where there is no D7 brane, and add a D3 brane probe at this point. The BPS particle at this point 
can be represented by the open strings ending between D3 brane and D7 branes. If we consider a string ends on $[p,q]$ type D7 brane and D3 brane, 
then this string has electric-magnetic charge $[p,q]$, which gives a stable BPS particle with charge $[p,q]$.  

Let's first focus on the $I_1$ type theories, namely 
the generic deformation has only type $I_1$  singular fibers. For each $I_1$ fiber, there is an associated BPS particle whose electric-magnetic charge 
is given by the type of branes $z_i$; One can 
find a quiver by using the Dirac paring for the charges associated with singular fibers.
The generalization to  $I_n$ type theory is straightforward.
For the $I_n$ fibers which is represented by the branes $(X_{[z]}^n)$,  the corresponding BPS particle has charge $\sqrt{n} z$.

One need to solve following two problems: firstly if the ordering of the branes are changed, the charge vectors would change (see the formula \ref{moverule}); but the BPS quiver does depend on 
the ordering (quivers from different ordering  generally are not related by quiver mutation), so to find out the BPS quiver one need to find a special ordering; Secondly,  the charge vectors for the brane could be either $z$ or $-z$, and one need a way to fix the choice. 

The BPS quiver for those theories are found in \cite{Caorsi:2019vex}, and the above two problems can be solved. First, the special ordering for the brane configurations are given as follows
\begin{align*}
&II^*,III^*, IV^*:A^{n}X_{[2,-1]}C,~n=8,7,6 \nonumber\\
 &I_n^*: ~~~~~~~~~~~~~~~~A^{n+4}X_{[1,-1]}C,~~n=0,1,2 \nonumber\\
&II,III, IV:~A^{n}X_{[0,-1]}C,~~n=0,1,2.
\end{align*}
There is a general pattern for above brane configurations: each sequence has a core of the type $X_{[z]} C$, and then a 
stack of $A$  branes are added on top of the core. The special ordering for $I_4$ and $I_2$ type theories are found by collapsing adjacent $A$  branes of $I_1$ theories.

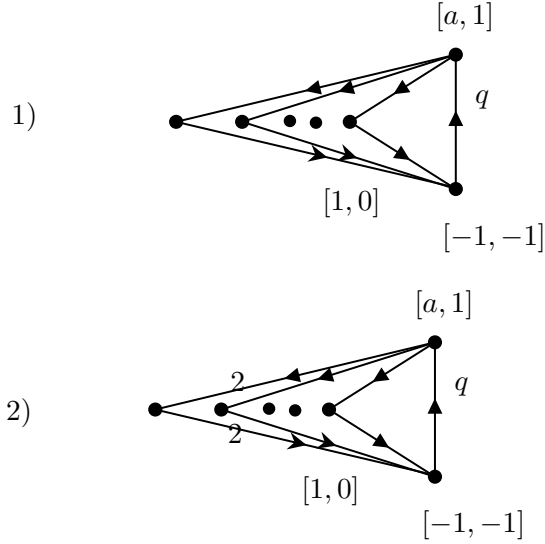
\begin{figure}
\begin{center}

\tikzset{every picture/.style={line width=0.75pt}} 

\begin{tikzpicture}[x=0.65pt,y=0.65pt,yscale=-1,xscale=1]

\draw  [fill={rgb, 255:red, 0; green, 0; blue, 0 }  ,fill opacity=1 ] (401,62.5) .. controls (401,60.57) and (402.57,59) .. (404.5,59) .. controls (406.43,59) and (408,60.57) .. (408,62.5) .. controls (408,64.43) and (406.43,66) .. (404.5,66) .. controls (402.57,66) and (401,64.43) .. (401,62.5) -- cycle ;
\draw  [fill={rgb, 255:red, 0; green, 0; blue, 0 }  ,fill opacity=1 ] (401,140.5) .. controls (401,138.57) and (402.57,137) .. (404.5,137) .. controls (406.43,137) and (408,138.57) .. (408,140.5) .. controls (408,142.43) and (406.43,144) .. (404.5,144) .. controls (402.57,144) and (401,142.43) .. (401,140.5) -- cycle ;
\draw    (404.5,62.5) -- (404.5,140.5) ;
\draw [shift={(404.5,95)}, rotate = 90] [fill={rgb, 255:red, 0; green, 0; blue, 0 }  ][line width=0.08]  [draw opacity=0] (8.93,-4.29) -- (0,0) -- (8.93,4.29) -- cycle    ;
\draw  [fill={rgb, 255:red, 0; green, 0; blue, 0 }  ,fill opacity=1 ] (340,101.5) .. controls (340,99.57) and (341.57,98) .. (343.5,98) .. controls (345.43,98) and (347,99.57) .. (347,101.5) .. controls (347,103.43) and (345.43,105) .. (343.5,105) .. controls (341.57,105) and (340,103.43) .. (340,101.5) -- cycle ;
\draw    (343.5,101.5) -- (404.5,140.5) ;
\draw [shift={(378.21,123.69)}, rotate = 212.59] [fill={rgb, 255:red, 0; green, 0; blue, 0 }  ][line width=0.08]  [draw opacity=0] (8.93,-4.29) -- (0,0) -- (8.93,4.29) -- cycle    ;
\draw    (343.5,101.5) -- (404.5,62.5) ;
\draw [shift={(368.52,85.5)}, rotate = 327.41] [fill={rgb, 255:red, 0; green, 0; blue, 0 }  ][line width=0.08]  [draw opacity=0] (8.93,-4.29) -- (0,0) -- (8.93,4.29) -- cycle    ;
\draw  [fill={rgb, 255:red, 0; green, 0; blue, 0 }  ,fill opacity=1 ] (240,101.5) .. controls (240,99.57) and (241.57,98) .. (243.5,98) .. controls (245.43,98) and (247,99.57) .. (247,101.5) .. controls (247,103.43) and (245.43,105) .. (243.5,105) .. controls (241.57,105) and (240,103.43) .. (240,101.5) -- cycle ;
\draw  [fill={rgb, 255:red, 0; green, 0; blue, 0 }  ,fill opacity=1 ] (278,101.5) .. controls (278,99.57) and (279.57,98) .. (281.5,98) .. controls (283.43,98) and (285,99.57) .. (285,101.5) .. controls (285,103.43) and (283.43,105) .. (281.5,105) .. controls (279.57,105) and (278,103.43) .. (278,101.5) -- cycle ;
\draw    (281.5,101.5) -- (404.5,62.5) ;
\draw [shift={(336.8,83.96)}, rotate = 342.41] [fill={rgb, 255:red, 0; green, 0; blue, 0 }  ][line width=0.08]  [draw opacity=0] (8.93,-4.29) -- (0,0) -- (8.93,4.29) -- cycle    ;
\draw    (281.5,101.5) -- (404.5,140.5) ;
\draw [shift={(347.77,122.51)}, rotate = 197.59] [fill={rgb, 255:red, 0; green, 0; blue, 0 }  ][line width=0.08]  [draw opacity=0] (10.72,-5.15) -- (0,0) -- (10.72,5.15) -- (7.12,0) -- cycle    ;
\draw    (243.5,101.5) -- (408,140.5) ;
\draw [shift={(330.62,122.15)}, rotate = 193.34] [fill={rgb, 255:red, 0; green, 0; blue, 0 }  ][line width=0.08]  [draw opacity=0] (10.72,-5.15) -- (0,0) -- (10.72,5.15) -- (7.12,0) -- cycle    ;
\draw    (243.5,101.5) -- (404.5,62.5) ;
\draw [shift={(317.68,83.53)}, rotate = 346.38] [fill={rgb, 255:red, 0; green, 0; blue, 0 }  ][line width=0.08]  [draw opacity=0] (8.93,-4.29) -- (0,0) -- (8.93,4.29) -- cycle    ;
\draw  [fill={rgb, 255:red, 0; green, 0; blue, 0 }  ,fill opacity=1 ] (306,101) .. controls (306,99.34) and (307.34,98) .. (309,98) .. controls (310.66,98) and (312,99.34) .. (312,101) .. controls (312,102.66) and (310.66,104) .. (309,104) .. controls (307.34,104) and (306,102.66) .. (306,101) -- cycle ;
\draw  [fill={rgb, 255:red, 0; green, 0; blue, 0 }  ,fill opacity=1 ] (321,102) .. controls (321,100.34) and (322.34,99) .. (324,99) .. controls (325.66,99) and (327,100.34) .. (327,102) .. controls (327,103.66) and (325.66,105) .. (324,105) .. controls (322.34,105) and (321,103.66) .. (321,102) -- cycle ;
\draw  [fill={rgb, 255:red, 0; green, 0; blue, 0 }  ,fill opacity=1 ] (389,229.5) .. controls (389,227.57) and (390.57,226) .. (392.5,226) .. controls (394.43,226) and (396,227.57) .. (396,229.5) .. controls (396,231.43) and (394.43,233) .. (392.5,233) .. controls (390.57,233) and (389,231.43) .. (389,229.5) -- cycle ;
\draw  [fill={rgb, 255:red, 0; green, 0; blue, 0 }  ,fill opacity=1 ] (389,307.5) .. controls (389,305.57) and (390.57,304) .. (392.5,304) .. controls (394.43,304) and (396,305.57) .. (396,307.5) .. controls (396,309.43) and (394.43,311) .. (392.5,311) .. controls (390.57,311) and (389,309.43) .. (389,307.5) -- cycle ;
\draw    (392.5,229.5) -- (392.5,307.5) ;
\draw [shift={(392.5,262)}, rotate = 90] [fill={rgb, 255:red, 0; green, 0; blue, 0 }  ][line width=0.08]  [draw opacity=0] (8.93,-4.29) -- (0,0) -- (8.93,4.29) -- cycle    ;
\draw  [fill={rgb, 255:red, 0; green, 0; blue, 0 }  ,fill opacity=1 ] (328,268.5) .. controls (328,266.57) and (329.57,265) .. (331.5,265) .. controls (333.43,265) and (335,266.57) .. (335,268.5) .. controls (335,270.43) and (333.43,272) .. (331.5,272) .. controls (329.57,272) and (328,270.43) .. (328,268.5) -- cycle ;
\draw    (331.5,268.5) -- (392.5,307.5) ;
\draw [shift={(366.21,290.69)}, rotate = 212.59] [fill={rgb, 255:red, 0; green, 0; blue, 0 }  ][line width=0.08]  [draw opacity=0] (8.93,-4.29) -- (0,0) -- (8.93,4.29) -- cycle    ;
\draw    (331.5,268.5) -- (392.5,229.5) ;
\draw [shift={(356.52,252.5)}, rotate = 327.41] [fill={rgb, 255:red, 0; green, 0; blue, 0 }  ][line width=0.08]  [draw opacity=0] (8.93,-4.29) -- (0,0) -- (8.93,4.29) -- cycle    ;
\draw  [fill={rgb, 255:red, 0; green, 0; blue, 0 }  ,fill opacity=1 ] (228,268.5) .. controls (228,266.57) and (229.57,265) .. (231.5,265) .. controls (233.43,265) and (235,266.57) .. (235,268.5) .. controls (235,270.43) and (233.43,272) .. (231.5,272) .. controls (229.57,272) and (228,270.43) .. (228,268.5) -- cycle ;
\draw  [fill={rgb, 255:red, 0; green, 0; blue, 0 }  ,fill opacity=1 ] (266,268.5) .. controls (266,266.57) and (267.57,265) .. (269.5,265) .. controls (271.43,265) and (273,266.57) .. (273,268.5) .. controls (273,270.43) and (271.43,272) .. (269.5,272) .. controls (267.57,272) and (266,270.43) .. (266,268.5) -- cycle ;
\draw    (269.5,268.5) -- (392.5,229.5) ;
\draw [shift={(324.8,250.96)}, rotate = 342.41] [fill={rgb, 255:red, 0; green, 0; blue, 0 }  ][line width=0.08]  [draw opacity=0] (8.93,-4.29) -- (0,0) -- (8.93,4.29) -- cycle    ;
\draw    (269.5,268.5) -- (392.5,307.5) ;
\draw [shift={(335.77,289.51)}, rotate = 197.59] [fill={rgb, 255:red, 0; green, 0; blue, 0 }  ][line width=0.08]  [draw opacity=0] (10.72,-5.15) -- (0,0) -- (10.72,5.15) -- (7.12,0) -- cycle    ;
\draw    (231.5,268.5) -- (396,307.5) ;
\draw [shift={(318.62,289.15)}, rotate = 193.34] [fill={rgb, 255:red, 0; green, 0; blue, 0 }  ][line width=0.08]  [draw opacity=0] (10.72,-5.15) -- (0,0) -- (10.72,5.15) -- (7.12,0) -- cycle    ;
\draw    (231.5,268.5) -- (392.5,229.5) ;
\draw [shift={(305.68,250.53)}, rotate = 346.38] [fill={rgb, 255:red, 0; green, 0; blue, 0 }  ][line width=0.08]  [draw opacity=0] (8.93,-4.29) -- (0,0) -- (8.93,4.29) -- cycle    ;
\draw  [fill={rgb, 255:red, 0; green, 0; blue, 0 }  ,fill opacity=1 ] (294,268) .. controls (294,266.34) and (295.34,265) .. (297,265) .. controls (298.66,265) and (300,266.34) .. (300,268) .. controls (300,269.66) and (298.66,271) .. (297,271) .. controls (295.34,271) and (294,269.66) .. (294,268) -- cycle ;
\draw  [fill={rgb, 255:red, 0; green, 0; blue, 0 }  ,fill opacity=1 ] (309,269) .. controls (309,267.34) and (310.34,266) .. (312,266) .. controls (313.66,266) and (315,267.34) .. (315,269) .. controls (315,270.66) and (313.66,272) .. (312,272) .. controls (310.34,272) and (309,270.66) .. (309,269) -- cycle ;

\draw (391,30.4) node [anchor=north west][inner sep=0.75pt]    {$[ a,1]$};
\draw (395,158.4) node [anchor=north west][inner sep=0.75pt]    {$[ -1,-1]$};
\draw (326,138.4) node [anchor=north west][inner sep=0.75pt]    {$[ 1,0]$};
\draw (414,81.4) node [anchor=north west][inner sep=0.75pt]    {$q$};
\draw (147,88.4) node [anchor=north west][inner sep=0.75pt]    {$1)$};
\draw (144,260.4) node [anchor=north west][inner sep=0.75pt]    {$2)$};
\draw (379,197.4) node [anchor=north west][inner sep=0.75pt]    {$[ a,1]$};
\draw (383,325.4) node [anchor=north west][inner sep=0.75pt]    {$[ -1,-1]$};
\draw (314,305.4) node [anchor=north west][inner sep=0.75pt]    {$[ 1,0]$};
\draw (402,248.4) node [anchor=north west][inner sep=0.75pt]    {$q$};
\draw (271.5,275.4) node [anchor=north west][inner sep=0.75pt]    {$2$};
\draw (273,245.4) node [anchor=north west][inner sep=0.75pt]    {$2$};

\end{tikzpicture}

\end{center}
\caption{The BPS quiver for $I_1$ and $I_4$  series. $a=-2$ and so $q=3$ for $E_n$ type theories; $a=-1$ and $q=2$ for $I_n^*$ type theories; $a=0$ and $q=1$ for $H_n$ type theories.}
\label{bps1}
\end{figure}

Secondly, the rule for fix the sign ambiguity of the brane charges are following: the charge is $[1,0]$ for the first sequence of $A^n$ branes;  the charge is $X_{[a,1]},~~a=-2,-1,0$ for second one; finally  the 
charge is $[-1,-1]$ for $C$ brane. The adjacency matrix $Q_{ij}$ of the quiver is given as follows:
\begin{equation}
Q_{ij}=z_i\cdot z_j,~~i<j,
\end{equation}
and the full matrix $Q$ is fixed by imposing  anti-symmetric condition. Using the charge vector $z_1=[1,0],~~z_2=[a,1],~~z_3=[-1,-1]$, we get the matrix
\begin{equation*}
Q=\left(\begin{array}{ccc}
0&1&-1\\
-1&0&-a+1\\
1&a-1&0
\end{array}\right).
\end{equation*}
The BPS quiver can be read from above matrix and take the form shown in figure. \ref{bps1}, which is exactly the same as that found in \cite{Caorsi:2019vex}.

\begin{figure}[h]
\begin{center}

\tikzset{every picture/.style={line width=0.75pt}} 

\begin{tikzpicture}[x=0.55pt,y=0.55pt,yscale=-1,xscale=1]

\draw  [fill={rgb, 255:red, 0; green, 0; blue, 0 }  ,fill opacity=1 ] (401,62.5) .. controls (401,60.57) and (402.57,59) .. (404.5,59) .. controls (406.43,59) and (408,60.57) .. (408,62.5) .. controls (408,64.43) and (406.43,66) .. (404.5,66) .. controls (402.57,66) and (401,64.43) .. (401,62.5) -- cycle ;
\draw  [fill={rgb, 255:red, 0; green, 0; blue, 0 }  ,fill opacity=1 ] (401,140.5) .. controls (401,138.57) and (402.57,137) .. (404.5,137) .. controls (406.43,137) and (408,138.57) .. (408,140.5) .. controls (408,142.43) and (406.43,144) .. (404.5,144) .. controls (402.57,144) and (401,142.43) .. (401,140.5) -- cycle ;
\draw    (404.5,62.5) -- (404.5,140.5) ;
\draw [shift={(404.5,95)}, rotate = 90] [fill={rgb, 255:red, 0; green, 0; blue, 0 }  ][line width=0.08]  [draw opacity=0] (8.93,-4.29) -- (0,0) -- (8.93,4.29) -- cycle    ;
\draw  [fill={rgb, 255:red, 0; green, 0; blue, 0 }  ,fill opacity=1 ] (340,101.5) .. controls (340,99.57) and (341.57,98) .. (343.5,98) .. controls (345.43,98) and (347,99.57) .. (347,101.5) .. controls (347,103.43) and (345.43,105) .. (343.5,105) .. controls (341.57,105) and (340,103.43) .. (340,101.5) -- cycle ;
\draw    (343.5,101.5) -- (404.5,140.5) ;
\draw [shift={(378.21,123.69)}, rotate = 212.59] [fill={rgb, 255:red, 0; green, 0; blue, 0 }  ][line width=0.08]  [draw opacity=0] (8.93,-4.29) -- (0,0) -- (8.93,4.29) -- cycle    ;
\draw    (343.5,101.5) -- (404.5,62.5) ;
\draw [shift={(368.52,85.5)}, rotate = 327.41] [fill={rgb, 255:red, 0; green, 0; blue, 0 }  ][line width=0.08]  [draw opacity=0] (8.93,-4.29) -- (0,0) -- (8.93,4.29) -- cycle    ;
\draw  [fill={rgb, 255:red, 0; green, 0; blue, 0 }  ,fill opacity=1 ] (278,101.5) .. controls (278,99.57) and (279.57,98) .. (281.5,98) .. controls (283.43,98) and (285,99.57) .. (285,101.5) .. controls (285,103.43) and (283.43,105) .. (281.5,105) .. controls (279.57,105) and (278,103.43) .. (278,101.5) -- cycle ;
\draw    (281.5,101.5) -- (404.5,62.5) ;
\draw [shift={(336.8,83.96)}, rotate = 342.41] [fill={rgb, 255:red, 0; green, 0; blue, 0 }  ][line width=0.08]  [draw opacity=0] (8.93,-4.29) -- (0,0) -- (8.93,4.29) -- cycle    ;
\draw    (281.5,101.5) -- (404.5,140.5) ;
\draw [shift={(347.77,122.51)}, rotate = 197.59] [fill={rgb, 255:red, 0; green, 0; blue, 0 }  ][line width=0.08]  [draw opacity=0] (10.72,-5.15) -- (0,0) -- (10.72,5.15) -- (7.12,0) -- cycle    ;
\draw  [fill={rgb, 255:red, 0; green, 0; blue, 0 }  ,fill opacity=1 ] (400,252.5) .. controls (400,250.57) and (401.57,249) .. (403.5,249) .. controls (405.43,249) and (407,250.57) .. (407,252.5) .. controls (407,254.43) and (405.43,256) .. (403.5,256) .. controls (401.57,256) and (400,254.43) .. (400,252.5) -- cycle ;
\draw  [fill={rgb, 255:red, 0; green, 0; blue, 0 }  ,fill opacity=1 ] (400,330.5) .. controls (400,328.57) and (401.57,327) .. (403.5,327) .. controls (405.43,327) and (407,328.57) .. (407,330.5) .. controls (407,332.43) and (405.43,334) .. (403.5,334) .. controls (401.57,334) and (400,332.43) .. (400,330.5) -- cycle ;
\draw    (403.5,252.5) -- (403.5,330.5) ;
\draw [shift={(403.5,285)}, rotate = 90] [fill={rgb, 255:red, 0; green, 0; blue, 0 }  ][line width=0.08]  [draw opacity=0] (8.93,-4.29) -- (0,0) -- (8.93,4.29) -- cycle    ;
\draw  [fill={rgb, 255:red, 0; green, 0; blue, 0 }  ,fill opacity=1 ] (339,291.5) .. controls (339,289.57) and (340.57,288) .. (342.5,288) .. controls (344.43,288) and (346,289.57) .. (346,291.5) .. controls (346,293.43) and (344.43,295) .. (342.5,295) .. controls (340.57,295) and (339,293.43) .. (339,291.5) -- cycle ;
\draw    (342.5,291.5) -- (403.5,330.5) ;
\draw [shift={(377.21,313.69)}, rotate = 212.59] [fill={rgb, 255:red, 0; green, 0; blue, 0 }  ][line width=0.08]  [draw opacity=0] (8.93,-4.29) -- (0,0) -- (8.93,4.29) -- cycle    ;
\draw    (342.5,291.5) -- (403.5,252.5) ;
\draw [shift={(367.52,275.5)}, rotate = 327.41] [fill={rgb, 255:red, 0; green, 0; blue, 0 }  ][line width=0.08]  [draw opacity=0] (8.93,-4.29) -- (0,0) -- (8.93,4.29) -- cycle    ;

\draw (390,30.4) node [anchor=north west][inner sep=0.75pt]    {$_{[ -2,1]}$};
\draw (394,154.4) node [anchor=north west][inner sep=0.75pt]    {$_{[ 1,1]}$};
\draw (414,81.4) node [anchor=north west][inner sep=0.75pt]    {$3$};
\draw (147,88.4) node [anchor=north west][inner sep=0.75pt]    {$1)$};
\draw (144,260.4) node [anchor=north west][inner sep=0.75pt]    {$2)$};
\draw (311,79.4) node [anchor=north west][inner sep=0.75pt]    {$2$};
\draw (312,106.4) node [anchor=north west][inner sep=0.75pt]    {$2$};
\draw (356,81.4) node [anchor=north west][inner sep=0.75pt]    {$2$};
\draw (356,105.4) node [anchor=north west][inner sep=0.75pt]    {$2$};
\draw (258,106.4) node [anchor=north west][inner sep=0.75pt]    {$_{2[ 1,0]}$};
\draw (389,220.4) node [anchor=north west][inner sep=0.75pt]    {$_{\sqrt{2}[ 0,-1]}$};
\draw (414,271.4) node [anchor=north west][inner sep=0.75pt]    {$2$};
\draw (355,271.4) node [anchor=north west][inner sep=0.75pt]    {$2$};
\draw (355,295.4) node [anchor=north west][inner sep=0.75pt]    {$2$};
\draw (291,273.4) node [anchor=north west][inner sep=0.75pt]    {$_{\sqrt{2}[ 1,0]}$};
\draw (382,330.4) node [anchor=north west][inner sep=0.75pt]    {$_{\sqrt{2}[ 1,-1]}$};
\draw (328,123.4) node [anchor=north west][inner sep=0.75pt]    {$_{2[ 1,0]}$};

\end{tikzpicture}

\end{center}
\caption{The BPS quiver for $(II,I_4^2I_1^2)$ theory and $(I_0^*, I_2^3)$ theory.}
\label{bps2}
\end{figure}

For the theories admitting undeformable $I_n^*,II^*,III^*,IV^*$ singularities, there is no BPS quiver as there is no  BPS hypermultiplet associated with the singularity (see next subsection). The BPS information can be found using the folding trick of 
of the BPS quiver of the parent theory though. The brane configuration and string junctions would be useful to study the BPS states.

\newpage
\subsection{BPS states associated with local singularities}
Unlike undeformable $I_n$ singularity, the physics associated with other undeformable singularities are quite unfamiliar.  A $(\sqrt{n},0)$ (in suitable EM frame) massless hypermultiplet is assumed at $I_n$ singularity, but 
the particle contents for other singularities are much less clear.

 Let's now try to understand the BPS particles associated with un-deformable singularities by using the string junctions.  There is no gauge algebra associated with them, and 
so string loops around the singularity should be used, see figure. \ref{string}. The asymptotical charges for these junctions are shown in the formula. \ref{stringloop}, and one
can also use fractional strings as long as the asymptotical charge is integral.  Now the self-intersection number of such a string loop with charge $(p,q)$ is given as:
\begin{equation*}
(J,J)=f_K(p,q).
\end{equation*}
where $f_K(p,q)$ is the quadratic form listed in table. \ref{kodairabrane}. The condition for $J$ to represent a BPS particle \cite{DeWolfe:1998bi} is
 \begin{equation*}
 (J,J)\geq -2+gcd(p,q).
 \end{equation*}
The self-intersection number of $J$ must be $-1$ to get a hypermultiplet, and this is only possible for
$I_n$ singularity (the charge is $(\sqrt{n},0)$). 
\begin{table}[htp]
\begin{center}
\resizebox{3in}{!}{
\begin{tabular}{|c|c|c|}
\hline 
Label & Quadratic form $f_K(p,q)$ & Charge vector $(p,q)$ \\ \hline
$I_n$ & $-{1\over n}p^2$ &$(\sqrt{n},0)$,~$(J,J)=-1$  \\ \hline
$I_n^*$&${n\over 4}q^2$ &   $q=\sqrt{4\over n}$,~~$(J,J)=1$ \\ \hline
$IV^*$ & ${1\over 3}p^2-pq+q^2$ &$(0,1),(3,1), (3,2)$,~~$(J,J)=1$\\ \hline
$III^*$ & ${1\over 2}p^2-2pq+{5\over2}q^2$ &$(1,1),(3,1),(\sqrt{2},0)$~,$(J,J)=1$ \\ \hline
$II^*$ & $p^2-5pq+7q^2$ &$(2,1),(3,1),(1,0),~~~(J,J)=1$ \\ \hline
\end{tabular}}
\end{center}
\caption{The $(p,q)$ string loop around ADE singularity and their intersection numbers.}
\label{default}
\end{table}

\begin{figure}
\begin{center}

\tikzset{every picture/.style={line width=0.75pt}} 

\begin{tikzpicture}[x=0.40pt,y=0.40pt,yscale=-1,xscale=1]

\draw   (234,218) .. controls (234,199.77) and (248.77,185) .. (267,185) .. controls (285.23,185) and (300,199.77) .. (300,218) .. controls (300,236.23) and (285.23,251) .. (267,251) .. controls (248.77,251) and (234,236.23) .. (234,218) -- cycle ;
\draw    (283,189) -- (351,135.72) ;
\draw [shift={(320.94,159.28)}, rotate = 141.92] [fill={rgb, 255:red, 0; green, 0; blue, 0 }  ][line width=0.08]  [draw opacity=0] (8.93,-4.29) -- (0,0) -- (8.93,4.29) -- cycle    ;
\draw  [fill={rgb, 255:red, 0; green, 0; blue, 0 }  ,fill opacity=1 ] (263,222) .. controls (263,219.79) and (264.79,218) .. (267,218) .. controls (269.21,218) and (271,219.79) .. (271,222) .. controls (271,224.21) and (269.21,226) .. (267,226) .. controls (264.79,226) and (263,224.21) .. (263,222) -- cycle ;
\draw  [fill={rgb, 255:red, 0; green, 0; blue, 0 }  ,fill opacity=1 ] (357,221) .. controls (357,218.79) and (358.79,217) .. (361,217) .. controls (363.21,217) and (365,218.79) .. (365,221) .. controls (365,223.21) and (363.21,225) .. (361,225) .. controls (358.79,225) and (357,223.21) .. (357,221) -- cycle ;
\draw  [fill={rgb, 255:red, 0; green, 0; blue, 0 }  ,fill opacity=1 ] (418,223) .. controls (418,220.79) and (419.79,219) .. (422,219) .. controls (424.21,219) and (426,220.79) .. (426,223) .. controls (426,225.21) and (424.21,227) .. (422,227) .. controls (419.79,227) and (418,225.21) .. (418,223) -- cycle ;
\draw  [fill={rgb, 255:red, 0; green, 0; blue, 0 }  ,fill opacity=1 ] (176,221) .. controls (176,218.79) and (177.79,217) .. (180,217) .. controls (182.21,217) and (184,218.79) .. (184,221) .. controls (184,223.21) and (182.21,225) .. (180,225) .. controls (177.79,225) and (176,223.21) .. (176,221) -- cycle ;
\draw  [fill={rgb, 255:red, 208; green, 2; blue, 27 }  ,fill opacity=1 ] (347,135.72) .. controls (347,133.51) and (348.79,131.72) .. (351,131.72) .. controls (353.21,131.72) and (355,133.51) .. (355,135.72) .. controls (355,137.93) and (353.21,139.72) .. (351,139.72) .. controls (348.79,139.72) and (347,137.93) .. (347,135.72) -- cycle ;
\draw    (315,431) -- (311,342.72) ;
\draw [shift={(312.77,381.86)}, rotate = 87.41] [fill={rgb, 255:red, 0; green, 0; blue, 0 }  ][line width=0.08]  [draw opacity=0] (8.93,-4.29) -- (0,0) -- (8.93,4.29) -- cycle    ;
\draw  [fill={rgb, 255:red, 0; green, 0; blue, 0 }  ,fill opacity=1 ] (223,429) .. controls (223,426.79) and (224.79,425) .. (227,425) .. controls (229.21,425) and (231,426.79) .. (231,429) .. controls (231,431.21) and (229.21,433) .. (227,433) .. controls (224.79,433) and (223,431.21) .. (223,429) -- cycle ;
\draw  [fill={rgb, 255:red, 0; green, 0; blue, 0 }  ,fill opacity=1 ] (311,431) .. controls (311,428.79) and (312.79,427) .. (315,427) .. controls (317.21,427) and (319,428.79) .. (319,431) .. controls (319,433.21) and (317.21,435) .. (315,435) .. controls (312.79,435) and (311,433.21) .. (311,431) -- cycle ;
\draw  [fill={rgb, 255:red, 0; green, 0; blue, 0 }  ,fill opacity=1 ] (378,430) .. controls (378,427.79) and (379.79,426) .. (382,426) .. controls (384.21,426) and (386,427.79) .. (386,430) .. controls (386,432.21) and (384.21,434) .. (382,434) .. controls (379.79,434) and (378,432.21) .. (378,430) -- cycle ;
\draw  [fill={rgb, 255:red, 0; green, 0; blue, 0 }  ,fill opacity=1 ] (136,428) .. controls (136,425.79) and (137.79,424) .. (140,424) .. controls (142.21,424) and (144,425.79) .. (144,428) .. controls (144,430.21) and (142.21,432) .. (140,432) .. controls (137.79,432) and (136,430.21) .. (136,428) -- cycle ;
\draw  [fill={rgb, 255:red, 208; green, 2; blue, 27 }  ,fill opacity=1 ] (307,342.72) .. controls (307,340.51) and (308.79,338.72) .. (311,338.72) .. controls (313.21,338.72) and (315,340.51) .. (315,342.72) .. controls (315,344.93) and (313.21,346.72) .. (311,346.72) .. controls (308.79,346.72) and (307,344.93) .. (307,342.72) -- cycle ;
\draw    (400,622) -- (329,534.72) ;
\draw [shift={(361.34,574.48)}, rotate = 50.87] [fill={rgb, 255:red, 0; green, 0; blue, 0 }  ][line width=0.08]  [draw opacity=0] (8.93,-4.29) -- (0,0) -- (8.93,4.29) -- cycle    ;
\draw  [fill={rgb, 255:red, 0; green, 0; blue, 0 }  ,fill opacity=1 ] (241,621) .. controls (241,618.79) and (242.79,617) .. (245,617) .. controls (247.21,617) and (249,618.79) .. (249,621) .. controls (249,623.21) and (247.21,625) .. (245,625) .. controls (242.79,625) and (241,623.21) .. (241,621) -- cycle ;
\draw  [fill={rgb, 255:red, 0; green, 0; blue, 0 }  ,fill opacity=1 ] (329,623) .. controls (329,620.79) and (330.79,619) .. (333,619) .. controls (335.21,619) and (337,620.79) .. (337,623) .. controls (337,625.21) and (335.21,627) .. (333,627) .. controls (330.79,627) and (329,625.21) .. (329,623) -- cycle ;
\draw  [fill={rgb, 255:red, 0; green, 0; blue, 0 }  ,fill opacity=1 ] (396,622) .. controls (396,619.79) and (397.79,618) .. (400,618) .. controls (402.21,618) and (404,619.79) .. (404,622) .. controls (404,624.21) and (402.21,626) .. (400,626) .. controls (397.79,626) and (396,624.21) .. (396,622) -- cycle ;
\draw  [fill={rgb, 255:red, 0; green, 0; blue, 0 }  ,fill opacity=1 ] (154,620) .. controls (154,617.79) and (155.79,616) .. (158,616) .. controls (160.21,616) and (162,617.79) .. (162,620) .. controls (162,622.21) and (160.21,624) .. (158,624) .. controls (155.79,624) and (154,622.21) .. (154,620) -- cycle ;
\draw  [fill={rgb, 255:red, 208; green, 2; blue, 27 }  ,fill opacity=1 ] (325,534.72) .. controls (325,532.51) and (326.79,530.72) .. (329,530.72) .. controls (331.21,530.72) and (333,532.51) .. (333,534.72) .. controls (333,536.93) and (331.21,538.72) .. (329,538.72) .. controls (326.79,538.72) and (325,536.93) .. (325,534.72) -- cycle ;
\draw    (126,220) -- (460,222.72) ;
\draw    (93,428) -- (427,430.72) ;
\draw    (112,619) -- (446,621.72) ;

\draw (163,236.4) node [anchor=north west][inner sep=0.75pt]   [font=\tiny] {$A^{4} BC$};
\draw (256,229.4) node [anchor=north west][inner sep=0.75pt]     [font=\tiny]{$A^{4}$};
\draw (291,130.4) node [anchor=north west][inner sep=0.75pt]   [font=\tiny]  {$( 2,0)$};
\draw (123,443.4) node [anchor=north west][inner sep=0.75pt]    [font=\tiny] {$A^{4} BC$};
\draw (213,439.4) node [anchor=north west][inner sep=0.75pt]    [font=\tiny] {$A^{4}$};
\draw (320,357.4) node [anchor=north west][inner sep=0.75pt]    [font=\tiny] {$( 1,-1)$};
\draw (338,234.4) node [anchor=north west][inner sep=0.75pt]   [font=\tiny]  {$\ \ \ \ B$};
\draw (402,235.4) node [anchor=north west][inner sep=0.75pt]   [font=\tiny]  {$\ \ \ \ C$};
\draw (141,635.4) node [anchor=north west][inner sep=0.75pt]   [font=\tiny]  {$A^{4} BC$};
\draw (231,631.4) node [anchor=north west][inner sep=0.75pt]    [font=\tiny] {$A^{4}$};
\draw (365,539.4) node [anchor=north west][inner sep=0.75pt]    [font=\tiny] {$( 1,1)$};
\draw (288,444.4) node [anchor=north west][inner sep=0.75pt]   [font=\tiny]  {$\ \ \ \ B$};
\draw (363,444.4) node [anchor=north west][inner sep=0.75pt]    [font=\tiny] {$\ \ \ \ C$};
\draw (306,634.4) node [anchor=north west][inner sep=0.75pt]    [font=\tiny] {$\ \ \ \ B$};
\draw (375,636.4) node [anchor=north west][inner sep=0.75pt]    [font=\tiny] {$\ \ \ \ C$};

\end{tikzpicture}

\end{center}
\caption{The string junctions represent the stable BPS hypermultiplets of $(I_0^*, I_4I_1^2)$ theory.}
\label{starjunction}
\end{figure}
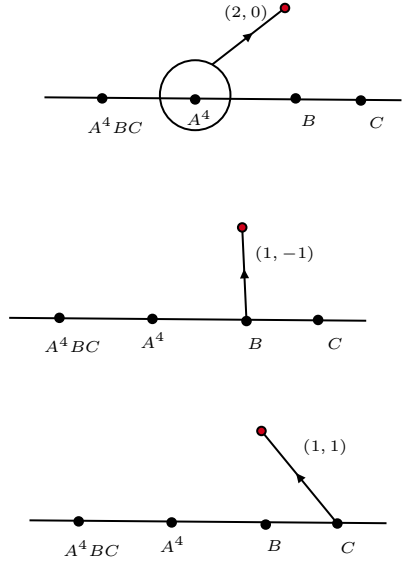

\textbf{Example}: Let's consider the the brane configuration for $SU(2)$ $\mathcal{N}=2^*$ theory, and so the brane configuration is $(A^4BC)(A^4) B C$.  The three basic string junction is shown in figure. \ref{starjunction},  which give the stable BPS hypermultiplet for the theory. 
These BPS particles form the BPS quiver.

Next let's consider the local singularity which is formed by merging $I_1$ singularities, and so there is ADE gauge algebra on collapsed D7 branes. The BPS particle contents are studied in \cite{DeWolfe:1998zf,DeWolfe:1998bi}. For example, one get BPS string junctions transforming
in fundamental representation of the $su(n)$ gauge algebra if the asymptotical charge is $(1,0)$ and the singularity is $I_n$. 
This is consistent with the fact that the low energy theory is given by $U(1)$ gauge theory coupled with $n$ massless hypermultiplets with charge one.
One might do similar computations for other singularities. 

\newpage

\subsection{Stratification of the Coulomb branch}
The brane configuration is quite useful to find the stratification of the Coulomb branch. We first start with brane configuration for the generic configuration, and then merge  
the branes to get strata with non-generic configuration.

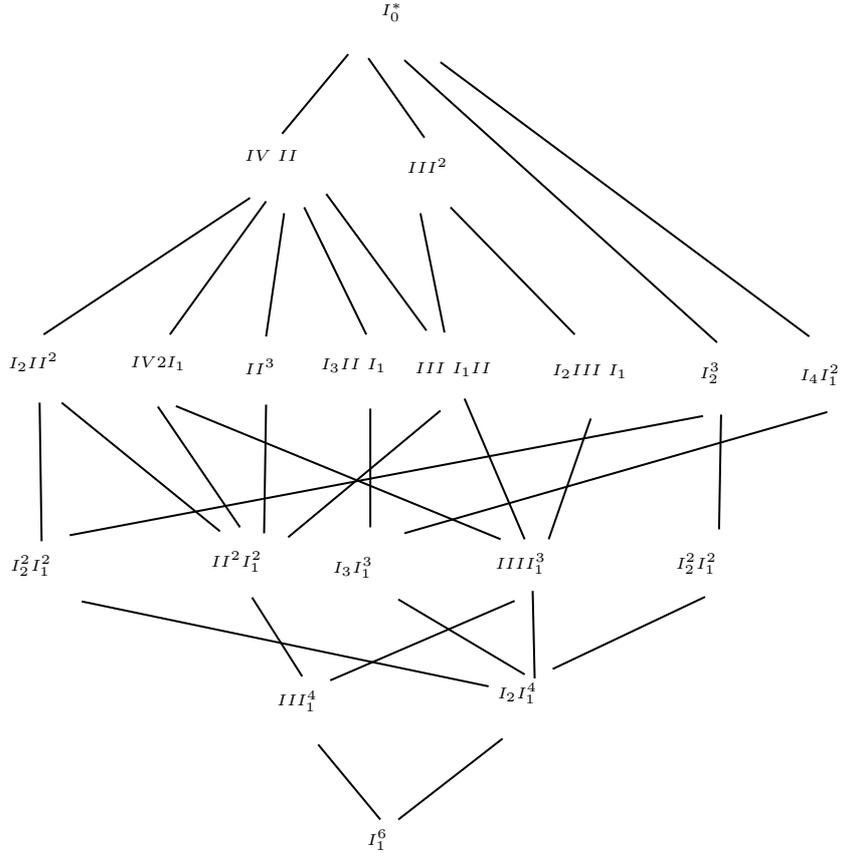
\begin{figure}
\begin{center}
\tikzset{every picture/.style={line width=0.75pt}} 

\begin{tikzpicture}[x=0.75pt,y=0.75pt,yscale=-1,xscale=1]

\draw    (278,76.72) -- (245,116.72) ;
\draw    (288,78.72) -- (316,118.72) ;
\draw    (306,79.72) -- (462,221.72) ;
\draw    (229,149) -- (126,217.72) ;
\draw    (237,150.72) -- (189,217.72) ;
\draw    (246,156.72) -- (237,218.72) ;
\draw    (256,153.72) -- (287,217.72) ;
\draw    (314,156.72) -- (326,216.72) ;
\draw    (329,153.72) -- (391,217.72) ;
\draw    (324,80.72) -- (508,218.72) ;
\draw    (267,147) -- (317,215.72) ;
\draw    (289,255) -- (289,314.72) ;
\draw    (336,250) -- (366,320.72) ;
\draw    (399,260) -- (378,320.72) ;
\draw    (464,258) -- (463,315.72) ;
\draw    (306,317.72) -- (517,256.72) ;
\draw    (237,253) -- (236,317.72) ;
\draw    (183,254) -- (224,314.72) ;
\draw    (135,252) -- (214,316.72) ;
\draw    (192,253.72) -- (354,320.72) ;
\draw    (124,252) -- (125,321.72) ;
\draw    (303,351) -- (366,388.72) ;
\draw    (269,391.72) -- (361,351.72) ;
\draw    (380,386) -- (456,349.72) ;
\draw    (230,350) -- (255,389.72) ;
\draw    (145,352) -- (348,394.72) ;
\draw    (371,390.72) -- (370,346.72) ;
\draw    (263,424) -- (294,461.72) ;
\draw    (355,421) -- (303,461.72) ;
\draw    (324,256) -- (248,319.72) ;
\draw    (455,258.72) -- (139,318.72) ;

\draw (293,49.4) node [anchor=north west][inner sep=0.75pt]  [font=\tiny]     {$I_{0}^{*}$};
\draw (225,123.4) node [anchor=north west][inner sep=0.75pt]  [font=\tiny]     {$IV\ II$};
\draw (306,127.4) node [anchor=north west][inner sep=0.75pt]    [font=\tiny]   {$III^{2}$};
\draw (107,225.4) node [anchor=north west][inner sep=0.75pt]   [font=\tiny]  {$I_{2} II^2$};
\draw (168,227.4) node [anchor=north west][inner sep=0.75pt]  [font=\tiny]   {$IV2I_{1}$};
\draw (225,228.4) node [anchor=north west][inner sep=0.75pt]  [font=\tiny]   {$II^3$};
\draw (263,228.4) node [anchor=north west][inner sep=0.75pt]   [font=\tiny]   {$I_{3} II\ I_{1}$};
\draw (310,230.4) node [anchor=north west][inner sep=0.75pt]  [font=\tiny]  {$III\ I_{1} II\ $};
\draw (375,231.4) node [anchor=north west][inner sep=0.75pt]  [font=\tiny]  {$\ I_{2} III\ I_{1}$};
\draw (452,230.4) node [anchor=north west][inner sep=0.75pt]   [font=\tiny]    {$I_{2}^3$};
\draw (502,231.4) node [anchor=north west][inner sep=0.75pt]     [font=\tiny]  {$I_{4} I_{1}^2$};
\draw (269,328.4) node [anchor=north west][inner sep=0.75pt]     [font=\tiny]  {$I_{3} I_{1}^3$};
\draw (350,326.4) node [anchor=north west][inner sep=0.75pt]   [font=\tiny]    {$IIII_{1}^3$};
\draw (440,326.4) node [anchor=north west][inner sep=0.75pt]    [font=\tiny]   {$I_{2}^2 I_{1}^2$};
\draw (208,325.4) node [anchor=north west][inner sep=0.75pt]    [font=\tiny]   {$II^2 I_{1}^2$};
\draw (108,327.4) node [anchor=north west][inner sep=0.75pt]     [font=\tiny]  {$I_{2}^2 I_{1}^2$};
\draw (241,395.4) node [anchor=north west][inner sep=0.75pt]    [font=\tiny]   {$III_{1}^4$};
\draw (351,391.4) node [anchor=north west][inner sep=0.75pt]    [font=\tiny]   {$I_{2} I_{1}^4$};
\draw (286,465.4) node [anchor=north west][inner sep=0.75pt]   [font=\tiny]    {$I_{1}^6$};

\end{tikzpicture}
\end{center}
\caption{The stratification for the $(I_0, I_1^6)$ theory. The top one describes the SCFT point, and the bottom one describes the configuration of generic deformations.}
\label{generic}
\end{figure}

\textbf{Example}:  Let's consider theories labeled as $(I_0^*, I_1^6)$. The brane configuration for it is  $(A^4BC)A^4BC$. The brane system for SCFT point is $(A^4BC)$ (from now on, the brane configuration at $\infty$ is ignored), and all the branes are on top of each other. The total number of deformations for this singularity is $6$ (the flavor symmetry is $D_4$, and there is a Coulomb branch operator.). 
We then look at the deformation of  above collapsed brane systems:
\begin{enumerate}
\item First, there are configurations with 5 deformations: 1) $IV II$ \footnote{The way of counting the number of deformations is: there is only one common Coulomb branch deformation, and the mass deformation and relevant deformation for the local singularity would be added.}; 2) $III^2$.
\item Secondly, there are  configurations with four deformations: 1): $I_4I_1^2$; 2): $IV^2I_1$; 3): $I_3 II~ I_1$; 4): $I_2III~I_1$; 5): $III~II~I_1$; 6):~$I_2^3$; 7): $I_2 II^2$; 8): $II^3$.
\item Thirdly,  there are  configurations with three deformations: 1) $ I_3 I_1^3$; 2): $IIII_1^3$; 3) $I_2^2 I_1^2$; 4): $I_2II^2I_1$; 5): $II^2 I_1^2$.
\item Fourthly, there are  configurations with two deformations: 1) $ I_2 I_1^4$; 2): $II~ I_1^4$.
\item Finally, there is only one configuration with one deformation: $I_1^6$.
\end{enumerate}
The basic brane configurations are: 1): $I_0^*=A^4BC$; 2): $III^2: (A^2C)(A^2C)$; 3): $IVII: (A^3 X_{[0,-1]}) (AC)$; 4): $I_2^3:(A^2)(B^2)(X_{[0,1]}^2)$; 5): $I_4 I_1^2:(A^4)BC$. The rest of the brane configurations can be found by further splitting the branes for $II, III, IV, I_4$ singularity. 
The result of the full stratification is shown in figure. \ref{generic}.

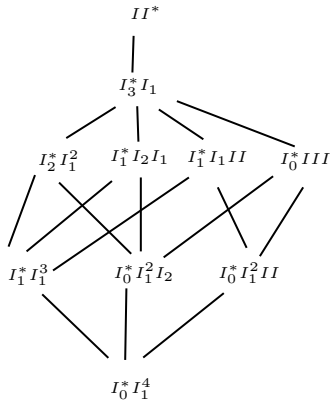
\begin{figure}
 \begin{center}

\tikzset{every picture/.style={line width=0.75pt}} 

\begin{tikzpicture}[x=0.45pt,y=0.45pt,yscale=-1,xscale=1]

\draw    (222,305.72) -- (277,359) ;
\draw    (292,300.72) -- (291,360.72) ;
\draw    (373,304.72) -- (306,358.72) ;
\draw    (298,89) -- (297,119.72) ;
\draw    (242,174.72) -- (283,148.72) ;
\draw    (280,210.72) -- (209,269.72) ;
\draw    (301,148) -- (302,177.72) ;
\draw    (319,150.72) -- (356,175.72) ;
\draw    (334,143.72) -- (432,181.72) ;
\draw    (439,216) -- (403,273.72) ;
\draw    (414,206.72) -- (323,274.72) ;
\draw    (344,207.72) -- (231,285.72) ;
\draw    (236,212) -- (296,271.72) ;
\draw    (216,205.72) -- (194,265.72) ;
\draw    (368,216) -- (392,266.72) ;
\draw    (304,207.72) -- (304,269.72) ;

\draw (276,373.4) node [anchor=north west][inner sep=0.75pt] [font=\tiny]   {$I_{0}^{*} I_{1}^{4}$};
\draw (191,278.4) node [anchor=north west][inner sep=0.75pt]    [font=\tiny] {$I_{1}^{*} I_{1}^{3}$};
\draw (279,276.4) node [anchor=north west][inner sep=0.75pt]   [font=\tiny]  {$I_{0}^{*} I_{1}^{2} I_{2}$};
\draw (366,276.4) node [anchor=north west][inner sep=0.75pt]  [font=\tiny]   {$I_{0}^{*} I_{1}^{2} II$};
\draw (293,61.4) node [anchor=north west][inner sep=0.75pt]   [font=\tiny]  {$II^{*}$};
\draw (283,121.4) node [anchor=north west][inner sep=0.75pt]    [font=\tiny] {$I_{3}^{*} I_{1}$};
\draw (216,182.4) node [anchor=north west][inner sep=0.75pt]  [font=\tiny]   {$I_{2}^{*} I_{1}^{2}$};
\draw (276,180.4) node [anchor=north west][inner sep=0.75pt]   [font=\tiny]  {$I_{1}^{*} I_{2} I_{1}$};
\draw (340,181.4) node [anchor=north west][inner sep=0.75pt]  [font=\tiny]   {$I_{1}^{*} I_{1} II$};
\draw (417,183.4) node [anchor=north west][inner sep=0.75pt]  [font=\tiny]   {$I_{0}^{*} I II$};

\end{tikzpicture}

 \end{center}
 \caption{The stratification for $(II,I_0^*I_1^4)$ theory. The SCFT point is at the bottom of the figure.}
 \label{nongeneric}
 
 \end{figure}

One can use the brane configuration to find out the stratification of other type of theories. In other cases,
we have sub-brane system which can not be separated. In the  $I_1$ series, we start with the most singular 
configuration and studies its all possible deformation pattern. For other series, we go from the other direction, namely,
 we start with the singular fibers for generic deformation, and combining the singularities. The rule is that: 1): in the process of merging,
 we can not split the collapsed branes; 2) The result of merging will just produce the brane systems for the Kodaira's singularity. Using 
 these rules, it is then possible to find out the stratification of these general theories.
 The stratification for theory $(II, I_0^*I_1^4)$ is shown in figure. \ref{nongeneric}.

\newpage

\section{Classification of  rank one 5d $\mathcal{N}=1$ SCFT}
Let's  turn our attention to the classification of 5d $\mathcal{N}=1$ theory. We'd like to first review some basic facts of 5d theories and 
the main difference from the 4d $\mathcal{N}=2$ theory will be discussed, see \cite{Xie:2017pfl} for more details. Firstly, the only relevant SUSY preserving deformations for 5d theory are the mass deformation;
Secondly, the 5d theory also has a Coulomb branch, but it is parameterized by the real numbers and is not described by the expectation values of 
 the protected operators; Thirdly, there are massive BPS particles (Instanton particle) at the generic point of Coulomb branch, but massive string-like objects (monopole string) also exist.
 
Let's now compactify a 5d theory on a circle with finite size, and the resulting theory is often called 5d KK theory. The Coulomb branch of 5d KK theory is parameterized by the complex numbers (as one of the component of
 5d abelian gauge field is combined with the real field of  original 5d Coulomb branch into a complex number). The generalized Coulomb branch is parameterized by 
 above complex number together with the mass deformations (There are no relevant and marginal UV deformation parameters). Another difference is that the BPS particles of KK theory now 
 carry not only the electric-magnetic charge, flavor charge, but they also carry the winding number change (KK charge), so the central charge of these particles take the form
 \begin{equation*}
 Z=na+n_Da^D+\sum S_i m_i+n S_0.
 \end{equation*}
Therefore the charge lattice $\Gamma$ has dimension 
\begin{equation}
2r+f+3.
\end{equation}
Here $r$ is the rank of theory, and $f$ the number of mass parameters.

The Coulomb branch solution of 5d KK theory is also described by the mixed Hodge module, and so the classification method of 5d rank one theory is the same as that of 4d theory.
 The only difference is the type of fibers one can put at the infinity over the compactified Coulomb branch $\mathbb{P}^1$: 

$\bullet$: For the Coulomb branch solution of 5d KK theory, the fiber at infinity can only be  type $I_n,~~n\geq 1$.

The classification strategy is the same: we first classify theory with only undeformable $I_n$ type fibers at the bulk, and then  use the base change methods 
to find the theory with $I_n^*, II^*, III^*, IV^*$ type fibers at the bulk. The result of the classification is listed in table. \ref{5d}.

\textbf{Symmetry}:
One can compute the flavor symmetry of the theory by finding the root system associated with the singular fibers. The computation 
is exactly the same as what we did for four dimensional theory, and the results are listed in table. \ref{5dfull}.  We do need to point out that: in general the rank of the flavor symmetry 
is reduced under base change. This is different from the 4d theory. The one form symmetry might be also identified with the torsion part of Mordell-Weil lattice, and 
see also table. \ref{5dfull}.

\begin{table}
\begin{center}
\resizebox{6in}{!}{
\begin{tabular}{|l|l|l|l| l| l|l|}
\hline 
$I_1$ series&$(I_9, I_1^3)$ &$(I_8, I_1^4)_a$&$(I_8,I_1^4)_b$& $(I_7, I_1^5)$ & $(I_6, I_1^6)$& $(I_5,I_1^7)$ \\ \hline
~&$(I_4,I_1^8)$&$(I_3, I_1^9)$ &$(I_2, I_1^{10})$& $(I_1,I_1^{11})$ & ~& ~ \\ \hline
$I_2$ series&  $(I_2, I_2^5)$ & $(I_4, I_2^4)$\\ \hline
$I_3$ series&  $(I_3, I_3^3)$  \\ \hline
$I_4$ series&  $(I_1, I_4I_1^7)$&$(I_2, ~I_4I_1^6)$ & $(I_3, I_4I_1^5)$&$(I_4, I_4I_1^4)_a$& $ (I_4, I_4I_1^4)_b$ & $(I_5, I_4I_1^3)$\\ \hline
~&$(I_1,I_4^2I_1^3)_a,(I_1, I_4^2I_1^3)_b$& $(I_2,I_4^2I_1^2)$ \\ \hline
$Z_6$ covering & $(I_6, I_1^6)\to (I_1, II^* I_1)$ \\ \hline 
$Z_4$ covering & $(I_8, I_1^4)\to (I_2, III^* I_1)$& $(I_4, I_2^4)\to (I_1, III^*_{Q=\sqrt{2}} I_2)$ & $(I_4, I_1^8)\to (I_1, III^* I_1^2)$ \\ \hline 
$Z_3$ covering & $(I_9, I_1^3)\to (I_3, IV^* I_1)$& $(I_3, I_3^3)\to (I_1, IV_{Q=\sqrt{3}}^* I_3)$ & $(I_6, I_1^6)\to (I_2, IV^* I_1^2)$ & ~ \\ \hline 
 ~&$(I_3, I_1^9)\to (I_1, IV^* I_1^3)$& \\ \hline
$Z_2$ covering & $(I_8, I_1^4)\to (I_4, I_0^* I_1^2)$& $(I_6, I_1^6)\to (I_3, I_0^*I_1^3)$ & $(I_4, I_2^4)\to (I_2, I_0^*{~}_{Q=\sqrt{2}} I_2^2)$ \\ \hline 
~&$(I_2, I_4I_1^6)\to (I_1, I_2^* I_1^3)$& $(I_4, I_4I_1^4) \to (I_2, I_2^* I_1^2) $    \\ \hline
~&$(I_4, I_1^8)\to (I_2, I_0^*I_1^4)$ & $(I_2, I_2^5)\to (I_1, I_1^* I_2^2)$ & $(I_2, I_1^{10})\to (I_1, I_0^*I_1^5)$\\ \hline 
\end{tabular}}
\end{center}
\caption{Singular fiber configurations for  rank one 5d KK theory.}
\label{5d}
\end{table}

\textbf{Discrete gauging}: In 4d case, the discrete gauging acts on the flavor symmetry by the outer automorphism group of $G_F$ \cite{Argyres:2016yzz}. In the 5d case, 
the discrete gauging seems to act by the outer automorphism group of the corresponding affine Lie algebra. For example: the affine $E_6$ Lie algebra has 
a $Z_3$ outer automorphism group  from which we can get the affine $G_2$ Lie algebra. Curiously, the theory $(I^3, I_1^9)$ has 
$E_6$ flavor symmetry (from which we actually get an affine $E_6$ algebra, see the discussion in next subsection), and under the $Z_3$ discrete gauging, we get a theory $(I_1, III^* I_1^3)$ which does have $G_2$ flavor symmetry (and an associated $G_2$ affine Lie algebra). 
Affine $E_7$ Lie algebra has a $Z_2$ automorphism group and we get affine $F_4$ Lie algebra by using the outer automorphism. Interestingly,
by using the $Z_2$ discrete gauging of theory $(I_1^2, I_1^{10})$ (whose flavor symmetry is $E_7$), we get a theory $(I_1,I_0^*I_1^5)$ whose 
flavor symmetry is $F_4$. 

Discrete gauging of 5d $\mathcal{N}=1$ theory has been studied in \cite{Acharya:2021jsp,Kim:2021fxx}. In particular, they propose a relation between the prepotential as follows
\begin{equation}
F(T/Z_n)={1\over n}F(T).
\label{5dbase}
\end{equation}
Here $Z_n$ is the discrete gauging group. This relation can be regarded as a 5d version of the dimension formula used in the study of 4d theory, see formula. \ref{4dbase}. This condition simply means that if the singular fiber at $\infty$ of $B$ 
is $I_{nk}$, the singular fiber at $B^{'}$ should be $I_k$, here we assume a $Z_n$ action is used in doing base change.

\begin{table}[htp]
\begin{center}
\resizebox{6in}{!}{
\begin{tabular}{|l|l|l|l| l|l| l|l|l|}
\hline 
Theory & $G_F$ &$h$&Torsion&Theory&$G_F$& $h$&Torsion\\ \hline
\textbf{$I_1$ series}    \\ \hline
$(I_9,I_1^3)$& $\emptyset$ &0&$Z_3$&$(I_8,I_1^4)_a$ & $\mathfrak{u}_1$&0&0 \\ \hline
$(I_8,I_1^4)_b$&$\mathfrak{su}(2)$ &0& $Z_2$&$(I_7, I_1^5)$ & $\mathfrak{su}(2)\times \mathfrak{u}(1)$ & 0&0 \\ \hline
 $(I_6, I_1^6)$& $\mathfrak{su}(3)\times \mathfrak{su}(2)$ & 0&0& $(I_5, I_1^7)$ & $\mathfrak{su}(5)$ &0&0 \\ \hline
  $(I_4,I_1^8)$ &$\mathfrak{so}(10)$ & 0&0 & $(I_3, I_1^9)$ & $\mathfrak{e}_6$&0&0 \\ \hline
  $(I_2, I_1^{10})$& $\mathfrak{e}_7$ & 0& 0&$(I_1,I_1^{11})$ & $\mathfrak{e}_8$&0&0\\ \hline
\textbf{$I_2$ series}  \\ \hline
$(I_2, I_2^5)$ & $\mathfrak{sp}(4)$ & 2&$Z_2\times Z_2$&  $(I_4, I_2^4)$ & $\mathfrak{su}(2)$ &1& $Z_2\times Z_2$\\ \hline
\textbf{$I_3$ series} & \\ \hline
 $(I_3, I_3^3)$&  $\emptyset$ &0 &$Z_3\times Z_3$&&&& \\ \hline
\textbf{$I_4$ series}&  \\ \hline
$(I_1, I_4I_1^7)$& $\mathfrak{sp}(10)$&5&0&$(I_2, ~I_4I_1^6)$ &$\mathfrak{sp}(6)\times \mathfrak{sp}(2)$&3&0\\ \hline
 $(I_3, I_4I_1^5)$&$\mathfrak{sp}(4)\times \mathfrak{u}(1)$ & 2 &0&$(I_4, I_4I_1^4)_a$& $\mathfrak{sp}(4)$ &2&$Z_2$ \\ \hline
 $(I_4, I_4I_1^4)_b$ &   $\mathfrak{sp}(4)$  &2  &$Z_2$& $(I_5, I_4I_1^3)$ & $\mathfrak{u}(1)$&0&0\\ \hline
$(I_1,I_4^2I_1^3)_a$& $\mathfrak{sp}(4)$ & 2&$Z_2$&$(I_1, I_4^2I_1^3)_b$& $\mathfrak{sp}(4)$&2&0 \\ \hline
$(I_2,I_4^2I_1^2)$ &$\mathfrak{su}(2)$&1&$Z_4$ \\ \hline
\textbf{$Z_6$ covering} \\ \hline
$ (I_1, II^* I_1)$ & $\emptyset$ &0&0&&&&  \\ \hline 
\textbf{$Z_4$ covering} &  \\ \hline
$ (I_2, III^* I_1)$&  $\emptyset$ &0&$Z_2$ &$(I_1, III^*_{Q=\sqrt{2}} I_2)$ & $\emptyset$  &0 &$Z_2$\\ \hline
 $ (I_1, III^* I_1^2)$ & $\mathfrak{su}(2)$&0&0&&&&\\ \hline 
\textbf{$Z_3$ covering} & \\ \hline
 $ (I_3, IV^* I_1)$&$\emptyset$ & 0& $Z_2$&$ (I_1, IV_{Q=\sqrt{3}}^* I_3)$ & $\emptyset$ &0 &$Z_2$\\ \hline
 $ (I_2, IV^* I_1^2)$ & $\mathfrak{su}(2)$ &0&0&$(I_1, IV^* I_1^3)$& $\mathfrak{g}_2$ &0&0\\ \hline
\textbf{$Z_2$ covering} & \\ \hline
 $(I_4, I_0^* I_1^2)$& $\mathfrak{su}(2)$ &0&$Z_2$&$(I_3, I_0^*I_1^3)$ & $ \mathfrak{su}(3)$&0&0 \\ \hline
 $ (I_2, I_0^*{~}_{Q=\sqrt{2}} I_2^2)$  & $\mathfrak{su}(2)$&0&$Z_2$&$ (I_1, I_2^* I_1^3)$& $\mathfrak{sp}(4)$&2&0 \\ \hline
 $(I_2, I_2^* I_1^2) $   & $\mathfrak{su}(2)$& 1 &$Z_2$&$ (I_2, I_0^*I_1^4)$ & $ \mathfrak{spin}(7) $&0&0 \\ \hline
  $ (I_1, I_1^* I_2^2)$ & $\mathfrak{su}(2)$ &0&0 &$ (I_1, I_0^*I_1^5)$ &$\mathfrak{f}_4$&0&0\\ \hline 
\end{tabular}}
\end{center}
\caption{Physical data for rank one 5d KK theory. Here $G_F$ is the flavor symmetry, $h$ denotes the number of free hypermultiplets at the generic point of Coulomb branch. Torsion means the torsion subgroup of 
the full Mordell-Weil group, which could be identified with the one-form symmetry of the theory.}
\label{5dfull}
\end{table}

\newpage
\subsection{D7 brane configuration  and affine  Lie algebra}

One can also associate D7 brane configurations for the 5d KK theory, and the results are listed in table. \ref{5dbrane}, see earlier
results for the $I_1$ series \cite{Yamada:1999xr}. The $I_1$ series should be identified with the rank one theory studied in \cite{Morrison:1996xf}.

\textbf{Affine Lie algebra}: There are affine Lie algebra associated with $E_n$ type of theories. For the $I_1$ series, the
brane configuration takes the general form $(A^{8-n}) A^{n} BCBC$. The $(A^{8-n})$ branes describe the singular fiber at infinity. The generic deformations at the bulk are described by 
the brane system
\begin{align*}
&E_n::~~A^{n-1} BCBC,~n=1,\ldots,8,~~ \nonumber\\
&\tilde{E}_n:~~A^{n}X_{[2,-1]}X_{[1,-2]}C,~~n=0,1,\ldots 8.
\end{align*}
These two sequences are equivalent for $n\geq 2$. $E_1$ and $\tilde{E}_1$ gives different brane configuration.
Now one can associate an affine $E_n$ Lie algebra from these branes \cite{DeWolfe:1998bi}. The $E_n$ algebra is just 
the flavor symmetry for the corresponding 5d theory. 
 \begin{table}
\begin{center}
\resizebox{3in}{!}{\begin{tabular}{|l|l|l|l|}
\hline  
Name & Brane configuration &~&~ \\ \hline
$(I_1,I_1^{11})$ &$(A) A^7BCBC$ &$(I_2,I_2^5)$ &$(A^2) (A^2)(B^2)(B^2)(X_{[0,1]}^2) (X_{[2,1]}^2)$  \\ \hline
$(I_2,I_1^{10})$ &$(A^2)A^6BCBC$& $(I_4,I_2^4 )$ &$(A^4) (B^2) (X_{[0,1]}^2) (B^2)(X_{[0,1]}^2)$  \\ \hline
$(I_3,I_1^{10})$ &$(A^3)A^5BCBC$& $(I_1,I_4 I_1^7)$ &$(A) (A^4)A^3BCBC$  \\ \hline
$(I_4,I_1^{10})$ &$(A^4)A^4BCBC$&$(I_2,I_4 I_1^6)$ &$(A^2) (A^4)A^2BCBC$ \\ \hline
$(I_5,I_1^{10})$ &$(A^5)A^3BCBC$&$(I_3,I_4 I_1^5)$ &$(A^3) (A^4)ABCBC$ \\ \hline
$(I_6,I_1^{10})$ &$(A^6)A^2BCBC$&$(I_4,I_4 I_1^4)_a$ &$(A^4) (A^4)BCBC$   \\ \hline
$(I_7,I_1^{10})$ &$(A^7)ABCBC$ &$(I_4,I_4 I_1^4)_b$ &$(A^4) (A^4)AX_{[2,-1]}X_{[1,-2]}C$  \\ \hline
$(I_8,I_1^{10})_a$ &$(A^8)BCBC$ &$(I_5,I_4 I_1^3)$ &$(A^5) (A^4)X_{[2,-1]}X_{[1,-2]}C$ \\ \hline
$(I_8,I_1^{10})_b$ &$(A^8)AX_{[2,-1]}X_{[1,-2]}C$&$(I_1,I_4^2 I_1^3)_a$ &$(C)(A^4)(A^4)BCB$  \\ \hline
$(I_9,I_1^{10})$ &$(A^9)X_{[2,-1]}X_{[1,-2]}C$ &$(I_1,I_4^2 I_1^3)_b$ &$(A) (A^4)(A^4)X_{[2,-1]}X_{[1,-2]}C$  \\ \hline
~&~&$(I_3,I_3^3)$ &$(A^3)(B^3)(X_{[0,1]}^3)(C^3))$ \\ \hline
\end{tabular}}
\quad
\resizebox{3in}{!}{\begin{tabular}{|l|l|l|l|}
\hline  
Name & Brane configuration &~&~ \\ \hline
$ (I_1, II^* I_1)$ &$(A^7BC^2)X_{[3,1]}A$ & $(I_4, I_0^* I_1^2)$ & $(A^4BC)(A^4)BC$   \\ \hline
$ (I_2, III^* I_1)$ &$(A^2) (A^6BC^2)X_{[3,1]}$&  $ (I_2, I_0^*{~}_{Q=\sqrt{2}} I_2^2)$ & $(A^4BC)(A^2)(B^2)(X_{[0,1]}^2)$\\ \hline
$(I_1, III^*_{Q=\sqrt{2}} I_2)$ &$(A^2) (A^6BC^2)X_{[3,1]}$& $(I_2, I_2^* I_1^2) $ & $(A^6BC)(A^2)BC$  \\ \hline
 $ (I_1, III^* I_1^2)$ &$ (A^6BC^2)X_{[3,1]}A^2$ &  $ (I_1, I_1^* I_2^2)$ & $(A^5BC)(B^2)(X_{[0,1]}^2)$\\ \hline
  $ (I_3, IV^* I_1)$ &$(A^3)(A^5BC^2)X_{[3,1]}$& $(I_3, I_0^*I_1^3)$& $(A^4BC)(A^3)ABC$\\ \hline
  $ (I_1, IV_{Q=\sqrt{3}}^* I_3)$ &$(A^5BC^2)X_{[3,1]}A^3$&$ (I_1, I_2^* I_1^3)$ & $(A^7BC)ABC$ \\ \hline
   $ (I_2, IV^* I_1^2)$ &$(A^2)(A^5BC^2)X_{[3,1]}A$& $ (I_2, I_0^*I_1^4)$ &$(A^4BC)(A^2)A^2BC$\ \\ \hline
   $(I_1, IV^* I_1^3)$ &$A(A^5BC^2)X_{[3,1]}A^2$&$ (I_1, I_0^*I_1^5)$& $(A^4BC)A^4BC$\\ \hline
\end{tabular}}

\end{center}
\caption{Brane configurations for 5d $\mathcal{N}=1$ theory. The branes within the parenthesis can not be separated.}
\label{5dbrane}
\end{table}

The main difference from 4d theory is that: one can find a string loop carrying trivial asymptotic $(p,q)$ charge and flavor charge.
The string is then represented by a loop around all the fundamental D7 branes. The trivial $(p,q)$ charge condition for the loop implies the relation on the charge vector $z_0$
\begin{equation*}
z_0=Kz_0.
\end{equation*}
Here $K$ is the monodromy matrix around the branes. One can find one independent solution for the brane system of the 5d theory. 
For example, for the $I_1$ series, the monodromy around the $E_N$ or $\tilde{E}_N$ brane systems  is 
\begin{equation*}
K=\left[\begin{array}{cc}
1& 9-N\\
0&1\\
\end{array}\right].
\end{equation*}
and so the string loop has charge $z_0=[1,0]$. This null string is denoted as $\delta=[1,0]$, and play the role of imaginary root of affine Lie algebra. The full string junction lattice then gave a lattice for affine $E_N$ algebra. For all the 5d configurations shown in table. \ref{5dfull},
we always get an imaginary root, and so an affine Lie algebra is defined. The detailed computation for the affine Lie algebra will be left for elsewhere.

\subsection{BPS quiver }

 \textbf{$I_1$ series}: One can also associate a BPS quiver for all the 5d theory with only $I_1$ type fibers at the bulk.  The idea is the same as what we did for 4d theory: 
 there is one massive BPS particle associated with each $I_1$ singularity represented by a D7 brane $X_{[p_i,q_i]}$, and so the electric-magnetic charge for it is $(p_i,q_i)$. 
 To find the BPS quiver, we need to find a special ordering of the bulk D7 brane configuration, and the sign for each D7 brane charges. Fortunately, such an ordering and the choice of sign
 have already been found  through the map to the 5 brane web \cite{Hanany:2001py}, see the ordering of branes in table. \ref{5dbps}. The BPS quiver is found by computing the Dirac paring between the charges of the D7 brane, and see the quivers in \cite{Hanany:2001py,Closset:2019juk}
 
 \textbf{$I_n$ series}: Let's make a couple of remarks about the BPS quiver. First of all, the special ordering is not unique and so 
 we get different BPS quivers, and those BPS quivers should be related by quiver mutation. Second,  if 
 the special $I_1$ configuration can be collapsed to find a $I_n$ series (for instance, the configuration for $I_1$ series looks like $\ldots X_{z}^4\ldots$, then
 one can collapse the $X_z^4$ brane to form a $I_4$ fiber), one can find the BPS quiver for $I_n$ series as follows: the charge vector for the $I_4$ singularity is $2z$, and then we can form
 a BPS quiver by using Dirac pairing. For example, using the brane configuration for $(I_4, I_1^8)$ shown in table. \ref{5dbps}, we can find 
 the BPS quiver for $(I_4, I_2^4)$ theory: the charge vectors are $\sqrt{2}[1,0], \sqrt{2}[0,1],\sqrt{2}[-1,0],\sqrt{2}[0,-1]$. The resulting BPS quiver  is the same as the $(I_8,I_1^4)$ theory, so it should give the same 
 theory (the local physics is the same), but it seems that the one form symmetry is different.

\textbf{Flavor charge and sheafs on Del Pezzo surfaces}: In the above representation of BPS quiver,  only the electric-magnetic charge for the BPS particle is known. Here we give a method to determine 
the flavor charge for the basic BPS particles.
Let's use the brane configuration $A^n X_{[2,-1]} C X_{[4,1]}$ for the bulk singularities of $\tilde{E}_n$ theory. A string junction is represented as
 \begin{equation}
 J=\sum_{i=1}^r \lambda_i w^i+pw^p+q w^q+n \delta^{(-1,0)}, 
 \end{equation}
 here the basis is the following: 
first s string basis $\alpha_i$ for the simple roots of $E_n$ Lie algebra  is constructed in \cite{DeWolfe:1998zf}, and $w_i$ is the dual basis.  Then $w^p,w^q$ carries no flavor charge and 
is represented by  the string loop: $w^p$ carries charge $(1,0)$ and $w^q$ carries charge $(0,1)$. Finally, there is an imaginary root $\delta^{(-1,0)}$ which is represented by the string loop  with charge $(-1,0)$, and this string loop winds around the whole brane configuration, and it is given as
\begin{equation*}
\delta=x_{[2,-1]}+2c-x_{[4,1]}.
\end{equation*}
Here $x_{[2,-1]}, c, x_{[4,1]}$ are basic string for the corresponding basic $D7$ branes.
The self-intersection number of a string junction $J$  is given as 
\begin{equation*}
(J,J)=-\lambda^2_{E_r}+2nq+f_{E_r}(p,q);
\end{equation*}
Here $\lambda^2$ is the inner product defined using the Cartan matrix of $E_n$ Lie algebra, and $f_{E_r}(p,q)$ is the quadratic form for $E_n$ brane configuration. Using this basis, one can find the flavor charges and KK charge for any BPS particle represented by the 
string junctions.

There exists a useful map between the string junction of 5d $E_r$ theory and sheaves over del Pezzo surface $X_r$. These maps may be useful for the study of the BPS spectrum, i.e. the special ordering of D7 branes might have following interpretation: the fundamental strings
in the special ordering give an exceptional collection over Del Pezzo surface.  Let's review the map between sheafs and string junction \cite{Hauer:1999pt}. The Picard group of $X_r$ is generated by $(l,l_1,\ldots, l_r)$ with the intersection form
\begin{equation*}
l^2=1,~~l_i^2=-1,~~l_i\cdot l_j=0.
\end{equation*}
The canonical class is given as 
\begin{equation*}
K_r=-3l+\sum_{i=1}^r l_i.
\end{equation*}
Notice choose $C_i=l_i-l_{i+1},~i=1,\ldots, r-1$ and $C_r=l-l_1-l_2-l_3$, and the intersection form of them yield the $E_r$ Cartan matrix. We can then easily find the dual basis $w_i$ which is defined by the relation 
\begin{equation*}
w^i\cdot C_j=-\delta^i_j.
\end{equation*} 

The topological data for a coherent sheaf $F$ is characterized by the data $(r, ch_1, ch_2)$, with $r$ the rank of the sheaf, $ch_1$ the first Chern class $c_1(F)$,  and $ch_2(F)={1\over 2} (c_1^2-c_2)$. The degree is defined as $d(F)=-K_S\cdot c_1(F)$.
The Euler number for two coherent sheafs  $E,F$ is defined as
\begin{equation*}
\chi(E,F)=\sum (-1)^iExt^i(E,F).
\end{equation*}
For two coherent sheafs on del Pezzo surface, we have 
\begin{equation*}
\chi(E,F)=r(E)r(F)+{1\over 2}(r(E)d(F)-r(F)d(E))+r(E)ch_2(F)+r(F)ch_2(E)-c_1(E)\cdot c_2(F).
\end{equation*}

It was shown in \cite{Hauer:1999pt} that one can map a coherent sheaf to 
a string junction as follows. We start with a coherent sheaf whose  Chern class is given as $(r,c_1(F), k(F))$, here $k=\int ch_2(F)$, and $c_1$ is given as (in the basis of $w_i, K_r$):
\begin{equation*}
c_1(F)=\sum_{i=1}^r\lambda_i w_i-{d(F)\over 9-r}K_r.
\end{equation*}
with the coefficient computed as 
\begin{equation*}
\lambda_i=c_1(F)\cdot C_i,~~~d(F)=-F \cdot K_r.
\end{equation*}
and string junction is given as 
\begin{equation*}
J_F=\sum_{i=1}^r\lambda_i w^i+d(F) w^p+rw^q-(r+k+{1\over 2} d(F) \delta^{(-1,0)}.
\end{equation*}
The above formula gives a map between string junction and coherent sheaf on Del Pezzo surfaces, and then it is possible to link the basis of the BPS quiver to exceptional collection 
of Del Pezzo surface. We leave the details to elsewhere.

 \begin{table}[htp]
\begin{center}
\resizebox{2in}{!}{\begin{tabular}{| c| c|}
\hline  
Name & Brane configuration for BPS quiver  \\ \hline
$(I_9, I_1^3)$ & $X_{[2,-1]}X_{[-1,2]}X_{[-1,-1]}$ \\ \hline
$(I_8, I_1^4)_a$ &$X_{[1,0]}X_{[-1,2]}X_{[-1,-1]}X_{[1,-1]}$ \\ \hline
$(I_8, I_1^4)_b$ &$X_{[1,-1]}X_{[1,1]}X_{[-1,1]}X_{[-1,-1]}$ \\ \hline
$(I_7, I_1^5)$ & $X_{[1,0]}X_{[-1,2]}X_{[-1,0]}X_{[0,-1]}X_{[1,-1]}$ \\ \hline
$(I_6, I_1^6)$ & $X_{[1,0]}X_{[0,1]}X_{[-1,1]}X_{[-1,0]}X_{[0,-1]}X_{[1,-1]}$ \\ \hline
$(I_5, I_1^7)$ & $X_{[1,0]}^2X_{[0,1]}X_{[-1,1]}X_{[-1,0]}X_{[0,-1]}^2$ \\ \hline
$(I_4, I_1^8)$ & $X_{[1,0]}^2X_{[0,1]}^2X_{[-1,0]}^2X_{[0,-1]}^2$ \\ \hline
$(I_3, I_1^9)$ & $X_{[1,0]}^3X_{[-1,1]}X_{[0,1]}X_{[-1,0]}^2X_{[0,-1]}^2$ \\ \hline
$(I_2, I_1^{10})$ & $X_{[1,0]}^4X_{[-1,1]}X_{[0,1]}X_{[-1,0]}^2X_{[0,-1]}X_{[-1,-1]}$ \\ \hline
$(I_1, I_1^{11})$ & $X_{[1,0]}^4X_{[-1,1]}X_{[0,1]}X_{[-1,0]}^3X_{[1,-1]}X_{[-1,-1]}$  \\ \hline
\end{tabular}}
\end{center}
\caption{Brane configuration for 5d KK theory from which one can find BPS quiver by using Dirac pairing.}
\label{5dbps}
\end{table}

\subsection{Special theory at Coulomb branch}
\textbf{IR free gauge theory}: It is possible to find out IR theory at arbitrary special vacua formed by merging the singularities of the generic deformation. This can be done by using the brane configurations. For example, the brane configuration takes 
the form $(A) A^7BCBC$ for $(I_1, I_1^{11})$  theory; the maximal singular configuration is $(A^7BC) B C$, and there is a $I_3^*$ singularity (represented by$ (A^7BC)$ configuration) in the bulk, whose low energy theory  is $SU(2)$ theory coupled with $N_f=7$ fundamental flavors.
This agrees with the result first found in \cite{Morrison:1996xf}: 5d $E_8$ $\mathcal{N}=1$ SCFT could be the UV limit of the $SU(2)$ gauge theory coupled with seven fundamental hypermultiplets.

\textbf{Pure 4d theory}: One can get purely 4d theory on the Coulomb branch of a 5d KK theory.  The brane construction is quite useful. The idea is following:  simply merge branes for the singularities 
so that  \textbf{bulk} brane configurations for a purely 4d theory can be found. This method is useful to find the number of free hypermultiplets at the generic point of the 5d theory.\
For example, there is a 5d theory $(I_1, III^* I_1^2)$, and a special configuration  $(I_1, (III^* I_1) I_1)$ can be found, namely one of the $I_1$ fiber is merged with $III^*$ fiber so that 
the total monodromy around $(III^*I_1)$ fiber is that of $II^*$ fiber. The low energy theory on that particular singularity is a pure 4d theory, see table. \ref{4dfull}.  Using this method, it is then possible to find all 
 purely 4d theory at the Coulomb branch of 5d KK theory.

\newpage
\section{Classification of rank one  6d $(1,0)$ SCFT}
Finally, let's consider rank one 6d $(1,0)$ SCFT. 6d $(1,0)$ theory has no relevant SUSY preserving deformation. 
The 6d theory could have a tensor branch (parameterized by real scalar) along which the low energy theory could be  described by  IR free non-abelian gauge theory. We then compactify the 6d theory on a torus $T^2$ to  get an effective 4d $\mathcal{N}=2$ theory. 
The tensor branch now becomes a coulomb branch which is now a complex variety.  Let's now assume the theory is rank one \footnote{Notice that in the literature people usually called a 6d theory rank one if the dimension of the tensor branch is one dimensional; In compactifying down to 4d, the dimension of the Coulomb branch would be increased if there is a gauge theory along the 6d tensor branch.}.  
There are  two differences of Coulomb branch of the 6d KK theory 
from that of 4d theory: a) The 6d theory has no relevant deformation,  but we may turn on flux line for the flavor symmetry along the torus so that there 
are still mass deformations; b) The massive BPS particle could carry two extra winding charges, and so the charge lattice $\Gamma$ for rank one 6d KK theory has dimension
\begin{equation*}
 \Gamma=f+4.
\end{equation*}\
Here $f$ is the rank of the flavor symmetry.

We can also classify the 6d KK theory by using the Coulomb branch solution, and the method is the same as we did for 4d and 5d KK theory. The only difference is following:

$\bullet$: The singular fiber at  infinity is smooth, namely, there is a $I_0$ fiber at $\infty$. 

The result of the classification is shown in table. \ref{6d}. 
The physical properties such as the brane configuration, flavor symmetry, free hypermultiplets at generic points, and the one-form symmetry is shown in table. \ref{6dfull}.

\begin{table}
\begin{center}
\resizebox{3in}{!}{
\begin{tabular}{|l|l|l|}
\hline 
$I_1$ series  & $I_1^{12}$&  \\ \hline
$I_2$ series&  $I_2^6$&  \\ \hline
$I_3$ series&  $I_3^4$&  \\ \hline
$I_4$ series&  $I_4I_1^8$ &$(I_4^2I_1^4)_a, (I_4^2I_1^4)_b$ \\ \hline
$Z_6$ covering & $ I_1^{12}\to II^*I_1^2$& \\ \hline 
$Z_4$ covering & $I_1^{12}\to III^* I_1^3$& \\ \hline 
$Z_3$ covering & $I_1^{12}\to IV^* I_1^4$ &\\ \hline
$Z_2$ covering & $I_4I_1^8\to I_2^* I_1^4$& $I_2^6\to I_0^*I_2^3$  \\ \hline
~& $I_4^2I_1^4 \to I_0^*I_4I_1^2$& $I_1^{12}\to I_0^* I_1^6$ \\ \hline 
\end{tabular}}
\end{center}
\caption{Singular fiber configuration for 6d KK theory.}
\label{6d}
\end{table}

\begin{table}[htp]
\begin{center}
\resizebox{3in}{!}{
\begin{tabular}{|l|l|l|l|l|l|}
\hline 
~&~&branes & $G_F$ &$h$ & Torsion  \\ \hline
$I_1$ series  & $I_1^{12}$& $A^8BCBC$ & $\mathfrak{e}_8$ & 0&0 \\ \hline
$I_2$ series&  $I_2^6$& $(A^2)(A^2)(B^2)(B^2)(X_{[0,1]}^2)(X_{[2,1]}^2)$& $\mathfrak{sp}(4)$ & 2&$Z_2\times Z_2$ \\ \hline
$I_3$ series&  $I_3^4$& $(A^3) (B^3) (X_{[0,1]}^3) (B^3)$ & $\emptyset$ &0&$Z_3\times Z_3$\\ \hline
$I_4$ series&  $I_4I_1^8$ & $(A^4) A^4  BCBC$  & $\mathfrak{sp}(10)$&5&0 \\ \hline
~&$(I_4^2I_1^4)_a$ & $(A^4) (A^4)  BCBC $ & $\mathfrak{sp}(4)$&  2&$Z_2$\\ \hline
~&$(I_4^2I_1^4)_b$ & $(A^4)  (A^4) AX_{[2,-1]}X_{[1,-2]}C $ & $\mathfrak{sp}(4)$&2&0\\ \hline
$Z_6$ covering & $ II^*I_1^2$& $(A^7BC^2)X_{[3,1]}A$&$\emptyset$ &0 &0\\ \hline 
$Z_4$ covering & $ III^* I_1^3$&$(A^6BC^2)X_{[3,1]}A^2$ &$\mathfrak{su}(2)$&1&0 \\ \hline 
$Z_3$ covering & $ IV^* I_1^4$ &$(A^5BC^2)X_{[3,1]}A^3$ &$\mathfrak{g}_2$&0&0\\ \hline
$Z_2$ covering & $I_2^* I_1^4$& $(A^6BC)A^2BC$&$\mathfrak{sp}(4)$&2&0 \\\hline
~&$I_0^*I_2^3$ & $(A^4BC)(A^2)(B^2)(X_{[0,1]})^2$& $\mathfrak{su}(2)$&1&$Z_2$  \\ \hline
~& $I_0^*I_4I_1^2$& $(A^4BC)(A^4)BC$&$\mathfrak{su}(2)$&1&$Z_2\times Z_2$ \\\hline
~& $I_0^* I_1^6$ & $(A_4BC)A^4BC$& $\mathfrak{f}_4$ &0&0\\ \hline 
\end{tabular}}
\end{center}
\caption{Physical data for 6d KK theory.}
\label{6dfull}
\end{table}

\textbf{Algebra on the branes}:  Let's look at $I_1^{12}$ theory, and the UV theory is  the famous 6d $E_8$ $(1,0)$ theory. This theory has $E_8$ flavor symmetry, and has the brane configuration $A^8BCBC$.
The rank of the charge lattice is $12$, and the intersection form can be found from the string junctions. The charge lattice has two imaginary roots $\delta_1, \delta_2$, which are represented by 
the $(1,0)$ and $(0,1)$ string loop around the brane configurations \cite{DeWolfe:1998pr}. The resulting algebra seems quite interesting.

\textbf{Low energy theory at the singularities}: One can merge various un-deformable singularities to get new singularities on the Coulomb branch, and the low energy theory can be found 
by looking at the table. \ref{4dfull} and \ref{simple}.  Let's look at the theory whose brane configuration is $A^8BCBC$ (6d $E_8$ $(1,0)$ theory), and configuration with maximal singular point is $(A^8BC)BC$, 
and so there is a $I_4^*$ singularity, and the low energy theory is $SU(2)$ gauge theory coupled with $N_f=8$ fundamental flavors. 

\newpage
\section{Conclusion}
In this paper, we classify rank one 5d $\mathcal{N}=1$ and 6d $(1,0)$ SCFT by using the effective 4d Coulomb branch solution. 
Each such solution is described by a mixed Hodge module (MHM), and one get a rational elliptic surface by looking at weight one part of MHM. Geometric results about rational elliptic surface such as the classification, Mordell-Weil lattice, 
Weierstrass  model are quite useful in studying these theories. 

In the process of classification, the following physical constraints are used:  a): For 4d theory, the fiber $F_\infty=I_n^*, II^*, III^*, IV^*$; b): For 5d theory,
the fiber $F_\infty= I_n,~n\geq 1$; c) For 6d theory, the fiber $F_\infty= I_0$; The generic deformation condition ensures that $II, III, IV$ singularity can not appear at the bulk; Finally, 
the Dirac quantization condition put constraints on the combination of $I_n$ type fibers. The data base of rational elliptic surface \cite{persson1990configurations,miranda1990persson} and  the base change maps \cite{karayayla2012classification}
are crucial for the classification. 

We also find the D7 brane configurations and string junctions are very useful in studying these theories, in particular, the low energy theory at \textbf{every} vacua can be determined. In this paper, we also showed how various physical 
properties of these theories can be understood using D7 brane configurations. It would be interesting to further study those theories.

The approach adopted in this paper can be generalized to the higher rank case theory (4d $\mathcal{N}=2$, 5d $\mathcal{N}=1$ KK and 6d $(1,0)$ KK theories all included). The analysis is much more complicated than rank one theory as there are 
several technical problems that one need to overcome:
\begin{enumerate}
\item \textbf{Local singularity}: The first step would be to study the local singularities and the associated physics. Unlike rank one case where there are only eight classes of singularities, the type of singular fibers 
increased increased greatly. For example, there are 126 types \cite{namikawa1973complete} for genus two singular fiber. One need to analyze the low energy theory associated with those singular fibers.

\item \textbf{Global constraints}:  The map between rank one Coulomb branch solution and rational elliptic surface is crucial for  the classification as the data base for such surfaces is available. Such classification for higher rank case is not available.

\item \textbf{Base change}:  Another important ingredients for rank one Coulomb branch solution is the classification of base change map of rational elliptic surface. Such classification is also not available for higher rank case. 
\end{enumerate}
To complete the classification for higher rank case, one need to solve above three problems. Those problems are largely solved for rank two case by the author, and the result will appear in \cite{danxiegenustwo}, (see \cite{Argyres:2022lah, Argyres:2022puv,Argyres:2022fwy} for the attempt in classifying 4d rank two superconformal theories). 

\section*{Acklowledgement}
DX would like to thank W.B Yan, D.X Zhang, and Jie Zhou for helpful discussions. DX is supported by Yau mathematical science center at Tsinghua University. 

\newpage
\appendix
\section{Hodge structure and Mixed Hodge structure}
Let's first review the Hodge structure and variation of Hodge structure.  By definition, a Hodge structure of weight \(n\) consists
of a lattice \(H_{\mathbb{Z}}\) ($\mathbb{Z}$ module) together with a decomposition:
\begin{equation*}
H_{\mathbb{C}} := H_{\mathbb{Z}} \otimes \mathbb{C} = \bigoplus_{p+q=n} H^{p,q},
\end{equation*}
such that \begin{equation}
H^{p,q}=\overline{H^{q,p}}.
\end{equation}
The Hodge structure can be reformulated by a pair $(H_{\mathbb{Z}}, F^\cdot)$, here $F^{\cdot}$ is a decreasing filtration
\begin{equation}
F^0\supset F^1\supset \ldots \supset F^{n-1}\supset F^n
\end{equation}
satisfying the condition $H_C=F^p \oplus \overline{F^{n-p+1}}$ for  every $p$. The Hodge subspaces are given as $H^{p,q}= F^p \cap \overline{F^q}$.  One also need a polarization $S$ on $H_Q$ such
that it satisfies following condition
\begin{align}
& S(F^p,F^{n-p+1})=0,~~\nonumber\\
& S(C\psi, \bar{\psi})>0~~\text{for}~~\psi\neq 0
\end{align}
Here $C$ is the Weil operator: $C|_{H^{p,q}}=i^{p-q}$.

\textbf{Example 1}: The most basic example is the Hodge structure of weight one (which is relevant for the cohomology of the compact Riemann surface $\Sigma$):  Here we have $H_Z=H^1(
\Sigma, \mathbb{Z})$, and the Hodge decomposition is given by $H^1(\Sigma, \mathbb{C})=H^{1,0}\oplus H^{0,1}$, where $H^{1,0}$ is given by holomorphic differential. The polarization is 
induced by cup product and the condition for the polarization is just the Riemann bilinear relation.

We could consider a family of Hodge structures, and this will lead to the definition of variation of Hodge structure. 
A \textbf{variation} of  Hodge
structure on a complex manifold \(B\) then consists of
\begin{itemize}
\item a local system \(\mathbb{V}\) of \(\mathbb{Q}\)-vector spaces, whose
  associated vector bundle with connection is denoted by \((\mathcal{V},\nabla)\);
\item holomorphic sub-bundles \(F^{\bullet}\mathcal{V} \subset \mathcal{V}\).
\end{itemize}
These data satisfy the following requirements:
\begin{itemize}
\item the infinitesimal period relation holds:
  \(\nabla_{v}F^{i}\mathcal{V}\subset F^{i-1}\mathcal{V}\);
\item for each \(b \in B\), the triple
  \((\mathbb{V}_b, F_{k,b})\) is a  Hodge structure.
\end{itemize}
 A \emph{polarization} of a variation of Hodge
structure \((\mathbb{V},F^{\bullet})\) on a complex manifold \(B\) is a
horizontal map
\begin{equation*}
S: \mathbb{V} \otimes \mathbb{V} \to \mathbb{C}_B
\end{equation*}
that induces a polarization on each fiber. In this case, we get a
variation of polarizable Hodge structure.

\textbf{Example 2}: The basic example for variation of Hodge structure is to consider a family of Riemann surface. Here is an example: $y^2=x(x-1)(x-t)$, here $t$ parameterizes the base manifold $B$.

In the case of Riemann surface, let's fix a point of homology cycles $\gamma_A, A=1,\ldots, g; \gamma_B, B=1,\ldots, g$ which is locally constant. 
The intersection form is chosen as $\gamma_A\cdot \gamma_B=\delta_{AB}$.
We then choose a basis of section of holomorphic one forms $\omega_1,\ldots, \omega_g$, and form the period integral
\begin{equation*}
\int_{\gamma_A} \omega_i,~~~\int_{\gamma_B} \omega_i,
\end{equation*}
each integral is called period and we get a $g\times 2g$ matrix called period matrix.

\textbf{Mixed Hodge structure}: The MHS is a far-reaching generalization of HS and was defined by Deligne. MHS consists of a triple $(H_{\mathbb{Z}}, F^\bullet,W_{\bullet})$, here $F^\bullet$ is an increasing filtration called Hodge filtration, 
and $W_{\bullet}$ is a decreasing filtration called the weight filtration.  The Hodge filtration is such that the the induced filtration on the quotient space $Gr_kH=W_k/W_{k-1}$ defines a weight $k$ Hodge structure. 
We also need a polarization so that its restriction on each quotient space gives a polarized Hodge structure. 

\textbf{Example 3}: There are two sets of basic examples for MHS. The first one is the smooth plane algebraic curve $C$ which is defined as: $f(x,y)=0$.  $C$ could be described by removing several points $p_i$ on a compact  Riemann surface $\Sigma$.
The first cohomology group $H^1(C, \mathbb{Z})$ admits a MHS: it has a weight one part and a weight two part. 
 which is essentially given by the first homology group of the compact Riemann surface $\Sigma$, and weight two part is given by the homology group associated with point $p_i$. The second one is the singular plane algebraic curve, and 
 in this case the MHS on $H^{1}(C,\mathbb{Z})$ consists of weight zero, weight one and weight two part.

\section{Variation of weight one Hodge structure}
In this section, we review the details of variation of weight one Hodge structure.  Let $H_Z$ be a lattice of rank $2g$, $Q$ a non-degenerate, skew symmetric 
bilinear form on $H_Z$, and $H=H_Z\otimes C$. A polarized Hodge structure of weight one is a decomposition of the complex vector space
\begin{equation*}
H=H^{1,0}\oplus H^{0,1},~~H^{0,1}=\overline{H^{1,0}}
\end{equation*}
satisfying the bilinear relations
\begin{enumerate}
\item $Q(u,u)=0$~if~$u\in H^{1,0}$, and
\item $iQ(u,\overline{u})>0$~if~$0\neq u \in H^{1,0}$
\end{enumerate}
The Hodge structure is equivalent to a Hodge filtration, and in this case, the Hodge filtration takes the form
\begin{equation*}
H=F^0\supset F^1\supset \{0\},~~F^1=H^{1,0}
\end{equation*}
Here $F^1$ is subspace of $H$.
The classifying space of all polarized Hodge structures of weight one on $H$ is then
\begin{equation*}
D=\{F^{1} \in G(g,H):~Q(F^1, F^1)=0;~iQ(F^1,\overline{F^1})>0\}
\end{equation*}
The subvariety of $G(g,H)$ consists of the maximal Q-isotropic subspaces of $H$ is the compact dual $\hat{D}$ of $D$.

Let's choose the symplectic basis, i.e. a rational basis ${\cal E}=\{e_1,\ldots, e_g,f_1,\ldots, f_g\}$ relative to which 
\begin{equation*}
Q=\left[\begin{array}{cc}
0&$I$ \\
$-I$&0
\end{array}\right]
\end{equation*}
Using this basis, we get the usual realization fo $D$ as Siegel's upper-half space. 

Let $F^1\in D$ and $w_1,\ldots, w_g$ a basis of $F^1$. Using the basis ${\cal E}$, we get a $g\times 2g$ matrix $\Omega=[\begin{array}{cc} \Omega_1 & \Omega_2 \end{array}]$. 
Here the row is the coefficients in the expansion of the basis of ${\cal E}$. The bilinear relations ensure that $F^1$ has a unique basis of the form
\begin{equation*}
\omega_i=\sum_j z_{ji} e_j+f_i
\end{equation*}
The bilinear relations ensure that the coefficients $Z=(z_{ji})$ satisfy the condition
\begin{equation*}
Z^t=Z,~~~Im(Z)~~\text{is~positive~definite}
\end{equation*}
$Z$ is then called normalized period matrix. 

\textbf{Monodromy weight filtration}: We now consider a variation of polarized Hodge structure of weight one defined on a punctured disk $\Delta^*$. The corresponding period mapping is 
\begin{equation*}
\phi:\Delta^*\to \Gamma / D,~~~\Gamma=G_Z=Sp(g,Z)
\end{equation*}

Let $\tilde{\phi}:U\to D$ be a global lifting of $\phi$ to the upper half plane $u$, and $\gamma \in \Gamma$ is the Picard-Lefschetz transformation. Then 
\begin{equation*}
\tilde{\phi}(z+1)=\gamma \tilde{\phi}(z)
\end{equation*}
Here $\gamma$ is a quasi unipotent monodromy transformation. We now assume that $\gamma$ is unipotent without losing the generality.  Let $N=\log \gamma$, 
and $N$ is then a rational element in the Lie algebra
\begin{equation}
\mathfrak{g}_0=\{X\in Hom(H_R, H_R):~Q(X\cdot,\cdot)+Q(\cdot, X\cdot)=0\}
\end{equation}
By the monodromy theorem $N^2=0$, and we have 
\begin{equation*}
N=\gamma-I
\end{equation*}
The monodromy  weight filtration defined by $N$ is the filtration
\begin{equation*}
\{0\}\subset W_0 \subset W_1 \subset W_2=H_C
\end{equation*}
Given by 
\begin{equation*}
W_0=Im(N),~~~W_1=Ker(N)
\end{equation*}
Since $N\in \mathfrak{g}_0$ and $N^2=0$, so $W_0$ is an isotropic subspace (here we use the facts $Q(N x, y)+Q (x,Ny)=0$, so $Q(Nx,Ny)+Q(x,N^2y)=Q(Nx,Ny)=0$, and $W_0=\{Nx\}$;
$W_1$ is the Q-annihilator of $W_0$:
\begin{equation*}
W_1=W_0^{\perp}=\{u\in H:Q(u,v)=0~\text{for all $u\in W_0$} \}
\end{equation*}
here we use the fact $Q(Nx,y)=0$ if $Ny=0$, namely $y\in W_1$ if $y\in W_0^\perp$.

Let $\Omega$ be a maximal totally isotropic subspace of $H$, and ${\cal E}$ is a symplectic basis such that $W_0= Span\{e_1, \ldots, e_\nu\}$, 
and $\Omega=  Span\{e_1, \ldots, e_g\}$. In this basis, the nilpotent operator can be written as 
\begin{equation*}
N=[\begin{array}{cc}
0&\eta \\
0&0 
\end{array}],~~\eta=[\begin{array}{cc}
\eta_{11}&0 \\
0&0 
\end{array}]
\end{equation*}
Here $\eta_{11}$ is a $\nu \times \nu$ symmetric matrix.

\textbf{Nilpotent orbit theorem and limit Hodge filtration}:
We now state the nilpotent orbit theorem. With the period mapping $\phi, \tilde{\phi}, N$ as above. Let 
\begin{equation*}
\tilde{\psi}:~U\to \hat{D}
\end{equation*}
be the map $\tilde{\psi}(z)=e^{-zN}\tilde{\phi}(z)$. Since $\tilde{\psi}(z+1)=\tilde{\psi}(z)$, we get a map
$\psi:\Delta^*\to \hat{D}$ which is given by $\psi(t)=\tilde{\psi}({1\over 2\pi i} \log t)$.  The Nilpotent orbit theorem states that 
\begin{enumerate}
\item The map $\psi: \Delta^*\to \hat{D}$ has a removable singularity at the orgin.
\item Let $F_a=\psi(0)\in \hat{D}$, then for $Im(z)$ sufficiently large $exp(zN)F_a \in D$. 
\item Relative to a $G_R$ invariant distance in $D$, we have 
\begin{equation*}
d(exp(zN)F_a,\tilde{\phi}(z))=o(e^{-2\pi Im(z)}),~~as~Im(z)\to \infty
\end{equation*}
\end{enumerate}
The filtration $F_a$ is called limit Hodge filtration, and $\exp(zN)F_a$ is called Nilpotent orbit which gives 
an asymptotical expansion for the period mapping. The nilpotent orbit theorem implies that the normalized periods has the expansion
\begin{equation*}
Z(t)=W(t)+{1\over 2\pi i} (\log t )\eta
\end{equation*}
and $W(0)$ is a symmetric matrix which takes the following form
\begin{equation*}
W(0)=\left[\begin{array}{cc}
W_{11}&W_{12}\\
W_{12}^t & W_{22}\\
\end{array}\right]
\end{equation*}
Here $W_{22}$ is a $(g-\nu)\times (g-\nu)$ symmetric matrix with positive definite imaginary part.

It is possible to generalize the nilpotent orbit to several varieties. In fact, let 
\begin{equation*}
\phi:(\Delta^*)^r\to \Gamma/D
\end{equation*}
be the period map, and let $\gamma_1,\ldots, \gamma_r$ denote the corresponding Picard-Lefschetz transformations, and 
we assume they all to be unipotent. Let $N_i=log \gamma_i=\gamma_i-I$. These are commuting nilpotent elements of $g_Q$ and form the Nilpotent
 orbit theorem we have that the symmetric forms $Q(\cdot, N_i \cdot)$ are positive semidefinite. 
 
 Let $\sigma \in \mathfrak{g}_0$ be the monodromy cone 
 \begin{equation*}
 \sigma=\{\sum_{i=1}^r \lambda_i N_i:\lambda_i \in R; \lambda_i>0\}
 \end{equation*} 
 For any element $N\in \sigma$ we have 
 \begin{equation*}
 Ker(N)=\cap_{i=1}^r Ker(N_i)
 \end{equation*}
 By duality, we also have 
 \begin{equation*}
 Im(N)=\sum_{i=1}^r Im(N_i)
 \end{equation*}
 Therefore, all elements in the monodromy cone $\sigma$ define the same weight filtration.

\textbf{Limit mixed Hodge structure}:  The limiting Hodge structure $F_a=\psi(0)$ is an element of the dual space and 
does not, in general, define a polarized Hodge structure. But, together with the weight filtration $W(N)$, the triple $(H, F_a, W(N))$ defines a polarized
mixed Hodge structure:
\begin{enumerate}
\item The filtration $F_a$ defines a Hodge structure of weight $l$ on the graded quotient $Gr_l^{W(N)}=W_l(N)/W_{l-1}(N)$.
\item The Hodge structure induced by $F_a$ on $Gr_{l+1}^{W(N)},l\geq 0$ is polarized by the bilinear form
\begin{equation*}
Q_l=Q(\cdot,N^l\cdot),
\end{equation*}
\end{enumerate}

One can show above mixed Hodge structure using nilpotent orbit theorem. The important fact is that since $exp(zN)$ acts trivially on $Gr^{W(N)}$,
we may assume that $F_a\in D$. 

For $l=0$, the above fact is equivalent to $F_a^1\cap W_0(N)=0$, and this can be proven using bilinear relation.
 
 For $l=1$, the fact that $F_a$ defines a polarized Hodge structure on $Gr_1^{W(N)}$ is equivalent to the factorization
\begin{equation*}
W_1(N)=W_0(N)\oplus (F_a^1\cap W_1(N)) \oplus (\overline{F_a^1}\cap W_1(N))
\end{equation*}
if $f_1,f_2 \in F_a^1\cap W_1(N)$ are such that $f_1+\overline{f_2} \in W_0(N)$, then
\begin{equation*}
iQ(f_1,\overline{f_1})=iQ(f_1,\overline{f_1}+f_2)=0
\end{equation*} 
We use the fact that if $f_1\in W_1(N)$, then $f_1\perp W_0(N)$, and also the fact $Q(f_1, f_2)=0$ for $f_1, f_2 \in F_a^1$.  The above 
equation implies that $f_1=0$ and similarly $f_2=0$.  We can similarly prove the above direct sum decomposition.
 \section{Admissible variation of mixed Hodge structure}
 Let's now review the variation of mixed Hodge structure.  Basically, it involves a local system $V$, and a  Weight filtration $W$ and Hodge filtration ${\cal F}$. They satisfy the 
 condition
 \begin{enumerate}
 \item $V$ is a local system over the base manifold $S$.
 \item $W=\{W_k\}$ is an increasing weight filtration  of $V$ by local subsystems.
 \item $F=\{F^p\}$ is a decreasing filtration by holomorphic sub-bundles of $V\otimes \mathbb{C}$.
 \item $\Delta F^p\subset \Omega_S^1\otimes F^{p-1}$.
 \item The data $(Gr_k^WV, F(W_k/W_{k-1}))$ defines a weight $k$ variation of Hodge structure. This is equivalent to the fact that $(V_s, W_s, F_s)$ is a mixed Hodge structure at a point $s\in S$. 
 \end{enumerate}
If there is a polarization such that there is an induced polarization on graded part, then the structure is called graded polarizable variation of mixed Hodge structure.
 
We may also consider the limit of the variation of mixed Hodge structure. Similarly, we have the monodromy $\gamma$ of local system, which we also 
assume to be unipotent. The nilpotent part of the monodromy group is denoted as $N$. There is a definition of relative filtration $M$ of $N$ with respect to $W$. Given the pair $(W, N)$, the relative filtration $M$ satisfies following conditions:
\begin{enumerate}
\item $N(M_i)\subset M_{i-2}$
\item $M Gr_k^W=~N|_{Gr_k^W}$
\end{enumerate}
The second condition means that the induced filtration of $N$ on $Gr_k^W$ is the same as the induced filtration of $M$ on $Gr_k^W$.
The variation of MHS is called admissible if there is a relative filtration $M$ with respect the nilpotent monodromy $N$ and $W$.

\bibliographystyle{JHEP}
\bibliography{ADhigher}

\providecommand{\href}[2]{#2}\begingroup\raggedright\begin{thebibliography}{10}

\bibitem{Intriligator:1996ex}
K.~A. Intriligator and N.~Seiberg, {\it {Mirror symmetry in three-dimensional
  gauge theories}},  {\em Phys.Lett.} {\bf B387} (1996) 513--519,
  [\href{http://xxx.lanl.gov/abs/hep-th/9607207}{{\tt hep-th/9607207}}].

\bibitem{Seiberg:1994rs}
N.~Seiberg and E.~Witten, {\it {Electric - magnetic duality, monopole
  condensation, and confinement in N=2 supersymmetric Yang-Mills theory}},
  {\em Nucl. Phys.} {\bf B426} (1994) 19--52,
  [\href{http://xxx.lanl.gov/abs/hep-th/9407087}{{\tt hep-th/9407087}}].
  [Erratum: Nucl. Phys.B430,485(1994)].

\bibitem{Seiberg:1994aj}
N.~Seiberg and E.~Witten, {\it {Monopoles, duality and chiral symmetry breaking
  in N=2 supersymmetric QCD}},  {\em Nucl. Phys.} {\bf B431} (1994) 484--550,
  [\href{http://xxx.lanl.gov/abs/hep-th/9408099}{{\tt hep-th/9408099}}].

\bibitem{Seiberg:1996bd}
N.~Seiberg, {\it {Five-dimensional SUSY field theories, nontrivial fixed points
  and string dynamics}},  {\em Phys. Lett. B} {\bf 388} (1996) 753--760,
  [\href{http://xxx.lanl.gov/abs/hep-th/9608111}{{\tt hep-th/9608111}}].

\bibitem{Seiberg:1996qx}
N.~Seiberg, {\it {Nontrivial fixed points of the renormalization group in
  six-dimensions}},  {\em Phys. Lett. B} {\bf 390} (1997) 169--171,
  [\href{http://xxx.lanl.gov/abs/hep-th/9609161}{{\tt hep-th/9609161}}].

\bibitem{Seiberg:1997ax}
N.~Seiberg, {\it {Notes on theories with 16 supercharges}},  {\em Nucl. Phys. B
  Proc. Suppl.} {\bf 67} (1998) 158--171,
  [\href{http://xxx.lanl.gov/abs/hep-th/9705117}{{\tt hep-th/9705117}}].

\bibitem{Cordova:2016emh}
C.~Cordova, T.~T. Dumitrescu, and K.~Intriligator, {\it {Multiplets of
  Superconformal Symmetry in Diverse Dimensions}},  {\em JHEP} {\bf 03} (2019)
  163, [\href{http://xxx.lanl.gov/abs/1612.0080}{{\tt arXiv:1612.0080}}].

\bibitem{Nekrasov:1996cz}
N.~Nekrasov, {\it {Five dimensional gauge theories and relativistic integrable
  systems}},  {\em Nucl. Phys. B} {\bf 531} (1998) 323--344,
  [\href{http://xxx.lanl.gov/abs/hep-th/9609219}{{\tt hep-th/9609219}}].

\bibitem{Ganor:1996pc}
O.~J. Ganor, D.~R. Morrison, and N.~Seiberg, {\it {Branes, Calabi-Yau spaces,
  and toroidal compactification of the N=1 six-dimensional E(8) theory}},  {\em
  Nucl. Phys. B} {\bf 487} (1997) 93--127,
  [\href{http://xxx.lanl.gov/abs/hep-th/9610251}{{\tt hep-th/9610251}}].

\bibitem{Witten:1997sc}
E.~Witten, {\it {Solutions of four-dimensional field theories via M theory}},
  {\em Nucl. Phys. B} {\bf 500} (1997) 3--42,
  [\href{http://xxx.lanl.gov/abs/hep-th/9703166}{{\tt hep-th/9703166}}].

\bibitem{Martinec:1995by}
E.~J. Martinec and N.~P. Warner, {\it {Integrable systems and supersymmetric
  gauge theory}},  {\em Nucl. Phys. B} {\bf 459} (1996) 97--112,
  [\href{http://xxx.lanl.gov/abs/hep-th/9509161}{{\tt hep-th/9509161}}].

\bibitem{Donagi:1995cf}
R.~Donagi and E.~Witten, {\it {Supersymmetric Yang-Mills theory and integrable
  systems}},  {\em Nucl. Phys. B} {\bf 460} (1996) 299--334,
  [\href{http://xxx.lanl.gov/abs/hep-th/9510101}{{\tt hep-th/9510101}}].

\bibitem{Xie:2021hxd}
D.~Xie and D.~Zhang, {\it {Mixed Hodge structure and $\mathcal{N}=2$ Coulomb
  branch solution}},  \href{http://xxx.lanl.gov/abs/2107.1118}{{\tt
  arXiv:2107.1118}}.

\bibitem{Xie:2022}
D.~Xie, {\it {Mixed Hodge module and classification of theory with eight
  supercharges, Work in progress}}, .

\bibitem{saito1990mixed}
M.~Saito, {\it Mixed hodge modules},  {\em Publications of the Research
  Institute for Mathematical Sciences} {\bf 26} (1990), no.~2 221--333.

\bibitem{steenbrink1985variation}
J.~Steenbrink and S.~Zucker, {\it Variation of mixed hodge structure. i},  {\em
  Inventiones mathematicae} {\bf 80} (1985), no.~3 489--542.

\bibitem{barth2015compact}
W.~Barth, K.~Hulek, C.~Peters, and A.~Van~de Ven, {\em Compact complex
  surfaces}, vol.~4.
\newblock Springer, 2015.

\bibitem{carlson2017period}
J.~Carlson, S.~M{\"u}ller-Stach, and C.~Peters, {\em Period mappings and period
  domains}.
\newblock Cambridge University Press, 2017.

\bibitem{schutt2019elliptic}
M.~Sch{\"u}tt and T.~Shioda, {\em Mordell--Weil Lattices}.

\bibitem{persson1990configurations}
U.~Persson, {\it Configurations of kodaira fibers on rational elliptic
  surfaces},  {\em Mathematische Zeitschrift} {\bf 205} (1990), no.~1 1--47.

\bibitem{miranda1990persson}
R.~Miranda, {\it Persson's list of singular fibers for a rational elliptic
  surface},  {\em Mathematische Zeitschrift} {\bf 205} (1990), no.~1 191--211.

\bibitem{Caorsi:2018ahl}
M.~Caorsi and S.~Cecotti, {\it {Special Arithmetic of Flavor}},  {\em JHEP}
  {\bf 08} (2018) 057, [\href{http://xxx.lanl.gov/abs/1803.0053}{{\tt
  arXiv:1803.0053}}].

\bibitem{Argyres:2016yzz}
P.~C. Argyres and M.~Martone, {\it {4d $ \mathcal{N} $ =2 theories with
  disconnected gauge groups}},  {\em JHEP} {\bf 03} (2017) 145,
  [\href{http://xxx.lanl.gov/abs/1611.0860}{{\tt arXiv:1611.0860}}].

\bibitem{karayayla2012classification}
T.~Karayayla, {\it The classification of automorphism groups of rational
  elliptic surfaces with section},  {\em Advances in Mathematics} {\bf 230}
  (2012), no.~1 1--54.

\bibitem{DeWolfe:1998eu}
O.~DeWolfe, T.~Hauer, A.~Iqbal, and B.~Zwiebach, {\it {Uncovering the
  symmetries on [p,q] seven-branes: Beyond the Kodaira classification}},  {\em
  Adv. Theor. Math. Phys.} {\bf 3} (1999) 1785--1833,
  [\href{http://xxx.lanl.gov/abs/hep-th/9812028}{{\tt hep-th/9812028}}].

\bibitem{Fukae:1999zs}
M.~Fukae, Y.~Yamada, and S.-K. Yang, {\it {Mordell-Weil lattice via string
  junctions}},  {\em Nucl. Phys. B} {\bf 572} (2000) 71--94,
  [\href{http://xxx.lanl.gov/abs/hep-th/9909122}{{\tt hep-th/9909122}}].

\bibitem{Argyres:2015ffa}
P.~Argyres, M.~Lotito, Y.~L\"u, and M.~Martone, {\it {Geometric constraints on
  the space of $ \mathcal{N} $ = 2 SCFTs. Part I: physical constraints on
  relevant deformations}},  {\em JHEP} {\bf 02} (2018) 001,
  [\href{http://xxx.lanl.gov/abs/1505.0481}{{\tt arXiv:1505.0481}}].

\bibitem{Argyres:2015gha}
P.~C. Argyres, M.~Lotito, Y.~L\"u, and M.~Martone, {\it {Geometric constraints
  on the space of $ \mathcal{N} $ = 2 SCFTs. Part II: construction of special
  K\"ahler geometries and RG flows}},  {\em JHEP} {\bf 02} (2018) 002,
  [\href{http://xxx.lanl.gov/abs/1601.0001}{{\tt arXiv:1601.0001}}].

\bibitem{Argyres:2016xmc}
P.~Argyres, M.~Lotito, Y.~L\"u, and M.~Martone, {\it {Geometric constraints on
  the space of $ \mathcal{N}$ = 2 SCFTs. Part III: enhanced Coulomb branches
  and central charges}},  {\em JHEP} {\bf 02} (2018) 003,
  [\href{http://xxx.lanl.gov/abs/1609.0440}{{\tt arXiv:1609.0440}}].

\bibitem{Argyres:2016xua}
P.~C. Argyres, M.~Lotito, Y.~L\"u, and M.~Martone, {\it {Expanding the
  landscape of $ \mathcal{N} $ = 2 rank 1 SCFTs}},  {\em JHEP} {\bf 05} (2016)
  088, [\href{http://xxx.lanl.gov/abs/1602.0276}{{\tt arXiv:1602.0276}}].

\bibitem{Xie:2019vzr}
D.~Xie and W.~Yan, {\it {4d $\mathcal{N}=2$ SCFTs and lisse W-algebras}},  {\em
  JHEP} {\bf 04} (2021) 271, [\href{http://xxx.lanl.gov/abs/1910.0228}{{\tt
  arXiv:1910.0228}}].

\bibitem{Xie:2015rpa}
D.~Xie and S.-T. Yau, {\it {4d N=2 SCFT and singularity theory Part I:
  Classification}},  \href{http://xxx.lanl.gov/abs/1510.0132}{{\tt
  arXiv:1510.0132}}.

\bibitem{kulikov1998mixed}
V.~S. Kulikov, {\em Mixed Hodge structures and singularities}.
\newblock No.~132. Cambridge University Press, 1998.

\bibitem{Xie:2015xva}
D.~Xie and S.-T. Yau, {\it {Semicontinuity of 4d N=2 spectrum under
  renormalization group flow}},  {\em JHEP} {\bf 03} (2016) 094,
  [\href{http://xxx.lanl.gov/abs/1510.0603}{{\tt arXiv:1510.0603}}].

\bibitem{miranda}
R.~Miranda, {\it {The basic theory of elliptic
  surfaces,https://www.math.colostate.edu/~miranda/BTES-Miranda.pdf}}, .

\bibitem{Tachikawa:2015wka}
Y.~Tachikawa, {\it {Frozen singularities in M and F theory}},  {\em JHEP} {\bf
  06} (2016) 128, [\href{http://xxx.lanl.gov/abs/1508.0667}{{\tt
  arXiv:1508.0667}}].

\bibitem{Banks:1996nj}
T.~Banks, M.~R. Douglas, and N.~Seiberg, {\it {Probing F theory with branes}},
  {\em Phys. Lett. B} {\bf 387} (1996) 278--281,
  [\href{http://xxx.lanl.gov/abs/hep-th/9605199}{{\tt hep-th/9605199}}].

\bibitem{Landsteiner:1997vd}
K.~Landsteiner, E.~Lopez, and D.~A. Lowe, {\it {N=2 supersymmetric gauge
  theories, branes and orientifolds}},  {\em Nucl. Phys. B} {\bf 507} (1997)
  197--226, [\href{http://xxx.lanl.gov/abs/hep-th/9705199}{{\tt
  hep-th/9705199}}].

\bibitem{Hori:1998iv}
K.~Hori, {\it {Consistency condition for five-brane in M theory on R**5 / Z(2)
  orbifold}},  {\em Nucl. Phys. B} {\bf 539} (1999) 35--78,
  [\href{http://xxx.lanl.gov/abs/hep-th/9805141}{{\tt hep-th/9805141}}].

\bibitem{Xie:2012hs}
D.~Xie, {\it {General Argyres-Douglas Theory}},  {\em JHEP} {\bf 1301} (2013)
  100, [\href{http://xxx.lanl.gov/abs/1204.2270}{{\tt arXiv:1204.2270}}].

\bibitem{Xie:2017obm}
D.~Xie, {\it {$\mathcal{N}=2$ SCFT with minimal flavor central charge}},
  \href{http://xxx.lanl.gov/abs/1712.0324}{{\tt arXiv:1712.0324}}.

\bibitem{Caorsi:2019vex}
M.~Caorsi and S.~Cecotti, {\it {Homological classification of 4d $ \mathcal{N}
  $ = 2 QFT. Rank-1 revisited}},  {\em JHEP} {\bf 10} (2019) 013,
  [\href{http://xxx.lanl.gov/abs/1906.0391}{{\tt arXiv:1906.0391}}].

\bibitem{Argyres:2018taw}
P.~C. Argyres and M.~Lotito, {\it {Flavor symmetries and the topology of
  special K\"ahler structures at rank 1}},  {\em JHEP} {\bf 02} (2019) 026,
  [\href{http://xxx.lanl.gov/abs/1811.0001}{{\tt arXiv:1811.0001}}].

\bibitem{Gaiotto:2014kfa}
D.~Gaiotto, A.~Kapustin, N.~Seiberg, and B.~Willett, {\it {Generalized Global
  Symmetries}},  {\em JHEP} {\bf 02} (2015) 172,
  [\href{http://xxx.lanl.gov/abs/1412.5148}{{\tt arXiv:1412.5148}}].

\bibitem{Gaiotto:2009hg}
D.~Gaiotto, G.~W. Moore, and A.~Neitzke, {\it {Wall-crossing, Hitchin Systems,
  and the WKB Approximation}},  \href{http://xxx.lanl.gov/abs/0907.3987}{{\tt
  arXiv:0907.3987}}.

\bibitem{Aharony:2013hda}
O.~Aharony, N.~Seiberg, and Y.~Tachikawa, {\it {Reading between the lines of
  four-dimensional gauge theories}},  {\em JHEP} {\bf 08} (2013) 115,
  [\href{http://xxx.lanl.gov/abs/1305.0318}{{\tt arXiv:1305.0318}}].

\bibitem{Xie:2013vfa}
D.~Xie, {\it {Aspects of line operators of class S theories}},
  \href{http://xxx.lanl.gov/abs/1312.3371}{{\tt arXiv:1312.3371}}.

\bibitem{Minahan:1996cj}
J.~A. Minahan and D.~Nemeschansky, {\it {Superconformal fixed points with E(n)
  global symmetry}},  {\em Nucl. Phys. B} {\bf 489} (1997) 24--46,
  [\href{http://xxx.lanl.gov/abs/hep-th/9610076}{{\tt hep-th/9610076}}].

\bibitem{Argyres:1995xn}
P.~C. Argyres, M.~R. Plesser, N.~Seiberg, and E.~Witten, {\it {New N=2
  superconformal field theories in four-dimensions}},  {\em Nucl. Phys. B} {\bf
  461} (1996) 71--84, [\href{http://xxx.lanl.gov/abs/hep-th/9511154}{{\tt
  hep-th/9511154}}].

\bibitem{DeWolfe:1998bi}
O.~DeWolfe, T.~Hauer, A.~Iqbal, and B.~Zwiebach, {\it {Constraints on the BPS
  spectrum of N=2, D = 4 theories with A-D-E flavor symmetry}},  {\em Nucl.
  Phys. B} {\bf 534} (1998) 261--274,
  [\href{http://xxx.lanl.gov/abs/hep-th/9805220}{{\tt hep-th/9805220}}].

\bibitem{DeWolfe:1998pr}
O.~DeWolfe, T.~Hauer, A.~Iqbal, and B.~Zwiebach, {\it {Uncovering infinite
  symmetries on [p, q] 7-branes: Kac-Moody algebras and beyond}},  {\em Adv.
  Theor. Math. Phys.} {\bf 3} (1999) 1835--1891,
  [\href{http://xxx.lanl.gov/abs/hep-th/9812209}{{\tt hep-th/9812209}}].

\bibitem{Sen:1996sk}
A.~Sen, {\it {BPS states on a three-brane probe}},  {\em Phys. Rev. D} {\bf 55}
  (1997) 2501--2503, [\href{http://xxx.lanl.gov/abs/hep-th/9608005}{{\tt
  hep-th/9608005}}].

\bibitem{Mikhailov:1998bx}
A.~Mikhailov, N.~Nekrasov, and S.~Sethi, {\it {Geometric realizations of BPS
  states in N=2 theories}},  {\em Nucl. Phys. B} {\bf 531} (1998) 345--362,
  [\href{http://xxx.lanl.gov/abs/hep-th/9803142}{{\tt hep-th/9803142}}].

\bibitem{Alim:2011kw}
M.~Alim, S.~Cecotti, C.~Cordova, S.~Espahbodi, A.~Rastogi, and C.~Vafa, {\it
  {$\mathcal{N} = 2$ quantum field theories and their BPS quivers}},  {\em Adv.
  Theor. Math. Phys.} {\bf 18} (2014), no.~1 27--127,
  [\href{http://xxx.lanl.gov/abs/1112.3984}{{\tt arXiv:1112.3984}}].

\bibitem{Xie:2012gd}
D.~Xie, {\it {BPS spectrum, wall crossing and quantum dilogarithm identity}},
  {\em Adv. Theor. Math. Phys.} {\bf 20} (2016) 405--524,
  [\href{http://xxx.lanl.gov/abs/1211.7071}{{\tt arXiv:1211.7071}}].

\bibitem{DeWolfe:1998zf}
O.~DeWolfe and B.~Zwiebach, {\it {String junctions for arbitrary Lie algebra
  representations}},  {\em Nucl. Phys. B} {\bf 541} (1999) 509--565,
  [\href{http://xxx.lanl.gov/abs/hep-th/9804210}{{\tt hep-th/9804210}}].

\bibitem{Xie:2017pfl}
D.~Xie and S.-T. Yau, {\it {Three dimensional canonical singularity and five
  dimensional $ \mathcal{N} $ = 1 SCFT}},  {\em JHEP} {\bf 06} (2017) 134,
  [\href{http://xxx.lanl.gov/abs/1704.0079}{{\tt arXiv:1704.0079}}].

\bibitem{Acharya:2021jsp}
B.~Acharya, N.~Lambert, M.~Najjar, E.~E. Svanes, and J.~Tian, {\it {Gauging
  discrete symmetries of T$_{N}$-theories in five dimensions}},  {\em JHEP}
  {\bf 04} (2022) 114, [\href{http://xxx.lanl.gov/abs/2110.1444}{{\tt
  arXiv:2110.1444}}].

\bibitem{Kim:2021fxx}
H.-C. Kim, S.-S. Kim, and K.~Lee, {\it {S-foldings of 5d SCFTs}},  {\em JHEP}
  {\bf 05} (2022) 178, [\href{http://xxx.lanl.gov/abs/2112.1455}{{\tt
  arXiv:2112.1455}}].

\bibitem{Yamada:1999xr}
Y.~Yamada and S.-K. Yang, {\it {Affine seven-brane backgrounds and
  five-dimensional E(N) theories on S**1}},  {\em Nucl. Phys. B} {\bf 566}
  (2000) 642--660, [\href{http://xxx.lanl.gov/abs/hep-th/9907134}{{\tt
  hep-th/9907134}}].

\bibitem{Morrison:1996xf}
D.~R. Morrison and N.~Seiberg, {\it {Extremal transitions and five-dimensional
  supersymmetric field theories}},  {\em Nucl. Phys. B} {\bf 483} (1997)
  229--247, [\href{http://xxx.lanl.gov/abs/hep-th/9609070}{{\tt
  hep-th/9609070}}].

\bibitem{Hanany:2001py}
A.~Hanany and A.~Iqbal, {\it {Quiver theories from D6 branes via mirror
  symmetry}},  {\em JHEP} {\bf 04} (2002) 009,
  [\href{http://xxx.lanl.gov/abs/hep-th/0108137}{{\tt hep-th/0108137}}].

\bibitem{Closset:2019juk}
C.~Closset and M.~Del~Zotto, {\it {On 5d SCFTs and their BPS quivers. Part I:
  B-branes and brane tilings}},  \href{http://xxx.lanl.gov/abs/1912.1350}{{\tt
  arXiv:1912.1350}}.

\bibitem{Hauer:1999pt}
T.~Hauer and A.~Iqbal, {\it {Del Pezzo surfaces and affine seven-brane
  backgrounds}},  {\em JHEP} {\bf 01} (2000) 043,
  [\href{http://xxx.lanl.gov/abs/hep-th/9910054}{{\tt hep-th/9910054}}].

\bibitem{namikawa1973complete}
Y.~Namikawa and K.~Ueno, {\it The complete classification of fibres in pencils
  of curves of genus two},  {\em Manuscripta mathematica} {\bf 9} (1973), no.~2
  143--186.

\bibitem{danxiegenustwo}
D.~Xie, {\it {On rank two theories with eight supercharges, to appear}}, .

\bibitem{Argyres:2022lah}
P.~C. Argyres and M.~Martone, {\it {The rank 2 classification problem I: scale
  invariant geometries}},  \href{http://xxx.lanl.gov/abs/2209.0924}{{\tt
  arXiv:2209.0924}}.

\bibitem{Argyres:2022puv}
P.~C. Argyres and M.~Martone, {\it {The rank 2 classification problem II:
  mapping scale-invariant solutions to SCFTs}},
  \href{http://xxx.lanl.gov/abs/2209.0991}{{\tt arXiv:2209.0991}}.

\bibitem{Argyres:2022fwy}
P.~C. Argyres and M.~Martone, {\it {The rank-2 classification problem III:
  curves with additional automorphisms}},
  \href{http://xxx.lanl.gov/abs/2209.1055}{{\tt arXiv:2209.1055}}.

\end{thebibliography}\endgroup

\end{document}